\documentclass[a4paper,1za1pt]{article} 

\pdfoutput=1 

\usepackage{jcappub}
\usepackage{pgfplots}

\usepackage[normalem]{ulem}
\usepackage{graphicx,bm}
\usepackage{graphicx}
\usepackage{dcolumn}
\usepackage{bm}
\usepackage{hyperref}
\usepackage{multirow} 
\usepackage{amssymb}
\usepackage{amsmath}
\usepackage{verbatim}
\usepackage{mathrsfs}
\usepackage{amsfonts}
\usepackage{latexsym}
\usepackage{epsfig}
\usepackage{color}
\usepackage{graphicx,subfigure}
\usepackage{units}
\usepackage{overpic}
\usepackage{enumerate}

\usepackage{color}

\begin{document}
\title{Horndeski stars}

\author[1]{Juan Barranco,}
\author[2]{Javier Chagoya,}
\author[1]{Alberto Diez-Tejedor,}
\author[1]{Gustavo Niz,}
\author[1]{Armando A. Roque}

\affiliation[1]{Departamento de Física, División de Ciencias e Ingenierías, Campus León, Universidad de Guanajuato, C.P. 37150, León, México}
\affiliation[2]{Unidad Acad\'emica de F\'{\i}sica, Universidad Aut\'onoma de Zacatecas, Calzada Solidaridad esquina con Paseo a la Bufa S/N, C.P. 98060, Zacatecas, M\'exico}


\abstract{ We establish the existence of time-dependent solitons in a modified gravity framework, which is defined by the low energy limit of theories with a weakly broken galileon symmetry and a mass term. These are regular vacuum configurations of finite energy characterized by a single continuous parameter representing the amplitude of the scalar degree of freedom at the origin. When the central field amplitude is small the objects are indistinguishable from boson stars. In contrast, increasing the central value of the amplitude triggers the effect of higher derivative operators in the effective theory, leading to departures from the previous solutions, until the theory becomes strongly coupled and model-dependent. The higher order operators are part of the (beyond) Horndeski theory, hence the name of the compact objects. Moreover, a remnant of the galileon non-renormalization theorem guarantees that the existence and properties of these solutions are not affected by quantum corrections. Finally, we discuss the linear stability under small radial perturbations, the mass-radius relation, the compactness, the appearance of innermost stable circular orbits and photon spheres, and some astrophysical signatures (accretion disks, gravitational radiation and lensing) that may be relevant to falsify the model.}
	
\maketitle

\section{\label{sec:level1} Introduction}

The gravitational understanding of compact objects, such as black holes and neutron stars, has received increased attention since LIGO's first gravitational wave (GW) detection in 2016~\cite{Abbott:2016blz}. By the end of the first-half of the third observing run the LIGO-VIRGO collaboration has confirmed 39 merger events \cite{Abbott:2020niy}, and with the following major instrumental upgrade it is likely that this number will substantially increase in the upcoming years. The foreseeable future of new interferometers in a network with higher instrumental sensitivity, and a larger observing sample, will provide a unique opportunity for observing GW signatures of a large variety of astrophysical objects, including those not formed by standard matter and/or not predicted within the framework of General Relativity (GR)~\cite{Cardoso:2019rvt, Abbott:2020jks}. 

The motivations to test gravity are vast, from the simple idea of constraining the theories that we believe represent the fundamental laws of nature, to the incomplete picture of the dark sector in cosmology or the fact that GR needs an ultraviolet (UV) completion if canonical quantization is assumed correct. The space of alternative theories of gravity is infinite, however, one can use Lovelock's theorem assumptions~\cite{lovelock,Lovelock:1972vz} to classify GR extensions~\cite{Clifton:2011jh,Joyce:2014kja,Berti:2015itd,Heisenberg:2018vsk}. According to this theorem GR is the only local and Lorentz invariant theory of a non-trivially interacting massless helicity two particle in four dimensions (see~\cite{Gupta:1954zz,Kraichnan:1955zz,Weinberg:1965rz,Deser:1969wk,Boulware:1974sr,Feynman:1996kb,Weinberg:1995mt} for further details on the theorem). Using extra degrees of freedom is one of such branches of alternative theories, and the simplest case is to add an additional scalar mediator apart from the usual spin two field. In this context, the most general local and Lorentz invariant scalar-tensor theory with second order equations of motion is that of Horndeski's proposal~\cite{Horndeski:1974wa, Deffayet:horndeski, Deffayet:horndeski2}. Furthermore, if we relax the condition on the order of the equations of motion it is possible to extend this formulation to the Degenerate Higher Order Scalar Tensor (DHOST) theory~\cite{Langlois:2015cwa,Crisostomi:2016czh}, which in spite of the higher order derivatives does not propagate an additional (ghostly) scalar degree of freedom. A particular subset of the DHOST theory is the beyond Horndeski theory~\cite{Zumalacarregui:2013pma}, also known as the Gleyzes-Langlois-Piazza-Vernizzi (GLPV)  theory~\cite{Gleyzes:2014dya,Gleyzes:2014qga}, which we set as our departure point in this paper for reasons that we clarify later. This theory has six arbitrary functions of the scalar field and its first derivatives (contracted with the spacetime metric to provide a scalar), hence a rich phenomenology, and incorporates any of the known screening mechanisms~\cite{Brax:2013ida, Joyce:2014kja} which help to recover GR predictions at Solar System scales, where there are strong post-Newtonian constraints \cite{Will_2014}. The Vainshtein screening mechanism~\cite{Vainshtein:1972sx}, driven by derivative self-couplings~\cite{Kimura:2011dc, Narikawa:2013pjr, Koyama:2013paa}, has attracted recent attention on these models due to its interesting phenomenology, such as the relation with consistent non-linear massive gravity theories~\cite{Rubakov:2008nh,deRham:2014zqa,Hinterbichler:2011tt}, higher dimensional braneworlds~\cite{Dvali:2000hr}, and strong-coupled behavior~\cite{Dvali:2006su}.

\begin{figure}[t]
\centering
	\scalebox{0.65}{
	\input{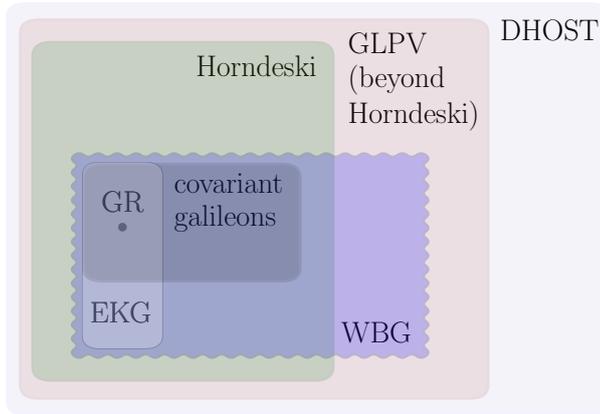}
	}  
	\caption{{\bf The landscape of scalar-tensor theories.} We focus on models where the weakly broken galileon (WBG) symmetry holds, purple area, leading to an EFT with higher derivative operators that is robust under quantum corrections. The EFT includes the covariant galileons, and reduces to what we call the Einstein-Klein-Gordon (EKG) model in the regime of weak coupling. General Gelativity (GR) is sketched as a single point in this figure, where only the two helicities of the massless tensor field propagate. See Section~\ref{sect2} for further details on this classification.}\label{Fig.Landscape}
\end{figure}

The GLPV theory encompasses a series of models that are, in general, non-renormalizable, and must be understood as a low energy effective field theory (EFT)~\cite{Georgi:1994qn,Pich:1998xt,Burgess:2003jk,Kaplan:2005es,Manohar:2018aog,Cohen:2019wxr,Burgess:2020tbq,Penco:2020kvy}. An interesting property of this theory is the possibility to self-consistently include higher derivative operators which do not appear in simpler realizations, such as e.g. the Brans-Dicke model~\cite{Brans:1961sx} or its most natural extensions~\cite{Fujii:2003pa,Capozziello:2010zz}. However, these operators are of mass dimension greater than four, and from the viewpoint of a quantum theory they typically show up when the derivative expansion is expected to break down and the EFT loses predictability. There are noticeable exceptions, the most prominent probably being the effective theory of a galileon scalar field~\cite{Nicolis:2008in}, where an internal symmetry $\phi\to\phi+b_\mu x^\mu + c$ protects the higher derivative operators from loop corrections and the previous statement does not necessarily hold true~\cite{Luty:2003vm, Goon:2016ihr}. Once we move to curved spaces, the couplings of the scalar with gravitons break the invariance under galileon transformations, nonetheless, there is a subset of the GLPV theory where the breakdown is weak (in a sense that we make more explicit in what follows) and the theory preserves as much as possible the quantum properties of the galileon~\cite{Pirtskhalava:2015nla,Santoni:2018rrx}; see Figure~\ref{Fig.Landscape} for illustrative purposes. The galilean structure limits the form of the arbitrary functions of the EFT, as we discuss in more detail in the next section, but still remains a rich phenomenology.

The idea of using compact objects to test modified gravity is not new. Analytical and numerical black hole and neutron star solutions have been reported before in the literature, either in GR (see e.g.~\cite{Shapiro:1983du} and references therein), or in alternative theories of gravity~\cite{damour2007binary,Sakstein:2015oqa,Chu:2012kz,deRham:2012fw,Babichev:2016rlq}. Neutron stars are regular configurations where a degenerate neutron gas sources gravity, defined by the pure metric tensor in GR or also by additional degrees of freedom in extended models.  Black holes also source the gravitational field(s), although they are vacuum singular objects.\footnote{Even if vacuum, in most astrophysical scenarios the black hole's singularity (which is expected to be softened in a more fundamental theory) appears as a result of the gravitational collapse of a progenitor star.} In this work, instead, we focus on solutions that are {\it vacuum} and {\it regular}, and that may coexist with black holes and neutron stars in certain gravity models. In particular, we are interested in solitary wave solutions~\cite{Rajaraman:1982is}, that is, regular field configurations that are bounded in space and maintain their shape (that depends harmonically on time in some cases) while they propagate at constant velocity, or remain at rest. The existence of these objects, also known for historical reasons as geons, i.e. gravitational electromagnetic entities, was proposed in the mid fifties in a pioneering work by Wheeler~\cite{Wheeler:1955zz}, and studied later in a systematic way by Brill, Hartle, Anderson and others~\cite{Brill:1964zz,Anderson:1996pu}. In GR they are introduced as (approximate) solutions to the vacuum Einstein's equations, but they turn out to be unstable, both to the leakage of GWs and to radial perturbations~\cite{Perry:1998hh}, so their potential impact phenomenological is rather limited. If the inclusion of additional gravitational modes made them possible, however, they could be a distinctive feature of modified gravity.

Any progress on these constructions must confront a {\it no-go} result which, under the name of Derrick's theorem \cite{Derrick:1964ww}, loosely states that there are no static and localized scalar field configurations, or in the event of existing (if the energy density is allowed to take negative values) they are unstable. Non-canonical kinetic terms might modify these conclusions,\footnote{Derrick's theorem assumes a canonical scalar field, where the Lagrangian density is given as the sum of the kinetic scalar $X=\partial_\mu\phi\partial^{\mu}\phi$ and a potential term $V(\phi)$. In addition it also assumes a flat spacetime background, although it can be generalized to the selfgravitating regime~\cite{Diez-Tejedor:2013sza,Hod:2018dij}.} and it is not difficult to convince oneself that they can circumvent the scaling argument that supports Derrick's theorem~\cite{Endlich:2010zj}. A well known example is that of the skyrmions~\cite{Skyrme:1962vh}, stable field configurations that exist in certain classes of non-linear sigma models, although they raise the natural question of whether new operators induced by loop corrections cannot affect their existence. In general, a UV completion is necessary to address this question, but it leads to new (or even different) degrees of freedom and symmetries that, ultimately, modify the original problem. This, however, does not affect the galileon, where the leading interactions are not renormalized, at least perturbatively, by loop corrections~\cite{Luty:2003vm, Goon:2016ihr}, and the derivative expansion remains under control even when coupled strongly. Although there is an extension of Derrick’s theorem that excludes the existence of soliton states in this model~\cite{Endlich:2010zj}, the invariance of the galileon cannot be realized exactly in nature, where every coupling of the field to gravity breaks it explicitly, and an analysis that includes symmetry breaking operators is not only pertinent but timely. In this paper we discuss under which conditions this negative result can be evaded in theories that present a weak breaking of the galileon symmetry (see Ref.~\cite{CarrilloGonzalez:2016lor} for previous work in this direction).

There are some ways to circumvent Derrick's theorem, which are classified broadly into two general types, depending on whether they rely on topological~\cite{Vilenkin:2000jqa,Weinberg:2012pjx}, or non-topological~\cite{Wilets:1990di,Lee:1991ax}, considerations. In both cases, and to avoid dispersion, one needs non-linear terms in the equations of motion. These may arise from self-interactions or couplings to other (gauge) fields, which counteract by mode mixing the unequal propagation velocities of the modes due to the linear terms. In addition to the non-linear operators, one needs conserved currents, that result in non-trivial configurations of less energy than any other one gathering the same charge, hence being stable under fission or evaporation. The main difference between topological and non-topological solitons relies on the nature of these currents: whereas in the former case they appear as a consequence of a non-trivial vacuum structure, the latter are possible through internal continuous global symmetries.

Examples of topological solitons include the global monopole (although it does not posses a finite total energy, leading to a deficit angle in presence of gravity~\cite{Barriola:1989hx}), or the 't Hooft~\cite{tHooft:1974kcl} and Polyakov~\cite{Polyakov:1974ek} monopole, which appear in theories with gauge fields. In both cases the fields are static. On the contrary, non-topological solitons consist on {\it time-dependent} field configurations, then avoiding Derrick's theorem, where a time-harmonic evolution of the scalar field results in a non-dispersive object. Decreasing the energy per unit charge (with respect to its value where the same amount of charge is far apart) requires an attraction between particles. In flat spacetime this is usually achieved through a self-interaction term, leading to configurations of non-vanishing Noether charge known as $Q$-balls, given their spatial symmetry~\cite{Coleman:1985ki} (the letter $Q$ is used as a shorthand for the charge). Gravity is attractive, and when it is included we can construct such compact objects without the need of self-interactions, where the resulting solutions are usually referred as boson stars (BSs)~\cite{Kaup:1968zz,Ruffini:1969qy,Jetzer:1991jr,Schunck:2003kk,Liebling:2012fv}. Apart form Q-balls and BSs, their close relatives the oscillons~\cite{Copeland:1995fq} (also known as oscillatons in presence of gravity~\cite{Seidel:1991zh}), are {\it long-lived} field configurations that do not require of a(n exact) conserved charge, being it topological or non-topological, to exist. The reason why these objects are possible is because of an {\it approximate} global symmetry that appears in the non-relativistic limit of the massive Klein-Gordon model~\cite{Mukaida:2014oza,Ibe:2019vyo}. This is nothing but the conservation of the particle number that we all experience in our everyday lives, where no particles can be created or annihilated (if there are no channels to decay in) at low energies. In this limit Q-balls (BSs) and oscillons (oscillatons) are indistinguishable from each other. Note the fine line that divides the classification of the solutions, which in some cases is blurred and ambiguous. 

Coming back to our effective model, the low energy regime of theories where the galileon symmetry is weakly broken is characterized by three energy scales: the Planck mass $M_{\textrm{Pl}}$, that controls the strength of graviton self-interactions, the (perturbative) cutoff scale $\Lambda_3$ of the EFT, that controls the non-trivial interactions of the scalar field that are invariant under galileon transformations, and $m$, the mass of the scalar particles. Operators induced by couplings of the scalar with gravitons are suppressed by powers of the scale $\Lambda_2\equiv (M_{\textrm{Pl}}\Lambda_3^3)^{1/4}$ and break the galileon symmetry down to the subgroup $\phi\to\phi+c$ (in this paper we always choose $\Lambda_3\ll M_{\textrm{Pl}}$, and then $\Lambda_3\ll \Lambda_2\ll M_{\textrm{Pl}}$). In addition, the mass term inclusion breaks explicitly the invariance under internal shifts, and makes possible the existence of solutions that decay to zero faster than $1/r^2$ at spatial infinity. Under those circumstances, and in absence of matter, our effective theory reduces to what we call the Einstein-Klein-Gordon (EKG) model in the regime in which the scalar field is coupled weakly, a limit where we can anticipate the existence of soliton states that are in fact indistinguishable from previously known BS/oscillaton solutions. Our motivation is to extend these solutions to the regime where the theory is coupled strongly, although in this paper we only do this perturbatively.

Before we conclude this section, it is important to make some clarifications regarding the scales. For convenience, and depending on the mass of the scalar field, we distinguish between infrared, $m\lesssim 10^{-3}\,$eV, and UV, $m\gtrsim 10^{-3}\,$eV, modifications of gravity. This classification is given by the characteristic range of the fifth forces, and depends on whether the length scales associated to the Yukawa interactions are attainable, or not, to local experiments, that limit any additional gravitational-strength force to the sub-millimeter distances~\cite{Lee:2020zjt}. If modified in the UV, $m\gtrsim 10^{-3}\,$eV, the Compton wavelength of the scalar field is shorter than about a millimeter, a distance that today is inaccessible to the gravitational observations. In a more technical language, the scalar degree of freedom is integrated out from the low energy spectrum, leading to higher order operators in the effective action that involve appropriate contractions of the Riemann tensor that are subdominant at accessible scales (these operators should be treated perturbatively and do not introduce ghost instabilities~\cite{Simon:1990ic,Simon:1991bm}). If on the contrary gravity is modified in the infrared, $m\lesssim 10^{-3}\,$eV, the fifth forces would leave signatures on e.g. Earth based torsion balance detectors. In this case, and to have a successful phenomenological behaviour, we need to guarantee a mechanism that screens any additional gravitational interaction at certain distances that are within the Compton wavelength of the scalar field~\cite{Brax:2013ida, Joyce:2014kja}. This is not in principle a problem for the GLPV theory, where higher derivative operators can induce a regime of strong coupling that modifies the effective range of the Yukawa interactions (see Figure~\ref{Fig:Regimes} as reference).

\subsection*{Outline of the paper and main results}

The organization of this work is as follows: in Section~\ref{sect2} we review the notion of {\it weakly broken galileon invariance}, leading to a low energy effective theory that includes operators with higher order derivatives. Connoisseurs of the EFT approach in modified gravity may want to jump to Section~\ref{sec3}, where we present the equations that describe the static and spherically symmetric regime of this model, and identify the soliton boundary conditions that we use to construct vacuum self-gravitating objects. In the regime where the scalar field is weakly coupled, our effective theory reduces to the EKG model, and this is the reason why we refer to the resulting configurations as Horndeski stars (HSs), given their similarity to BSs when field amplitudes are low. There is, however, an important difference between the two objects that we should make explicit: whereas the latter are sourced by matter, the former exist as a consequence of gravity itself. While the distinction might seem subtle at first view, it may have a significant impact given that studying the properties of HSs one could gain knowledge on fundamental aspects of the gravitational interaction that are otherwise unattainable from other observations.

In Section~\ref{sec:ph} we identify the most basic properties of HSs, where we establish the existence of a family of solutions that is linearly stable against small radial perturbations. This behaviour is similar to that of neutron and BSs, where the maximum mass configuration divides the solution curve into the stable and the unstable branches~\cite{Seidel:1990jh,Balakrishna:1997ej}. The state of maximum mass corresponds also to the most compact stable configuration, with compactness as large as $C\approx 0.35$ for certain choices of the model parameters. Even though similar numbers have been reported using canonical scalar fields with potentials constructed {\it ad hoc} to produce very compact configurations~\cite{PhysRevD.35.3658,Cardoso:2016oxy,Palenzuela:2017kcg}, the important difference is that our results are motivated from the viewpoint of an EFT, and then robust under quantum corrections. If gravity is modified in the UV, $m\gtrsim 10^{-3}\,$eV, the resulting objects are always lighter than $10^{-7}\,M_{\odot}$. We do not have access to individual objects of these masses using current astrophysical techniques, and for this reason we concentrate mainly to the case where gravity is modified in the infrared, $m\lesssim 10^{-3}\,$eV. Specifically, for masses of the scalar particle in the range of $10^{-13}\,\textrm{eV}\lesssim m \lesssim 10^{-10}\textrm{eV}$, we obtain objects of maximal mass between 1 and $1000$ solar masses, which may be accessible to current astrophysical observations. As a word of caution, notice that the range of scalar masses that we consider in this work cannot account for a perturbative EFT description of dark energy, for which it is required that $m\lesssim 10^{-33}\,$eV. If that is the case and for similar central amplitude values to the ones we use, the scalar particles should only develop structures that are larger than the observable universe, hence inaccessible to our experience. In contrast, our compact objects might constitute part of the dark matter component, close in spirit to the idea in Ref.~\cite{Cembranos:2008gj}, though we leave the details for a future paper. The last part of this work is devoted towards the phenomenology that could be used to differentiate HSs from more standard objects, such as black holes or neutron stars. In particular, under the most extreme of the conditions we have reached with our numerical code we have identified the formation of innermost stable circular orbits and photon spheres, which  have a relevance in the deflection of light rays and the emission spectra from accretion disks~\cite{Cardoso:2019rvt}, as we also discuss. In addition, we explore briefly the GW emission from the collision of two HSs. Finally, in Section~\ref{sec5} we present some concluding remarks.
For previous work with similar motivation see e.g.~\cite{Obregon:2004ty,Padilla:2010ir, Brihaye:2016lin,Sakstein:2018pfd}.

\begin{figure}[t]
\centering
	\scalebox{0.45}{
	\input{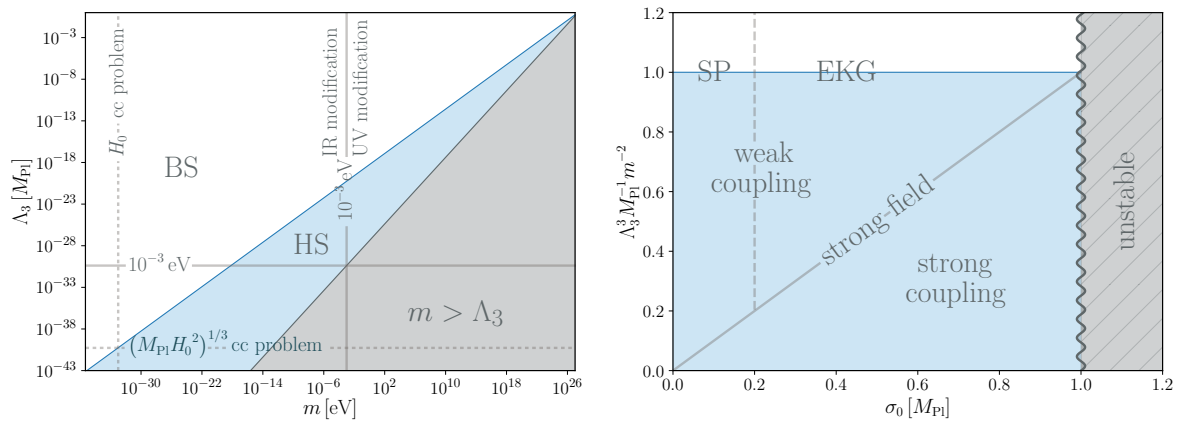}
	}  
	\caption{{\bf Horndeski stars.} {\it Left panel:} 
	Parameter space of the model. The region in blue ($\Lambda_3\lesssim M_{\textrm{Pl}}^{1/3}m^{2/3}$) corresponds to the combination of parameters where HSs of enough amplitude $\sigma_0$ can develop the distinctive features associated to the higher derivative operators. For theories in the white region ($\Lambda_3\gtrsim M_{\textrm{Pl}}^{1/3}m^{2/3}$), on the contrary, HSs are indistinguishable from BSs, no matter the value of the central amplitude, as long as we concentrate to stable configurations. In the shaded region, $m>\Lambda_3$, the scalar mode disappears from the low energy spectrum and no HSs are possible. {\it Right panel:} The regimes of a HS as a function of the central amplitude, for different combinations of the parameters of the model. The limit of very low amplitudes can be captured in terms of the much simpler Schr\"odinger-Poisson (SP) equations. The transverse continuous line that divides the two regions in blue corresponds to the critical amplitude at the transition to the Horndeski regime, $\sigma_0^{\textrm{s.f.}}\sim M_{\textrm{Pl}}^{1/2}\Lambda_3^{3/2}/m$, Eq.~(\ref{scale.self}), and indicates the shift to the to the (model-dependent) strong coupling behaviour. The horizontal border between the blue and the white regions is the same as the transverse one in the left panel and has associated $\sigma_0^{\textrm{s.f.}}\sim M_{\textrm{Pl}}$, signaling the onset of a gravitational instability. The two panels are for indicative purposes only and the numbers must be taken as orders of magnitude. See Section~\ref{sec3} for details. }
	\label{fig.parameter.space}
\end{figure}

The main results of this paper are summarized in Figure~\ref{fig.parameter.space}, where we classify HSs as function of the parameter space of the model. In practice, we restrict our attention to the stable branch of the solutions. For theories of modified gravity where $M_{\textrm{Pl}}^{1/3}m^{2/3}\lesssim\Lambda_3$ (white region in the left panel of this figure), the resulting objects are indistinguishable from BSs, whereas for theories with $M_{\textrm{Pl}}^{1/3}m^{2/3}\gtrsim\Lambda_3$ (blue region), there are some configurations, viz. those with larger central amplitudes, that present some distinctive features that make them distinguishable from previous solutions; see the right panel of Figure~\ref{fig.parameter.space} for details. In the present work we concentrate on theories that lie on the border between the two regimes, $M_{\textrm{Pl}}^{1/3}m^{2/3}\sim\Lambda_3$, where deviations with respect to the EKG theory are manifest and the differences with respect to the BS case can be parametrized in a model-independent way using the EFT language. In this approximation, the self-interaction terms are dominated by two operators of mass dimension six: one of them ---$c_4^{(1)}$--- is part of the Fab Four~\cite{Charmousis:2011bf, Charmousis:2011ea}, whereas the other ---$d_4^{(-1)}$--- lies in the quartic sector of the beyond Horndeski theory; see Table~\ref{Table.coefficients} for details. Surprising as it may seem the covariant galileons~\cite{Deffayet:horndeski} play no role in this scenario and this is in part the reason why it is possible to circumvent the result of~\cite{Endlich:2010zj} within the framework of a well motivated effective model. A general analysis that includes the more interesting region of the parameter space where $M_{\textrm{Pl}}^{1/3}m^{2/3}\gtrsim\Lambda_3$ will be presented elsewhere~\cite{Roque2020}.\footnote{In this paper we distinguish between the {\it strong field} and the {\it strong coupling} regimes. In the strong field regime the higher order operators may have a significant relevance in a particular calculation, although they can still be treated perturbatively. This is the regime to which we concentrate on this work. On the contrary, in the strong coupling regime all the operators have in principle the same weight in the calculation and a non-perturbative analysis is necessary. Note that we can think of the strong field regime as the border dividing the weak field approximation and the regime where the theory is coupled strongly. See the right panel of Figure~\ref{fig.parameter.space} as an illustration.}

{\bf Conventions.---} 
We follow the sign conventions of Misner, Thorne and Wheeler~\cite{Misner:1974qy}, and Wald~\cite{Wald:1984rg}. That is, we use the mostly plus signature for the spacetime metric, $(-,+,+,+)$, and the definitions $R_{\mu\nu\rho}{}^{\sigma}\equiv \partial_\nu\Gamma^{\sigma}_{\mu\rho}+\Gamma^{\alpha}_{\mu\rho}\Gamma^{\sigma}_{\alpha\nu}-(\nu\leftrightarrow \mu)$ for the Riemann tensor, $R_{\mu\nu}\equiv R_{\mu\alpha\nu}{}^{\alpha}$ for the Ricci tensor, and $R\equiv R_{\mu}{}^{\mu}$ for the Ricci scalar, according to Wald's notation. We also use the symmetric convention $(a,b)=\frac{1}{2}(ab+ba)$ and $[a,b]=\frac{1}{2}(ab-ba)$. Additionally, we work in natural units where $\hbar=c=1$ unless otherwise stated, and use either Newton's constant $G$ or the reduced Planck mass $M_{\textrm{Pl}}\equiv 1/\sqrt{8\pi G}=2.431\times 10^{18}\,$GeV at convenience. We assume a minimal coupling of matter to gravity (i.e. matter fields couple only to the Jordan spacetime metric, $g_{\mu\nu}^{\textrm{Jor.}}=g_{\mu\nu}$), although in most parts of the paper we are interested in vacuum solutions where this is not relevant. However, in Section~\ref{sec:ph}, and to make contact with observations, we need to determine the behaviour of light rays and point particles moving in the gravitational configuration where this assumption is important.

\section{Horndeski as an EFT}\label{sect2}

In this section we motivate the effective theory that we consider for the rest of this paper. Our starting point is an analysis carried out by Gleyzes, Langlois, Piazza and Vernizzi in Refs.~\cite{Gleyzes:2014dya,Gleyzes:2014qga}, where they introduce a new family of scalar-tensor theories that extend the Horndeski model~\cite{Horndeski:1974wa, Deffayet:horndeski, Deffayet:horndeski2}.\footnote{We sometimes omit the distinction with Horndeski if no confusion arises, hence the title of this section and the paper. There is a more general theory containing degenerate higher order models~\cite{Langlois:2015cwa,Crisostomi:2016czh} that even if interesting as extensions to Horndeski, do not inherit the remnant of the galileon symmetry that makes (a subset of) the GLPV theory appealing as an EFT, and this is the reason why we do not consider it here. See Figure~\ref{Fig.Landscape} for details.} These theories come into the gravitational and the matter pieces, described in terms of the following action:
\begin{equation}
S=S_{\textrm{grav}}+S_m = \int d^{4}x \sqrt{-g}\left(\sum_{i=2}^{5} \mathcal{L}_{i}[g_{\mu\nu},\phi]+\mathcal{L}_m[g_{\mu\nu},\Psi]\right),
\label{eq:lag}
\end{equation}
where $S_{\textrm{grav}}$ denotes the gravitational sector and is given as a linear combination of the Lagrangians:
\begin{subequations}\label{eq.Lagrangians}
	\begin{eqnarray}
	\mathcal{L}_{2}&\equiv& G_{2}(\phi, X),\label{BH1}\\
	\mathcal{L}_{3}&\equiv& G_{3}(\phi, X)\Box\phi,\\
	\mathcal{L}_{4}&\equiv& G_{4}(\phi, X)R-2G_{4X}(\phi,X)\left[(\Box\phi)^{2}-\phi^{\mu\nu}\phi_{\mu\nu}\right]+F_{4}(\phi,X){\epsilon^{\mu\nu\rho}}_{\sigma}\epsilon^{\mu'\nu' \rho'\sigma}\phi_{\mu}\phi_{\mu'}\phi_{\nu\nu'}\phi_{\rho\rho'},\label{eq.quartic}\\
	\mathcal{L}_{5}&\equiv& G_{5}(\phi, X)G_{\mu\nu}\phi^{\mu\nu}+\frac{1}{3}G_{5X}(\phi,X)\left[(\Box\phi)^{3}-3\Box\phi\phi_{\mu\nu}\phi^{\mu\nu}+2\phi_{\mu\nu}\phi^{\mu\sigma}{\phi^{\nu}}_{\sigma}\right]\nonumber\\
	&&+\,F_{5}(\phi,X)\epsilon^{\mu\nu\rho\sigma}\epsilon^{\mu'\nu' \rho'\sigma'}\phi_{\mu}\phi_{\mu'}\phi_{\nu\nu'}\phi_{\rho\rho'}\phi_{\sigma\sigma'}\,.\label{BH2}
	\end{eqnarray}
\end{subequations}

Some words are in place about the different quantities in Eqs.~(\ref{eq:lag}) and~(\ref{eq.Lagrangians}). As usual $R$ and $G_{\mu\nu}=R_{\mu\nu}-\frac{1}{2}Rg_{\mu\nu}$ are the Ricci scalar and Einstein tensor, respectively, defined with respect to the spacetime metric $g_{\mu\nu}$. The metric tensor is used to raise and lower spacetime indices, and to define the right measure of the spacetime integrals through its determinant $g\equiv \det(g_{\mu\nu})$. The six functions $G_{i}$ and $F_i$ are independent and {\it a priori} arbitrary, and can depend on the scalar field $\phi$ and its canonical kinetic term $X\equiv g^{\mu\nu}\phi_{\mu}\phi_{\nu}$, where $\phi_{\mu}\equiv\nabla_{\mu}\phi$. To simplify the writing we have used the definition $\phi_{\mu\nu}\equiv\nabla_{\mu}\nabla_{\nu}\phi$, together with the subindex notation in the functions $G_i$ and $F_i$ (e.g. $G_{i X}$) to denote partial differentiation with respect to their arguments. Note that with this convention the d'Alembert operator takes the form $\Box\phi\equiv\phi_{\mu}{}^{\mu}$. We have also used the totally antisymmetric Levi-Civita tensor $\epsilon^{\mu\nu\rho\sigma}$ to define the last term of the quartic $\mathcal{L}_4$ and quintic  $\mathcal{L}_5$ sectors, Eqs.~(\ref{eq.quartic}) and~(\ref{BH2}), respectively. The matter Lagrangian $\mathcal{L}_m$ contains all the standard model fields, and perhaps also further components that contribute to the dark sector in the Universe. We assume that these fields are minimally coupled to the spacetime (Jordan frame) metric $g_{\mu\nu}$, although this is not very relevant for most parts of this paper; remember that the core idea behind this study is to construct self-bounded compact configurations using the gravity fields $g_{\mu\nu}$ and $\phi$ alone. However, in the last section we will restore the matter sector when discussing some phenomenological aspects of the model.

If either functions $F_4$ or $F_5$ is different from zero, the higher order derivatives of the Lagrangian lead to higher order differential operators at the level of the equations of motion.\footnote{Although the $F_4$ and $F_5$ terms in the GLPV theory can be mapped to the pure Horndeski action via disformal transformations (see for example~\cite{Crisostomi:2016tcp}), in doing so the matter stops being minimally coupled to gravity. Here we always assume minimally coupled matter components, so that $F_4$ or $F_5$ are different sectors from the pure Horndeski theory.} 
Nonetheless, the true propagating degrees of freedom obey second order equations, avoiding Ostrogradski instabilities~\cite{Ostrogradsky:1850fid, Pais:1950za}. In contrast, setting both functions $F_4$ and $F_5$ to zero leads to the more familiar Horndeski theory, where the equations of motion remain second order. GR is recovered for the particular choice of $G_4=\frac{1}{2}M_{\textrm{Pl}}^2$, with all other functions set to zero ---except for $G_2=-M_{\textrm{Pl}}^2\Lambda$ if there is a nonvanishing cosmological constant---, hence the scalar field remains muted. If on the contrary $G_2$ is different from constant, the scalar degree of freedom propagates, but as long as we maintain the other functions fixed to those of GR the scalar field remains minimally coupled to gravity, irrespectively of the particular choice of $G_2$. A case of special interest is that of a canonical scalar field, where $G_2=-\frac{1}{2}X-V(\phi)$. We refer to the latter, when the potential is quadratic in the field, as the EKG model. One could naturally argue whether this represents or not a real modification of gravity, however, our general effective theory reduces to the EKG model in the regime where the scalar field is coupled weakly (see details below), and this is the reason why we use the latter as a fiducial model in many comparisons.

Even if viable as {\it classical} field theories (in the sense of no Ostrogradski instabilities), not all elements in the GLPV theory are as robust as low energy effective {\it quantum} theories. At the quantum level, loop Feynman diagrams lead to divergent quantities, and they require careful handling. This is usually done through the introduction of appropriate counterterms into action. The new terms remove the divergent contributions from the loops, giving rise to sensible finite outcomes, but modify the original theory and may invalidate the classical predictions. Related to this is the fact that EFTs are conceived as a low energy expansion, where the leading operators are expected to be those with least number of derivatives, hence some special conditions are required if we want the higher derivatives of the GLPV theory to give a real contribution to low energy observables. 

Symmetries may alleviate this situation. In the scalar sector, a well known example is that of the axion~\cite{Peccei:1977hh,Zyla:2020zbs}, where a(n approximate) shift symmetry $\phi\to\phi+c$ maintains radiative corrections under control, and in particular the mass term much lower than the EFT scale.\footnote{A less trivial realization (but in the tensor sector of the theory) is GR, where exact diffeomorphism invariance prevents the renormalization of the coefficients of the Eintein-Hilbert action (packaged conveniently into the Ricci scalar times the square root of the determinant of the spacetime metric), and organize in a systematic way the higher derivative operators that are induced at loop level, see e.g.~\cite{Hinterbichler:2011tt,Goon:2016ihr}. This is crucial for the separation of the scales in which non-linear (Schwarzschild) and quantum (Planck) physics operate in the large majority of relevant astrophysical scenarios, and allows us to trust the solutions to the classical Einstein's equations even when strongly coupled. More words about this later in this section.} However, the axion is coupled minimally to gravity, and of no interest as a modification of GR. To proceed, we need a prescription to couple in a non-trivial way a scalar to the gravitational field, Eqs.~(\ref{eq:lag}) and~(\ref{eq.Lagrangians}), but as important as the latter is to provide a mechanism that guarantees the quantum stability of the theory, similar in spirit to the shift symmetry of the axion field, but open to higher derivative operators and associated non-minimal couplings. Only in this way we could say without fear of being mistaken that our findings are not affected once quantum corrections are considered. 

\subsection{The galileons}

The simplest realization of a well motivated EFT where higher derivative operators can affect physical observables is probably that of the galileons~\cite{Nicolis:2008in}. The galileon is a scalar degree of freedom that is introduced in flat spacetime and is invariant under internal transformations of the form\footnote{We can look at this as a generalization of the shift symmetry of the axion field, where we have extended the invariance under shift transformations up to the field gradients. Notice the similarity with Galilean transformations, $x^i\to x^i+v^i_0t+x^i_0$, that connect internal reference frames in Classical Mechanics, hence the name of this theory.}
\begin{equation}\label{eq.galileon.symmetry}
\phi\to\phi+b_{\mu}x^{\mu}+c,
\end{equation}
where $b_{\mu}$ and $c$ are arbitrary constants and $x^{\mu}$ are the spacetime coordinates. In four dimensions, and in addition to a tadpole term that we do not include here, there are exactly four non-trivial operators, i.e. involving less than two derivatives per field, 
\begin{subequations}\label{eq.galileons}
\begin{eqnarray}
    \mathcal{L}_2^{\textrm{gal}} &=&-\frac{1}{2}(\partial\phi)^2,\\
    \mathcal{L}_3^{\textrm{gal}} &=& -\frac{c_3}{2\Lambda_3^3} (\partial\phi)^2\Box\phi,\\ 
    \mathcal{L}_4^{\textrm{gal}} &=& -\frac{c_4}{2\Lambda_3^6}(\partial\phi)^2\left[(\Box\phi)^{2}-\partial_{\mu}\partial_{\nu}\phi\partial^{\mu}\partial^{\nu}\phi\right],\\
    \mathcal{L}_5^{\textrm{gal}} &=& -\frac{c_5}{4\Lambda_3^9}(\partial\phi)^2\left[(\Box\phi)^{3}-3\Box\phi\partial_{\mu}\partial_{\nu}\phi\partial^{\mu}\partial^{\nu}\phi+2\partial_{\mu}\partial_{\nu}\phi\partial^{\mu}\partial^{\sigma}\phi\partial^{\nu}\partial_{\sigma}\phi\right],
\end{eqnarray}
\end{subequations}
that can contribute to the action and remain invariant under the transformations~(\ref{eq.galileon.symmetry}). The canonical kinetic term $\mathcal{L}_2^{\textrm{gal}}$, also known in this context as the quadratic galileon, is ubiquitous to any relativistic field theory and is present already in the Klein-Gordon model. Things get more interesting, however, at the level of the interactions. The cubic galileon  $\mathcal{L}_3^{\textrm{gal}}$ was identified for the first time in the decoupling limit of the Dvali-Gabadadze-Porrati (DGP) model~\cite{Dvali:2000hr}, and generalized in~\cite{Nicolis:2008in} to the quartic and quintic galileons, $\mathcal{L}_4^{\textrm{gal}}$ and $\mathcal{L}_5^{\textrm{gal}}$, respectively, that were found to describe the helicity-zero polarization of the decoupled ghost-free de Rham-Gabadadze-Tolley (dRGT) massive gravity model~\cite{deRham:2010ik}.

The coefficients $c_3$, $c_4$ and $c_5$ are arbitrary and of order one, with $\Lambda_3$ a characteristic scale that regulates the strength of the self-interactions.\footnote{This scale is usually associated to the cutoff of the EFT, where perturbative unitary breaks down and new physics shows up. However, if non-perturbative effects are under control, this scale may not be necessarily the real cutoff of the theory. A detailed discussion of this for the case of purely kinetic $k$-essence models $G_2=G_2(X)$, $G_4=\frac{1}{2}M_{\textrm{Pl}}^2$, $G_3=G_5=0$ can be found in Ref.~\cite{deRham:2014wfa}.} The galileons have some interesting features, including a theorem that guarantees that the leading order coefficients $c_3$, $c_4$ and $c_5$ are not renormalized, at least perturbatively, by loop corrections~\cite{Luty:2003vm, Goon:2016ihr}. This has far-reaching impacts, given that, in addition to~(\ref{eq.galileons}), there are other operators of the form $\Lambda_3^{4-n-3m}\partial^n(\partial^2\phi)^m$ that are trivially invariant under the transformations~(\ref{eq.galileon.symmetry}) and must be included in the EFT for self-consistency ---even if they were not present at tree level they are generated once we include loop diagrams. At low energies/large distances, however, these terms are subdominant with respect to those in Eqs.~(\ref{eq.galileons}) and do not play a significant role within the regime of validity of the theory,\footnote{To put it the other way around, when operators that are trivially invariant under galileon transformations come into play the EFT loses predictability and breaks down. Therefore, even if we do not write them down explicitly in~(\ref{eq.galileons}), they determine the regime of validity of the EFT.} that includes the strongly coupled regime (see Ref.~\cite{Burrage:2020bxp} for some complications in the massive case). However, any coupling of the scalar to gravity breaks, in general, the invariance under galileon transformations, and as a consequence the galileon symmetry cannot be realized exactly in nature.

\subsection{Weakly broken galileon symmetry}

Despite this negative result, there is a subset of the GLPV theory~(\ref{eq.Lagrangians}) that, even if not invariant under the galileon transformations~(\ref{eq.galileon.symmetry}), includes gravity and preserve as much as possible the quantum properties of the symmetric model. This introduces the notion of weakly broken galileon (WBG) invariance~\cite{Pirtskhalava:2015nla,Santoni:2018rrx}, leading to models that are largely insensitive to loop corrections, hence robust as EFTs. From now on we concentrate our attention to this subsector of the theory. 

The simplest choice of a model with WBG symmetry is that of the covariant galileons~\cite{Deffayet:horndeski}. They appear as a natural extension to the flat spacetime version of the model, Eqs.~(\ref{eq.galileons}), once we include the effects of gravity. In the language of Horndeski, Eqs.~(\ref{eq:lag}) and (\ref{eq.Lagrangians}), 
the covariant galileons are given by 
\begin{equation}\label{eq.covariant.galileons}
    G_2^{\textrm{gal}}=-\frac{1}{2}X,\quad G_3^{\textrm{gal}}=-\frac{c_3}{2\Lambda_3^3} X,\quad G_4^{\textrm{gal}}=\frac{1}{2}M_{\textrm{Pl}}^2+\frac{c_4}{8\Lambda_3^6}X^2,\quad 
    G_5^{\textrm{gal}}=-\frac{3c_5}{8\Lambda_3^9}X^2,    
\end{equation}
with $F_4^{\textrm{gal}}=F_5^{\textrm{gal}}=0$. These terms involve very particular combinations of non-minimal couplings to the Ricci scalar and the Einstein tensor, that not only render the equations of motion of second order but also control the size of the symmetry breaking operators. To better understand the relevance that this may have note that the breakdown of the galileon symmetry induces operators, e.g. $(\partial\phi)^{2n}$, that even if absent in the flat spacetime model, can be generated from loops with gravitons in the extended theory. Of course these operators have to be suppressed, at least, by one power of the Planck mass, but this may be sufficient to blur the effects of the higher derivative terms at low energies. The interesting point is that the scale suppressing the new operators can be estimated to be larger than one may have naively expected~\cite{Pirtskhalava:2015nla}. This is true, in principle, even at the level of the minimally coupled galileons ---those obtained by replacing partial derivatives by covariant ones in (\ref{eq.galileons}). The reason why it happens can be easily understood from the structure of the couplings that are codified in the covariant derivatives, resulting in at least one internal graviton line per three external single-derivatives $\partial\phi$ in any loop diagram. This structure induces EFT operators of the form e.g. $M_{\textrm{Pl}}^{2n}\Lambda_3^{10n-4}(\partial\phi)^{6n}$, rising the symmetry breaking scale from the galileon's $\Lambda_3$, to $(M_{\textrm{Pl}}^{2n}\Lambda_3^{10n-4})^{1/(12n-4)}$, which for large $n$ approaches asymptotically $(M_{\textrm{Pl}}\Lambda_3^5)^{1/6}$, otherwise being larger.\footnote{Remember that in this paper we always consider that $\Lambda_3\ll M_{\textrm{Pl}}$.} The situation is improved further by the particular choice of non-minimal couplings that in the case of the covariant galileons make the equations of second order, since they cancel out all vertices with three single-derivative operators, only surviving those with two $\partial\phi$ per graviton. As a consequence, the energy scale associated to the symmetry breaking increases to $(M_{\textrm{Pl}}^{n}\Lambda_3^{3n-4})^{1/(4n-4)}$, which approaches
\begin{equation}
    \Lambda_2\equiv(M_{\textrm{Pl}}\Lambda_3^3)^{1/4}
\end{equation}
only for asymptotically large $n$, where $\Lambda_3\ll \Lambda_2 \ll M_{\textrm{Pl}}$. Contrary to the flat spacetime model, Eq.~(\ref{eq.galileons}), the coefficients $c_i$ of the covariant galileons {\it are} affected by loop diagrams, but the renormalization is parametrically weak, with corrections suppressed by factors of $(\Lambda_2/\Lambda_3)^4$ with respect to tree level values.

This idea can be extended to a particular subclass of the GLPV theory that is, however, more general than just the covariant galileons (see Figure~\ref{Fig.Landscape}). The strategy is to include operators that, even if not invariant under galileon transformations (in the flat spacetime limit where these transformations are well defined), are suppressed by appropriate powers of the scale $\Lambda_2$. If we assume that the galileon symmetry Eq.~(\ref{eq.galileon.symmetry}) is broken down to $\phi\to\phi+c$, the new operators can be conveniently arranged in a compact form corresponding to the following choice of the arbitrary functions of the Lagrangian~(\ref{eq.Lagrangians})~\cite{Pirtskhalava:2015nla,Santoni:2018rrx}:\footnote{Up to our knowledge the quantum stability of the $F_5^{\textrm{WBG}}$ term has not been analyzed in the literature. Even so, this sector is suppressed at low energies (it includes operators of mass dimension nine or larger) and of no relevance for our purposes in this paper, see Eq.~(\ref{eq.selfinteractions}) for details. We can also eliminate the whole quintic (as well as the cubic) sector if we impose a discrete $\mathbb{Z}_2$ mirror symmetry $\phi\to -\phi$ on the theory~\cite{Diez-Tejedor:2018fue}. In that case the resulting effective action is the same as the one that we obtain here, which in fact is invariant under $\mathbb{Z}_2$ transformations, see Eq.~(\ref{eq.selfinteractions}).}
\begin{subequations}\label{eqs.WBG}
 \begin{eqnarray}
  && G_2^{\textrm{WBG}} = \Lambda_2^4\bar{G}_2[\Lambda_2^{-4}X],\\
  && G_3^{\textrm{WBG}} = \frac{\Lambda_2^4}{\Lambda_3^3}\bar{G}_3[\Lambda_2^{-4}X], \\
  && G_4^{\textrm{WBG}} = \frac{\Lambda_2^8}{\Lambda_3^6}\bar{G}_4[\Lambda_2^{-4}X], \quad
    F_4^{\textrm{WBG}} = \frac{1}{\Lambda_3^6}\bar{F}_4[\Lambda_2^{-4}X],\\
  && G_5^{\textrm{WBG}} = \frac{\Lambda_2^8}{\Lambda_3^9}\bar{G}_5[\Lambda_2^{-4}X], \quad
    F_5^{\textrm{WBG}} = \frac{1}{\Lambda_3^9}\bar{F}_5[\Lambda_2^{-4}X].   
 \end{eqnarray}
\end{subequations}
Note that in the previous expressions the barred functions can only depend on the dimensionless kinetic term $\Lambda_2^{-4}X$ but are otherwise arbitrary. Furthermore, in the following we assume that they can be expressed as series expansions, with the property that they admit the Minkowski spacetime ($g_{\mu\nu}=\eta_{\mu\nu}$ and $\phi=\textrm{const.}$)\footnote{At this point the value of $\phi$ is irrelevant, as long as it remains constant. This is a consequence of the shift symmetry of the theory. Later we will break this symmetry, giving a mass term to the scalar field, so that $\phi=0$ in that case.} as a possible solution, namely
\begin{subequations}\label{eq.functions:WBG}
 \begin{eqnarray}
  \bar{G}_i[\Lambda_2^{-4}X] &=& c_i^{(0)} + c_i^{(1)} \frac{X}{\Lambda_2^4} + c_i^{(2)} \frac{X^2}{\Lambda_2^8} + c_i^{(3)} \frac{X^3}{\Lambda_2^{12}}+\ldots,\\
  \bar{F}_i[\Lambda_2^{-4}X] &=& d_i^{(-1)} \frac{\Lambda_2^4}{X} + d_i^{(0)} + d_i^{(1)} \frac{X}{\Lambda_2^4} + d_i^{(2)} \frac{X^2}{\Lambda_2^8} + \ldots.
 \end{eqnarray}
\end{subequations}
The coefficients $c_i^{(j)}$ and $d_i^{(j)}$ are dimensionless and of order one, unless they vanish identically. They codify our ignorance about the UV physics, and at this point they are arbitrary. Notice that $c_2^{(0)}$ contributes to the EFT in the form of a cosmological constant, and we do not consider it here, $c_2^{(0)}=0$.\footnote{The cosmological observations determine the value of this number to be small, thus irrelevant for the solutions we are interested in this work.} Moreover, $c_2^{(1)}$ and $c_4^{(0)}$ set the normalization of the scalar and graviton kinetic terms, respectively, and therefore we fix them with no loss of generality to their standard values, $c_2^{(1)}=-c_4^{(0)}=-1/2$. This choice guarantees the absence of ghosts in the flat spacetime low energy limit. The operator $c_3^{(0)}$ can be absorbed into $c_2^{(1)}$ via an integration by parts, whereas $c_5^{(0)}$ can be expressed as a total derivative that does not contribute to the equations of motion, which is equivalent to consider $c_3^{(0)}=c_5^{(0)}=0$. Table~\ref{Table.coefficients} summarizes the role of the different coefficients of the EFT. The leading contributions to $\bar{F}_4$ and $\bar{F}_5$ may seem odd at first view, but they are consistent with flat spacetime (at least as a limit where $\phi\to \textrm{const.}$), and this is the reason why we have included inverse powers of the kinetic term in our analysis. Additionally, the operator $d_4^{(-1)}$ is necessary to maintain the speed of propagation of GWs equal to that of light, unless we set $c_4^{(1)}=0$ by hand. The inclusion of  $d_4^{(-1)}$ impacts the phenomenology of the solutions as we discuss below in the following section.

\begin{table}
\begin{center}
\begin{tabular}{l c c c c c c c c c c c c}
 \hline
 \hline
 \multicolumn{13}{c}{EFT coupling constants}\\
 \hline
  Quadratic Horndeski   && $-$ && $c_2^{(0)}=0$  && \textcolor{blue}{$c_2^{(1)}=-1/2$}    && $c_2^{(2)}$ && $c_2^{(3)}$ && $\ldots$\\
  Cubic Horndeski && $-$ && $c_3^{(0)}=0$  && \textcolor{green}{$c_3^{(1)}$}    && $c_3^{(2)}$ && $c_3^{(3)}$ && $\ldots$\\
  Quartic Horndeski   && $-$ && \textcolor{blue}{$c_4^{(0)}=1/2$}  && \textcolor{red}{$c_4^{(1)}=0,\pm 1/2$}    && \textcolor{green}{$c_4^{(2)}$} && $c_4^{(3)}$ && $\ldots$\\
  Quintic Horndeski   && $-$ && $c_5^{(0)}=0$  && $c_5^{(1)}$    && \textcolor{green}{$c_5^{(2)}$} && $c_5^{(3)}$ && $\ldots$\\
  Quartic beyond Horndeski   && \textcolor{red}{$d_4^{(-1)}$} && $d_4^{(0)}$  && $d_4^{(1)}$    && $d_4^{(2)}$ && $d_4^{(3)}$ && $\ldots$\\
  Quintic beyond Horndeski   && $d_5^{(-1)}$ && $d_5^{(0)}$  && $d_5^{(1)}$    && $d_5^{(2)}$ && $d_5^{(3)}$ && $\ldots$\\
 \hline
\end{tabular}
\end{center}
\caption{{\bf The coefficients of the EFT.} The infinite number of coupling constants $c_i^{(j)}$ and $d_i^{(j)}$ that characterize the EFT. The coefficients in blue determine the kinetic term of the scalar, $c_2^{(1)}=-1/2$, and tensor, $c_4^{(0)}=1/2$,  sectors, and are present already in the EKG model. The coefficients in red, $c_4^{(1)}$ and $d_4^{(-1)}$, provide the first deviation with respect to this model and play a central role in this paper. The coefficient $c_2^{(0)}$ contributes to the cosmological constant and we have set it to zero, whereas $c_3^{(0)}$ and $c_5^{(0)}$ are total derivatives that do not contribute to the equations of motion. The other coefficients are suppressed in the series expansion by higher powers of a mass scale and are of no interest for our current purposes. In particular, the coefficients in green, $c_3^{(1)}=-\frac{1}{2}c_3$, $c_4^{(2)}=\frac{1}{8}c_4$ and $c_5^{(2)}=-\frac{3}{8}c_5$, determine the interactions of the covariant galileons, Eq.~(\ref{eq.covariant.galileons}). See Eqs.~(\ref{eq.functions:WBG}), and the discussion below, for details.}
\label{Table.coefficients}
\end{table}

\subsection{Strong coupling}\label{sec.strong.coupling}

To get a deeper insight into the model we study briefly the limit in which the fields are weak. To proceed, we consider small deviations from the flat spacetime, which we already argued that is a solution of the theory. Infrared physics is dominated by operators of mass dimension less or equal than four. For our general effective model, Eqs.~(\ref{eqs.WBG})  and~(\ref{eq.functions:WBG}), these operators consist on the kinetic terms of the graviton, $h_{\mu\nu}$, and the scalar field, $\phi$, which for canonically normalized tensor modes corresponds to a metric of the form $g_{\mu\nu}=\eta_{\mu\nu}+M_{\textrm{Pl}}^{-1}h_{\mu\nu}$, with $\eta_{\mu\nu}$ the Minkowski metric. They are, indeed, the only two contributions to the effective action that survive in the limit where $M_{\textrm{Pl}},\,\Lambda_3\to\infty$. In addition, there is an infinite number of non-renormalizable operators of mass dimension greater than four that determine the interactions. Only a small fraction of these operators contribute at low energies, whose specific number is determined 
by the degree of precision that one desires in the observables. To lowest order in the series we have
\begin{equation}\label{eq.IRaction}
  \sqrt{-g}\mathcal{L}_{\textrm{grav}} = -\frac{1}{4}h^{\mu\nu}\mathcal{E}_{\mu\nu}^{\alpha\beta}h_{\alpha\beta}+\frac{1}{2}\phi\Box\phi +\frac{1}{2M_{\textrm{Pl}}}h_{\mu\nu}t^{\mu\nu}+\ldots,  
\end{equation}
where $\mathcal{E}_{\mu\nu}^{\alpha\beta}\equiv -\frac{1}{2}[\delta_\mu^\alpha\delta_\nu^\beta\Box-2\eta^{\alpha\gamma}\delta_{(\mu}^\beta\partial_{\nu)}\partial_\gamma+\eta^{\alpha\beta}\partial_\mu\partial_\nu-\eta_{\mu\nu}(\eta^{\alpha\beta}\Box-\eta^{\alpha\gamma}\eta^{\beta\sigma}\partial_\gamma\partial_\sigma)]$ is the Lichnerowicz Laplacian and $t_{\mu\nu}=\partial_\mu\phi\partial_\nu\phi-\frac{1}{2}(\partial_\alpha\phi\partial^{\alpha}\phi) \eta_{\mu\nu}$ is the stress-energy tensor of the scalar field, that at this point, is massless and without self-interactions. In the previous expression the ellipses make reference to operators that are suppressed by higher powers of a mass scale, that can be the Planck mass, the scale $\Lambda_3$, or a combination thereof. The first term in Eq.~(\ref{eq.IRaction}) is part of the operator $c_4^{(0)}=1/2$, whereas the second and the third one come from $c_2^{(1)}=-1/2$. At this level (and in absence of matter) the EFT is indistinguishable from the massless EKG model.\footnote{There is also a coupling of the graviton and the scalar mode to matter, but these operators vanish in vacuum and we have not included them explicitly in~(\ref{eq.IRaction}). The coupling of the scalar to matter may not be apparent at first view, but it appears after we diagonalize the kinetic term. In the context of the Brans-Dicke theory this is analogous to performing a conformal transformation to the Einstein frame; see e.g. Chapter~2 of Ref.~\cite{Fujii:2003pa} for details.}

Once field amplitudes get higher, there are two different ways in which the theory can suffer from strong coupling, i.e. a regime where relevant and irrelevant operators contribute in the same way to the effective action, although this does not necessarily signal the breakdown of the EFT~\cite{deRham:2014wfa}. At some point, the graviton could reach the Planck scale, namely
\begin{equation}
    h_{\mu\nu}\sim M_{\textrm{Pl}},
\end{equation}
so the irrelevant operators of the general form $M_{\textrm{Pl}}^{-n}h^n(\partial h)^2$, codified also in the Einstein-Hilbert term  $c_4^{(0)}=1/2$, compete on equal terms with those of the linear theory of Eq.~(\ref{eq.IRaction}). This typically occurs at the Schwarzschild radius and signals the onset of the strong gravity regime. A similar situation happens in the scalar sector, where as we approach 
\begin{equation}\label{eq.strong.coupling.scalar}
    X\sim \Lambda_2^4,\quad \Box\phi\sim \Lambda_3^3,
\end{equation}
the self-interaction terms start to compete at the same level as the linear operators. At this point the higher derivatives induce non-minimal couplings and modify the EKG model. Which field, tensor or scalar, is the first to become strongly coupled is something that depends on the scales of the theory, and also on the particular configurations of interest. For vacuum solutions, and in the static and spherically symmetric approximation we provide a classification of the possible scenarios in Section~\ref{subsec.eq.motion}.

Despite the similar pattern, there is an important difference between the regime of strong coupling in the scalar and tensor sectors that we should make explicit: whereas it is believed that a few general assumptions, such as difeomorphism and Lorentz invariance and quantum unitarity uniquely determine the self-interactions of a massless helicity two particle (which can be re-summed into the Einstein-Hilbert action, see e.g.~\cite{Deser:1969wk,Deser:2009fq}),
\begin{equation}\label{eq.free}
   \mathcal{L}_{\textrm{grav}}^{\Lambda_3\to\infty} = \frac{1}{2}M_{\textrm{Pl}}^2 R - \frac{1}{2}X,
\end{equation}
there is no such a guiding principle that establishes the self-interactions of a scalar degree of freedom. This ambiguity is codified in the coefficients of the EFT (apart from the linear ones $c_2^{(1)}=-c_4^{(0)}=-1/2$), that are arbitrary and become increasingly more important as the scalar sector couples strongly. This affects not only the self-interactions, but also the way in which the scalar couples to gravity, as it is evident from Eqs.~(\ref{eq:lag}) and~(\ref{eq.Lagrangians}). 

To make this apparent we include the contribution of a finite $\Lambda_3$, the energy scale that regulates the strength of the galileon self-interactions. In the first instance, and as we approach~(\ref{eq.strong.coupling.scalar}), an infinite number of terms contribute to the effective action, and apparently the theory loses predictability. At that point, two scenarios are most likely to unfold. On the one hand, we can consider that a finite number of coefficients $c_i^{(j)}$ and $d_i^{(j)}$ alone are different from zero, in such a way that the EFT remains predictive. On the other hand, there is also the possibility that an extra structure leads to a re-summation of the series, as it happens, for instance, in GR, or in the DBI model~\cite{Tseytlin:1999dj,Silverstein:2003hf,Alishahiha:2004eh}. In this paper, however, we take a conservative approach, and remain agnostic about the UV physics by restricting our attention to configurations that do not saturate the inequalities $X \lesssim \Lambda_2^4$, $\Box{\phi}\lesssim \Lambda_3^3$ and allow a perturbative analysis. That keeps the EFT under control and guarantees that, as a first approximation, it is sufficient to work Eq.~(\ref{eq.free}) to the next order in the series expansion. The resulting action contains two additional operators of mass dimension six, resulting in
\begin{equation}\label{eq.selfinteractions}
    \mathcal{L}_{\textrm{grav}}=\mathcal{L}_{\textrm{grav}}^{\Lambda_3\to\infty}+
    \frac{M_{\textrm{Pl}}}{\Lambda_3^3}\left[c_4^{(1)}XR - 2c_4^{(1)}[(\Box\phi)^2-\phi^{\mu\nu}\phi_{\mu\nu}]
 +d_4^{(-1)}\frac{1}{X}{\epsilon^{\mu\nu\rho}}_{\sigma}\epsilon^{\mu'\nu' \rho'\sigma}\phi_{\mu}\phi_{\mu'}\phi_{\nu\nu'}\phi_{\rho\rho'}
 \right]+ \ldots.
\end{equation}
As in (\ref{eq.IRaction}) the ellipses make reference to operators that are suppressed by higher powers of a mass scale. Interestingly, the covariant galileons $c_3^{(1)}$, $c_4^{(2)}$ and $c_5^{(2)}$ show up to the next order in the perturvative expansion, thus they can be neglected as a first approximation~\cite{Pirtskhalava:2015nla,Santoni:2018rrx}. 

\subsection{Infrared and ultraviolet modifications of gravity in our EFT framework}

We add some remarks on the coefficients $c_4^{(1)}$ and $d_4^{(-1)}$ of Eq.~(\ref{eq.selfinteractions}). In flat spacetime, the operator $c_4^{(1)}$ contributes to the effective action as a total derivative, but contrary to $c_3^{(0)}$ and $c_5^{(0)}$, it reappears as a dynamical contribution after the inclusion of gravity. This would not be the first time that this term has been proposed in the literature: it constitutes one of the Fab Four\footnote{The original definition of the Fab Four was using $G_{5}\sim\phi$, which by an integration by parts can be cast as $G_4\sim X$ ~\cite{Kobayashi:2014eva, Silva:2016smx, Diez-Tejedor:2018fue}.} of Charmousis {\it et al}~\cite{Charmousis:2011bf, Charmousis:2011ea} (a self-tuning mechanism to address the cosmological constant problem), and appears in the decoupling limit of the dRGT massive gravity theory~\cite{deRham:2010ik} (see e.g. Refs.~\cite{Starobinsky:2016kua, Salahshoor:2018plr, Babichev:2016jom, Brihaye:2016lin} for some gravitational and cosmological applications). For practical purposes, we can always absorb the coefficient $c_4^{(1)}$ into the scale $\Lambda_3$ and fix it with no loss of generality to $c_4^{(1)}=0,\,\pm1/2$, as we do in the following.

The operator $d_4^{(-1)}$, on the contrary, is arbitrary and beyond the Horndeski theory and has not received much attention before, as far as we are aware. The main reason why we have included this term into the effective action is that non-minimally coupled operators in the quartic and quintic sectors modify, in general, the speed of propagation of GWs~\cite{Lombriser:2015sxa, Lombriser:2016yzn}. In the framework of beyond Horndeski, however, the difference between the speed of light and the speed of GWs can be removed completely by demanding~\cite{Creminelli:2017sry}
\begin{equation}
    G_{5X}=0, \quad F_5=0,\quad 2G_{4X}-XF_4+G_{5\phi}=0. \label{constraint}
\end{equation}
If we concentrate on models where $G_5=0$, as is the case to leading order in Eq.~(\ref{eq.selfinteractions}), the condition $F_4=2G_{4X}/X$, for $X\neq0$, implies $c_{\textrm{GW}}=1$,\footnote{The opposite direction in the previous statements is not necessarily truth, and on-shell conditions may also lead to GW luminal propagation, see e.g.~\cite{Bordin:2020fww, Copeland:2018yuh}. However, in general models that do not satisfy the previous conditions are likely to require fine-tuning in order to satisfy $c_{\textrm{GW}}=1$.} in agreement with the recent bound on the speed of GWs in vacuum, $-3\times10^{-15}\leq c_{\text{GW}}/c-1\leq 7\times 10^{-16}$~\cite{TheLIGOScientific:2017qsa,Goldstein:2017mmi,Monitor:2017mdv}. This condition limits severely cosmological scenarios where the scalar field is assumed to be homogeneously distributed in space,\footnote{This is not the case for the purpose of this paper, where the scalar field is clumped in localized configurations.} as it happens in e.g. most dark energy models~\cite{Ezquiaga:2017ekz,Creminelli:2017sry,Baker:2017hug,Sakstein:2017xjx,Amendola:2017orw}. When interpreted most broadly, the impact of the observations may be less evident, and this is the reason why we consider different realizations depending on whether we want to remain in Horndeski, $c_4^{(1)}=\pm1/2$, $d_4^{(-1)}=0$ (at the expends of having $c_{\textrm{GW}}\neq c$ around some backgrounds), or prefer a more ``viable'' beyond Horndeski model, $c_4^{(1)}=1/2$, $d_4^{(-1)}= 1$, or $c_4^{(1)}=-1/2$, $d_4^{(-1)}= -1$.\footnote{The stability criteria of Refs.~\cite{Gleyzes:2014dya,Gleyzes:2014qga} shows a ghost in the spectrum for the choices $c_4^{(1)}=1/2$, $d_4^{(-1)}=0$ and $c_4^{(1)}=1/2$, $d_4^{(-1)}=1$. However, the instability only manifests itself when $X\sim \Lambda_2^4$ and the theory is strongly coupled. At this point, higher order operators in~(\ref{eq.selfinteractions}) come into play and can save the model.} Other values of $d_4^{(-1)}$ different from 0 and $\pm 1$ are possible but we will not consider them in this work. If both $c_4^{(1)}=d_4^{(-1)}=0$, it becomes necessary to identify the next order in the series expansion~(\ref{eq.selfinteractions}), where the covariant galileons appear. We discuss this possibility briefly in Appendix~\ref{app.higher.order}, but in the main text we use $c_4^{(1)}=d_4^{(-1)}=0$ as a shortcut for the EKG model.

\begin{figure}[t!]
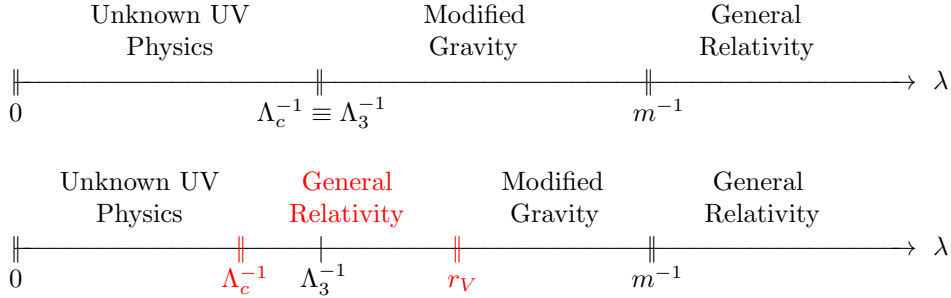


\hspace{2.6cm}Unknown UV \hspace{2.05cm} Modified \hspace{2.2cm} General \\
\vspace{-.2cm}
\hspace{2.96cm}Physics \hspace{2.6cm} Gravity \hspace{2.15cm} Relativity

\vspace{-0.2cm}
\[ \xrightarrow{\hspace*{11.7cm}} \; \lambda \]

\vspace{-.63cm}
\hspace{1.55cm}$|$\hspace{-0.05cm}$|$ 
\hspace{2.7cm} 
\hspace{0.8cm} $|$\hspace{-0.05cm}$|$ 
\hspace{1.65cm} 
\hspace{2.3cm}$|$\hspace{-0.05cm}$|$

\hspace{1.53cm}0 \hspace{2.85cm} 
$\Lambda_c^{-1}\equiv\Lambda_3^{-1}$ \hspace{1.cm} 
\hspace{2.cm}$m^{-1}$

\vspace{0.5cm}

\hspace{2.2cm}Unknown UV \hspace{0.85cm} {\color{red} General} \hspace{1.2cm} Modified \hspace{1.2cm} General \\
\vspace{-.2cm}
\hspace{2.56cm}Physics\hspace{1.28cm} {\color{red} Relativity} \hspace{1.16cm} Gravity \hspace{1.15cm} Relativity

\vspace{-0.2cm}
\[ \xrightarrow{\hspace*{11.7cm}} \; \lambda \]

\vspace{-.63cm}
\hspace{1.55cm}$|$\hspace{-0.05cm}$|$ 
\hspace{2.7cm}{\color{red} $|$\hspace{-0.05cm}$|$}
\hspace{0.7cm} $|$
\hspace{1.55cm}{\color{red} $|$\hspace{-0.05cm}$|$}
\hspace{2.3cm}$|$\hspace{-0.05cm}$|$

\hspace{1.53cm}0 \hspace{2.5cm}{\color{red} $\Lambda_c^{-1}$} \hspace{0.3cm}$\Lambda_3^{-1}$ \hspace{1.2cm}{\color{red} $r_V$}
\hspace{1.9cm}$m^{-1}$

  \caption{{\bf The EFT as a function of the scale.} {\it Upper panel:} Naive picture based on perturbative field theory. {\it Lower panel:} An improved interpretation considering the possibility of strong coupling. An intermediate GR regime enclosed between the cutoff scale $\Lambda_c^{-1}$ (that does not necessarily coincides with $\Lambda_3^{-1}$) and the Vainshtein radius $r_V$ emerges. In this paper we always assume that the mass $m$ of the scalar field is lower than the scale $\Lambda_3$.}
\label{Fig:Regimes}
\end{figure}

Before we proceed, let us look back at our general expressions~(\ref{eqs.WBG}) and~(\ref{eq.functions:WBG}). As long as the effective theory preserves the residual shift symmetry $\phi\to\phi+c$ of the galileon (which contrary to the transformations $\phi\to\phi+b_\mu x^{\mu}$ are not broken in general by couplings to gravitons), the scalar degree of freedom must necessarily remain massless, $m=0$, c.f. Eqs.~(\ref{eq.free}) and~(\ref{eq.selfinteractions}). This has far-reaching impacts given that a non-minimally coupled massless boson mediates a long-range fifth force of gravitational strength. Newtonian gravity has been tested down to submillimeter distances in local experiments~\cite{Lee:2020zjt}, where no  deviation from the inverse square law has been manifested up to the scale of sensitivity of current instruments. Therefore, if gravity is modified, it must be in a non-trivial way, and there should be a mechanism that screens fifth forces at certain length scales~\cite{Clifton:2011jh,Joyce:2014kja}.

One way this can be naturally accomplished is through strong coupling effects,\footnote{This illustrates how an apparent problem of perturbative quantum field theory (reflected as quantum fluctuations becoming strongly coupled at an unacceptable low energy scale), transforms into a virtue of the model in the form of the breakdown of classical perturbation theory, which is necessary to evade successfully current bounds on fifth forces at local scales. This idea is elaborated in Ref.~\cite{Dvali:2006su}. See Figure~\ref{Fig:Regimes} for illustrative purposes.} where non-linear operators in the effective action modify the naive $1/r^2$ force associated to the scalar mode. This opens the possibility of a regime starting at the Vainshtein~\cite{Vainshtein:1972sx} radius $r_V$ where fifth forces remain muted, moving the cutoff $\Lambda_c$ of the EFT to larger energy scales that do not coincide necessarily with $\Lambda_3$, as we sketch in the lower panel of Figure~\ref{Fig:Regimes}. In the language of our effective model a quantitative description of the Vainshtein mechanism requires knowledge of the coefficients $c_i^{(j)}$ and $d_i^{(j)}$. This analysis is model-dependent and beyond the scope of the present paper, but see e.g.~\cite{Koyama:2013paa} for a discussion of this topic.

Related with this is the fact that most theories exhibiting Vainshtein mechanism suffer from superluminalities and the wrong sign for analyticity, and this has led to believe that they cannot be extended to the UV in a standard local and analytic Wilsonian way~\cite{Adams:2006sv}. In recent years, the inclusion of a mass term $m\neq 0$ in the effective action has been shown to arrange the lowest order $Y$ positivity bounds (following the notation of~\cite{Tolley:2020gtv}), and a careful choice of successively higher derivative corrections can do the same with the higher orders~\cite{deRham:2017imi} (see also Ref.~\cite{deRham:2017avq} for additional work on the positivity bounds in the context of scalar theories with a mass gap). The mass term breaks the invariance under internal shift transformations, nonetheless, the galileon symmetry is preserved at loop level and the new operator does not affect the non-renormalization theorems~\cite{Pirtskhalava:2015nla}. As a by-product, the mass induces an exponential decay of the fifth force at large distances, which is otherwise necessary for the existence of vacuum configurations that are localized in space, as we argue in Section~\ref{sec.boundary}. However, the new  $PQ$ and $D$ positivity bounds derived in~\cite{Tolley:2020gtv} are more constraining than previous ones and exclude the region of the operator coefficients that survives the $Y$ bounds, even for the massive case. This points to the fact that, if physical, the galileons must appear as the low energy limit of a more fundamental theory of gravity which most probably admits a weaker notion of locality~\cite{Dvali:2012zc,Keltner:2015xda}, and where the validity of the assumptions that support the positivity bounds is not evident~\cite{Bellazzini:2019xts,Alberte:2020jsk}.

In view of the above, we make the following distinction between theories of modified gravity: if the mass of the particles that mediate the gravitational force are all larger than about $10^{-3}\,$eV (the mass scale associated to the submillimeter distances~\cite{Lee:2020zjt}), leaving the graviton aside, we say that gravity is modified in the UV. In this case the mass terms induce short-range fifth forces that are not accessible to local experiments, in agreement with current observations (note however that these massive degrees of freedom may contribute to the observations if clumped in the form of the compact objects that we analyze in this paper). If on the contrary the mass of any particle apart from the graviton is smaller than $10^{-3}\,$eV, we say that it constitutes a modification of gravity in the infrared, the reason why we have not observed any signal of new physics is that in this case the new degrees of freedom must be screened, somehow, at local scales.\footnote{This screening is specially relevant if the new infrared physics in the gravity sector is assumed to address the cosmological constant problem; see e.g. Ref.~\cite{Dvali:2000hr,ArkaniHamed:2002sp,deRham:2010ik} for some influential papers. If we want the scalar field to have a non-trivial impact on the cosmological evolution today we need to impose the Compton wavelength, $m^{-1}$, to be larger than the size of the Hubble horizon, $H_{0}^{-1}$, leading to an upper bound on the mass of the new particle of about $m\sim 10^{-33}\,$eV (being larger than this value it will contribute as dark matter). If, in addition, we want the higher derivatives to play a role in a model that deviates from a standard quintessence component we need the theory to be coupled strongly at cosmological scales, that is, $X\sim \Lambda_2^4$, $\Box\phi\sim \Lambda_3^3$ in the language of this paper. A simple dimensional argument, taking $\phi\sim M_{\textrm{Pl}}$ and $\partial \sim H_0$, leads to $\Lambda_3 \sim (M_{\textrm{Pl}}H_0^2)^{1/3}\sim 10^{-13}\,$eV as the characteristic scale associated to the EFT. Notice that the two energy scales lie at the border of $\Lambda_3\sim M_{\textrm{Pl}}^{1/3}m^{2/3}$ (see Figure~\ref{fig.parameter.space}, left panel).} We naturally expect $\Lambda_3$ to be larger than $m$, and then $\Lambda_3\gtrsim 10^{-3}\,$eV, if gravity is modified in the UV, although the converse is not necessarily true and there may be infrared modified gravity theories where $\Lambda_3\gtrsim 10^{-3}\,$eV.\footnote{Note that the strength of the higher derivative operators are mediated by inverse powers of $\Lambda_3$, and therefore a lower bound on this scale constrains the possible signatures that these terms may leave on low energy physical observables.} This classification, as well as most of the numbers we have come across in this paragraph are shown as reference in Figure~\ref{fig.parameter.space}.

\subsection{The low energy effective model}

Our main goal is to construct regular field configurations that remain bounded and survive in absence of matter. For this purpose, and to circumvent Derrick's theorem, we consider time-dependent solutions that are localized in space, but periodic in time~\cite{Wilets:1990di,Lee:1991ax}. They are usually referred to as non-topological solitons, and were presented briefly in the introductory section. The time-dependence can be accommodated in the internal space associated to an {\it approximate} global symmetry that appears in the low energy regime of the massive EKG model~\cite{Mukaida:2014oza,Ibe:2019vyo}, leading to configurations that even if not stable can be long-lived for practical purposes, or rather we can enlarge the internal space with an exact $U(1)$ symmetry, as we do next. Both theories posses the same low energy-limit, hence the same solutions to the field equations in the regime where the fields remain small, despite they may differ for configurations of size of the order of the Compton wavelength, where the relativistic effects become important.

A simple realization of such a picture is achieved by promoting $\phi$ to a complex field, and using the complex conjugate $\bar{\phi}$ to obtain real expressions in the Lagrangian, e.g. $X\equiv g^{\mu\nu}\phi_{\mu}\bar{\phi}_{\nu}$. This is easy to implement if we decide to keep terms with even powers of the scalar field, as it naturally happens in our general effective theory given the $\mathbb{Z}_2$ symmetry that appears at low energies. Notice that there is an apparent ambiguity in the quartic sector that is beyond the Horndeski theory, but we proved that all possible choices that make ${\epsilon^{\mu\nu\rho}}_{\sigma}\epsilon^{\mu'\nu' \rho'\sigma}\phi_{\mu}\phi_{\mu'}\phi_{\nu\nu'}\phi_{\rho\rho'}$ real are equivalent.
Combining Eqs.~(\ref{eq.free}) and~(\ref{eq.selfinteractions}), and introducing the mass term for the scalar field, we finally obtain:
\begin{eqnarray}\label{eq.interactions}
 \mathcal{L}_{\textrm{grav}} &=& \frac{1}{2}M_{\textrm{Pl}}^2 R - X -m^2\phi\bar{\phi} \\
 && + \frac{M_{\textrm{Pl}}}{\Lambda_3^3}\left[c_4^{(1)}XR - 2c_4^{(1)}[\Box\phi\Box\bar{\phi}-\phi^{\mu\nu}\bar{\phi}_{\mu\nu}]
 +d_4^{(-1)}\frac{1}{X}{\epsilon^{\mu\nu\rho}}_{\sigma}\epsilon^{\mu'\nu' \rho'\sigma}\phi_{\mu}\bar{\phi}_{\mu'}\phi_{\nu\nu'}\bar{\phi}_{\rho\rho'}
 \right]+ \ldots, \nonumber
\end{eqnarray}
where in the first line we have removed the $1/2$ factor in front of the kinetic term $X$ in order to get the standard normalization of a complex scalar field.

Equation~(\ref{eq.interactions}) is the final result of this section and the effective action that we consider for the rest of this paper. It consists of two dimensionless parameters, $c_4^{(1)}=0,\pm 1/2$ and $d_4^{(-1)}$, the last of which remains arbitrary, and two energy scales, $\Lambda_3$ and $m$, with the condition that $m<\Lambda_3\ll M_{\textrm{Pl}}$. The EKG model is recovered in the limit where $\Lambda_3\to\infty$, i.e. $c_4^{(1)}=d_4^{(-1)}=0$, whereas to obtain GR we need to impose further $m\to\infty$. We now proceed to compute the equations of motion and construct localized vacuum solutions in this theory.

\section{Horndeski stars}\label{sec3}

The first step to construct soliton solutions in our general effective model is to derive the equations that describe the static and spherically symmetric approximation. A more general study including rotation and time evolution could be possible, however, to make some initial progress we are mainly interested in the simplest asymptotic, highly symmetric and isolated case. (These configurations are ``sourced'' by a scalar degree of freedom and therefore any component of angular momentum, even if present in the initial data, would be probably radiated in the formation process~\cite{Palenzuela:2017kcg,Sanchis-Gual:2019ljs}.)

\subsection{Equations of motion}\label{subsec.eq.motion}

The equations of motion (in its covariant presentation) are obtained by varying the action with respect to the metric tensor and the scalar field. As we anticipated the low energy dynamics is dominated by the operators of Eq.~(\ref{eq.interactions}). A variation with respect to $g_{\mu\nu}$ results in
\begin{subequations}
\begin{equation}\label{fieleq}
    \left(M_{\textrm{Pl}}^2+2c_4^{(1)}\frac{M_{\textrm{Pl}}}{\Lambda_3^3}X\right)G_{\mu\nu} -  t_{\mu\nu}+\frac{M_{\textrm{Pl}}}{\Lambda_3^3}\left( c_4^{(1)}a_{\mu\nu}+
    d_4^{(-1)}\frac{Xb_{\mu\nu}-c_{\mu\nu}}{2X^2}\right)= 0,
\end{equation}
where $G_{\mu\nu}$ is the Einstein tensor and $t_{\mu\nu}$, $a_{\mu\nu}$, $b_{\mu\nu}$ and $c_{\mu\nu}$ are some functions of the metric tensor and the scalar field whose lengthily expressions are given in Eqs.~(\ref{eqs.app1}) of the Appendix~\ref{apendix1}. Notice that $t_{\mu\nu}$ is not suppressed by any factor of $\Lambda_3$, and its expression of Eq.~(\ref{eq.Cov}) coincides with the energy-momentum tensor of a scalar field without self-interactions. A variation with respect to $\phi$, on the contrary, yields the following equation for the scalar field:
\begin{equation}\label{scalareq}
    \Box\bar{\phi}-m^2\bar{\phi}+\frac{M_{\textrm{Pl}}}{\Lambda_3^3}\left(c_4^{(1)}a+d_4^{(-1)}\frac{X^2b-Xc+2d}{X^3}\right)=0,
\end{equation}
\end{subequations}
with the functions $a$, $b$, $c$ and $d$ given by Eqs.~(\ref{eqs.app2}). If we set $c_4^{(1)}=d_4^{(-1)}=0$, i.e. in the limit where $\Lambda_3\to\infty$, the higher derivative operators vanish and we recover the EKG theory, as it was otherwise expected.

To proceed, we assume a static and  spherically symmetric spacetime line-element, namely
\begin{equation}
 ds^{2}=-N^{2}(r)dt^{2}+g^{2}(r)dr^{2}+r^{2}d\theta^{2}+r^{2}\sin^2\theta d\varphi^{2} ,\label{metric}
\end{equation}
and impose a harmonic ansatz for the scalar field,
\begin{equation}
    \phi(t,r)=\sigma(r)e^{i \omega t} .\label{ansat}
\end{equation}
The functions $N(r)$, $g(r)$ and $\sigma(r)$ are real-valued and depend only on the areal radial coordinate $r$, whereas $\omega$, the angular frequency of oscillation of the scalar field, is real and constant. The explicit time-dependence of Eq.~(\ref{ansat}) reduces the equations of motion to a simpler time-independent system that is compatible with the static metric~\eqref{metric}, and that can be cast as an eigenvalue problem for the parameter $\omega$. After some manipulations, the equations of motion can be expressed in the following convenient form:
\begin{subequations}\label{eq.motion.spherical}
\begin{eqnarray}
    (1-\alpha)\frac{N'}{N}+(1-\beta)\frac{1-g^2}{2r} -\frac{r}{2M_{\textrm{Pl}}^2}\left[(1+\gamma_1)\sigma'^2-g^2\left(m^2-(1+\delta_1)\frac{\omega^2}{N^2}\right)\sigma^2\right] =0,\\
    (1-\alpha)\frac{g'}{g}-(1+\beta)\frac{1-g^2}{2r} -\frac{r}{2M_{\textrm{Pl}}^2}\left[(1-\gamma_2)\sigma'^2+g^2\left(m^2+(1-\delta_2)\frac{\omega^2}{N^2}\right)\sigma^2\right]  = 0,\\
    (1+\varepsilon)\sigma''+\left[\frac{2(1-\zeta)}{r}+(1-\eta)\left(\frac{N'}{N}-\frac{g'}{g}\right)\right]\sigma'-
    g^2\left(m^2-(1-\theta)\frac{\omega^2}{N^2}\right)\sigma= 0,\label{eq.KG.modified}
\end{eqnarray}
\end{subequations}
where $\alpha$, $\beta$, $\gamma_i$, $\delta_i$, $\epsilon$, $\zeta$, $\eta$ and $\theta$ are some dimensionless functions of $r$ that modify the EKG system and are reported in Appendix~\ref{apendix1}, Eqs.~(\ref{eq.parametersEKG}). In Section~\ref{sec.boundary} we identify the appropriate boundary conditions to solve this equations system, and refer to their solutions as Horndeski stars (HS), in close analogy to BSs sourced by a scalar field in the matter sector.

Despite its compact appearance, there are some important differences of the system~(\ref{eq.motion.spherical}) with respect to the EKG equations that might be worth discussing. On the one hand, second derivatives of the scalar field in the Lagrangian lead, in general, to higher order derivatives in the equations of motion. This is not what happens, however, for the Horndeski theory, where the equations of motion remain second order due to the very non-trivial internal structure of the action. In contrast, the operator $d_4^{(-1)}$ lies beyond Horndeski and breaks this nice property of the theory. Even if one naively expects derivatives of fourth order once away from Horndeski, this is not our case, where the resulting equations are third order due to the properties of the Levi-Civita tensor. Furthermore, some of the third derivatives can be also eliminated after manipulations using the properties of the Riemann tensor, $[\nabla_{\mu},\nabla_{\nu}]v_{\alpha}={R_{\mu\nu\alpha}}^{\beta}v_{\beta}$. Those that survive, however, vanish in spherical symmetry, and this is the reason why they do not appear explicitly in~(\ref{eq.motion.spherical}), which are of second order in $\sigma$. Regarding the metric functions, there appear second derivatives of $N$ and $g$ in the equations of motion, that even if not apparent from Eqs.~(\ref{eq.motion.spherical}), are codified in the function $\zeta$. Second order derivatives of the lapse function are common in GR, and they are in no way surprising. However, in spherical symmetry, second order derivatives of the radial component of the metric tensor are peculiar to the operators beyond Horndeski, and as a consequence they appear always multiplied by the coefficient $d_4^{(-1)}$, as we can appreciate in the Eq.~(\ref{eq.zeta}) of the appendix. It is important to stress that this does not affect the number of degrees of freedom that propagate in the theory~\cite{Gleyzes:2014dya,Gleyzes:2014qga}, and therefore the number of boundary conditions that we need to specify the solutions, as we discuss in more detail in Section~\ref{sec.boundary}.

As it is evident from the parametrization of Eqs.~(\ref{eq.motion.spherical}), we recover the EKG system in the limit where $\alpha$, $\beta$, $\ldots$, $\theta$ remain much smaller than one. This is what happens, modulo fine details,\footnote{As we clarify next in this paragraph, the contribution of the operator $c_4^{(1)}$ to the coefficients $\gamma_1$, $\gamma_2$ and $\eta$ is not small close to the centre of the configuration $r=0$, but this does not affect our conclusions.} when the scalar degree of freedom is coupled weakly, or in other words when
\begin{equation}\label{scale.self}
 \sigma_0\lesssim \sigma_0^{\textrm{s.f.}}\sim 
 \textrm{min.}\left\{\frac{M_{\textrm{Pl}}^{1/2}\Lambda_3^{3/2}}{m},\frac{\Lambda_3^3}{m^2}\right\}.
\end{equation}
To understand this condition on the central field, note that the absolute value of the kinetic term $X=-N^{-2}\omega^2\sigma^2+g^{-2}\sigma'^2$ is maximum at the origin, where $|X|\approx m^2\sigma_0^2$, and from this point onwards it decays smoothly to zero at spatial infinity; see Figure~\ref{fig.validity}, left panel, as an illustration. Something similar happens for the modulus of the d’Alembert operator acting on the scalar field, $\Box\phi=\left[N^{-2}\omega^2\sigma+g^{-2}\sigma''+g^{-2}(2/r+(\ln N)'-(\ln g)')\sigma'\right]e^{i\omega t}$, which also takes a maximum value $|\Box\phi|\approx m^2\sigma_0$ at the origin and then decays to zero at infinity; see the right panel of the same figure. If we demand the magnitude of the kinetic term to be lower than $\Lambda_2^4=M_{\textrm{Pl}}\Lambda_3^3$ {\it and} the modulus of d’Alembert operator acting on the scalar field to be lower than $\Lambda_3^3$, we obtain the inequality of Eq.~(\ref{scale.self}). This guarantees that the functions $\alpha$, $\beta$, $\delta_i$, $\epsilon$, $\zeta$ and $\theta$ in Eqs.~(\ref{eq.motion.spherical}) remain always smaller than one, and that they can be neglected in practical applications. The functions $\gamma_i$ and $\eta$, on the contrary, diverge as $1/r^2$ close to the origin.
However, they appear multiplied by factors of $\sigma'^2$ and $N'/N-g'/g$ in Eqs.~(\ref{eq.motion.spherical}), which for regular configurations scale as $r^2$ and $r$ near $r=0$, respectively [see e.g. Eqs.~(\ref{eq.expansion.origin})]. After some algebra, their contribution to the equations of motion can be estimated to be suppressed by factors of $X/\Lambda_2^4$ and/or $\Box\phi/\Lambda_3^3$, so they do not affect our conclusions and lead us to the same condition as in Eq.~(\ref{scale.self}). Compare this with the scale of strong gravity, that we identify to be of order $\sigma_0^{\textrm{s.g.}}\sim M_{\textrm{Pl}}$ in Appendix~\ref{app.SP}.

This leaves us with three possible scenarios, where $M^{1/3}_{\textrm{Pl}}m^{2/3}$ is the characteristic scale to compare with $\Lambda_3$:

$i)$ If $\Lambda_3\gg M^{1/3}_{\textrm{Pl}}m^{2/3}$ the scalar degree of freedom couples weakly, no matter the value of the central field, as long as the configurations are stable. This behaviour is due to the fact that in this case the strong gravity scale, $\sigma_0^{\textrm{s.g.}}\sim M_{\textrm{Pl}}$, is lower than the scale of the higher derivative operators, $\sigma_0^{\textrm{s.f.}} \sim M_{\textrm{Pl}}^{1/2}\Lambda_3^{3/2}/m$. Under this assumption, and as long as the central amplitude is much lower than the Planck scale, the configurations can be approximated by the solutions to the much simpler SP system, and in practice it is not necessary to solve the full system of Eqs.~(\ref{eq.motion.spherical}). We write the non-relativistic weak field limit of this theory in Appendix~\ref{app.SP}, and make a comparison between the solutions in Figure~\ref{fig.SP}. Once we increase the central amplitude up to $\sigma_0\sim M_{\textrm{Pl}}$ gravity couples strongly and the SP solutions cease to be physical. At this point the configurations develop a gravitational instability~\cite{Gleiser:1988rq,Gleiser:1988ih}, that for this particular case happens before we can see any signal of the non-canonical terms. As a consequence, the resulting objects are indistinguishable from BSs~\cite{Jetzer:1991jr,Schunck:2003kk,Liebling:2012fv}. 

$ii)$ If $\Lambda_3\sim M^{1/3}_{\textrm{Pl}}m^{2/3}$ the scalar degree of freedom approaches the strong coupling regime (in the limit of large fields), although it never reaches it for stable configurations. This behaviour is due to the fact that in this case the strong gravity scale, $\sigma_0^{\textrm{s.g.}}\sim M_{\textrm{Pl}}$, coincides with that of the higher derivative operators, $\sigma_0^{\textrm{s.f.}} \sim M_{\textrm{Pl}}^{1/2}\Lambda_3^{3/2}/m\sim \Lambda_3^3/m^2$. As in scenario~$i$ the non-relativistic regime is described in terms of the much simpler SP system. However, once we approach a central amplitude at the Planck scale, $\sigma_0\sim M_{\textrm{Pl}}$, gravity becomes strong, but with the important difference that the higher derivative operators show up before reaching the gravitational instability. As a consequence the most massive objects deviate with respect to BSs, leaving possible signatures of modified gravity in the configurations. In this paper we concentrate on this regime of the theory.

$iii)$ If $\Lambda_3\ll M^{1/3}_{\textrm{Pl}}m^{2/3}$ the scalar degree of freedom may couple strongly even when the central field is low. This behaviour is due to the fact that in this case the strong gravity scale, $\sigma_0^{\textrm{s.g.}}\sim M_{\textrm{Pl}}$, is higher than that of the higher derivative operators, $\sigma_0^{\textrm{s.f.}} \sim \Lambda_3^{3}/m^2$. This opens the very interesting possibility of objects that can be completely different to BSs, however, to make any progress we need to specify the particular form of the functions $G_i$ and $F_i$ that define the EFT, which results in a model-dependent analysis that requires further knowledge about the UV sector. We leave this study for a future work~\cite{Roque2020}.

The three scenarios are summarized in the left panel of Figure~\ref{fig.parameter.space}. The white region corresponds to scenario~$i$, where higher derivative operators are negligible no mater the amplitude of the central field, whereas the blue region to scenario $iii$, where configurations of enough amplitude are dominated by non-canonical terms. The scenario that we explore in this paper, $ii$, corresponds to the transverse line $\Lambda_3= M^{1/3}_{\textrm{Pl}}m^{2/3}$ (i.e. $\Lambda_3[10^{-13}\textrm{eV}]\sim m^{2/3}[10^{-33}\textrm{eV}]$ in physical units) that divides the blue and the white regions. Notice that we have excluded a region of the parameter space (shaded in grey) in scenario $iii$. This is because if the mass of the scalar particle is larger than the scale of the EFT, $m>\Lambda_3$, then no configuration exists in the low energy regime. In the right panel of this figure we show, for the combinations of parameters that make the higher derivative operators important, the transition from the EKG regime to the strongly coupled behaviour that appear as we increase the amplitude of the field at the origin, Eq.~(\ref{scale.self}). Note that the strong field regime serves as the border between the two behaviours. As we can appreciate from the figure for the configuration in scenario $ii$ it is necessary a central amplitude at the Planck scale to appreciate the differences with a BS.

\subsection{Boundary conditions}\label{sec.boundary}

\begin{figure}
\centering	
	\scalebox{0.45}{
	\input{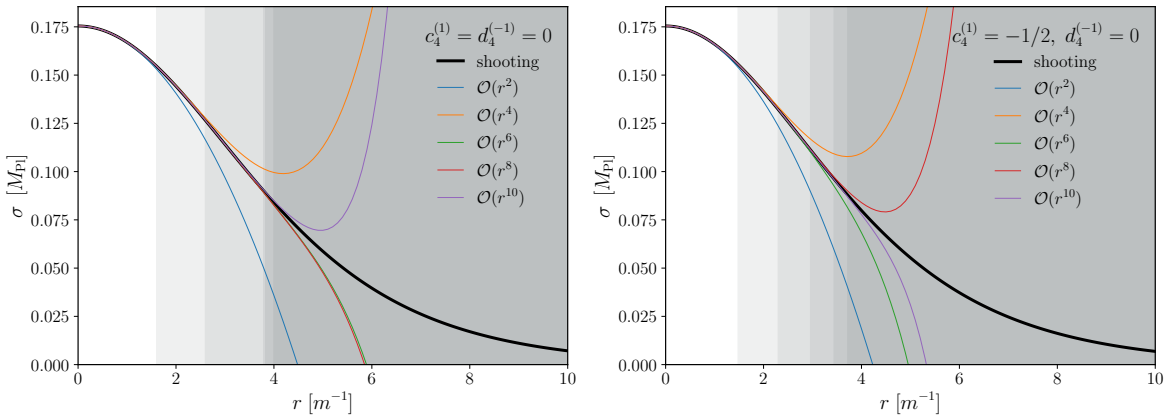}
	}
    \caption{{\bf A perturbative expansion close to the origin.} The profile of the wave function, $\sigma(r)$, to different orders in the series expansion. The colored lines on the left panel were obtained using Eq.~(\ref{eq.expansion.origin.1}), whereas those on the right panel using an expression that is not expanded in powers of $1/\bar{\Lambda}_3$. The solid lines were obtained integrating numerically Eqs.~(\ref{eq.motion.spherical}), as we describe in  Section~\ref{numerical}. For concreteness we fixed $\sigma_0=0.175M_{\textrm{Pl}}$, $\Lambda_3=1.5M_{\textrm{Pl}}^{1/3}m^{2/3}$, and considered two different choices of the dimensionless coupling constants: $c_4^{(1)}=d_4^{(-1)}=0$ (left panel), and $c_4^{(1)}=-1/2$, $d_4^{(-1)}=0$ (right panel). In each case the particular value of the frequency that lead to the ground state was obtained {\it a posteriori} after solving numerically the equations system, with $\bar{\omega}=1.1395$ (left panel) and $\bar{\omega}=1.1425$ (left panel). We shaded in grey the radii where the deviations with respect to the numerical solution are larger than 1\%. As we increase the order of the expansion the range of validity of the approximation grows, although the improvement is not significant from the eight order onward.}\label{fig.Origen}
\end{figure}	

For the numerical implementation, it is convenient to rewrite the equations in terms of the new dimensionless variables: 
\begin{eqnarray}
    \bar{r}\equiv m r, \quad \bar{\sigma}\equiv\frac{\sigma}{M_{\textrm{Pl}}},\quad \bar{\omega}\equiv\frac{\omega}{m}, \quad \bar{\Lambda}_3\equiv\frac{\Lambda_3}{M_{\textrm{Pl}}^{1/3}m^{2/3}}. \label{eq:dimless}
\end{eqnarray}
This choice drops out the dependence on the scalar field mass $m$ from the equations of motion, and leaves the combination in $\bar{\Lambda}_3$, together with the dimensionless coupling constants $c_4^{(1)}$ and $d_4^{(-1)}$, as the only free parameter of the model. Notice that the combination of energy scales in $\bar{\Lambda}_3$ is the same that we identified in Section~\ref{subsec.eq.motion}, with $\bar{\Lambda}_3 \gg 1$ ($\bar{\Lambda}_3 \ll 1$) corresponding to scenario $i$ ($iii$). The next step for the numerical analysis is to identify the boundary conditions that are compatible with solutions that are regular at the origin and localized in space, which then we decide to connect by presenting numerical solutions that interpolate between both regimes.  

To understand the solutions near the origin we perform a Taylor expansion of the equations around $\bar{r}=0$. The boundary conditions that lead to regular spacetime configurations (in the sense of no divergences of curvature scalars) are given by:
\begin{subequations}\label{boundayC}
\begin{align}
    & \bar{\sigma}(\bar{r}=0)= \bar{\sigma}_{0},\quad\quad \bar{\sigma}'(\bar{r}=0)=0,\\
    &N(\bar{r}=0)=N_{0},\hspace{0.5cm} N'(\bar{r}=0)=0,\\
    & g(\bar{r}=0)= 1,\hspace{0.95cm} \bar{g}'(\bar{r}=0)= 0,
\end{align}
\end{subequations}
with $\bar{\sigma}_0$ and $N_0$ two free and positive constants that one can choose arbitrarily.
The first coefficient represents the field amplitude at the origin, whereas the second one (the lapse function evaluated at the centre of the configuration), can always be reabsorbed in the definition of the time parameter and fixed to $N_0=1$ with no loss of generality. By introducing the ``initial'' conditions~(\ref{boundayC}) into the expanded equations of motion~(\ref{eq.motion.spherical}), and after some algebra, we obtain a perturbative solution of the form:
\begin{subequations}\label{eq.expansion.origin}
\begin{align}
    \bar{\sigma}(\bar{r}) = & \bar{\sigma}_0 \left\lbrace 1 + \bar{r}^2 \left[\frac{ 1-\bar{\omega }^2}{6}+\frac{c_4^{(1)}}{\bar{\Lambda}_3^3}\frac{  2\left(1- \bar{\omega }^2+2 \bar{\omega }^4\right)\bar{\sigma_0}^2}{6} \right.\right.  \nonumber\\
    & \hspace{1.9cm}\left.\left.+\frac{d_4^{(-1)}}{\bar{\Lambda}_3^3}\frac{\left(1-\bar{\omega }^2\right) \left[1+\bar{\omega }^2 \left(7-6  \bar\sigma _0^2\right) - \bar{\omega }^4\left(8-3 \bar{\sigma}_0^2\right)\right]}{27\bar{\omega }^2}+\ldots\right]
    +\dots\right\rbrace , \label{eq.expansion.origin.1} \\
    N^{2}(\bar{r}) = &  1 - \bar{r}^2 \left[\frac{\left(1-2\bar{\omega}^2\right)\bar{\sigma}_0^2}{3}-\frac{c_4^{(1)}}{\bar{\Lambda}_3^3}\frac{2\left[2\bar{\omega}^2-\bar{\omega}^4(2+3\bar{\sigma}_0^2)\right]\bar{\sigma}_0^2}{3} +\frac{d_4^{(-1)}}{\bar{\Lambda}_3^3}\frac{4\left(1-\bar{\omega}^2\right)\bar{\sigma}_0^2}{9} +\ldots\right]
    +\dots , \label{eq.expansion.origin.2}\\ 
    g^{2}(\bar{r}) = & 1 + \bar{r}^2 \left[\frac{\left(1+\bar{\omega }^2\right)\bar{\sigma}_{0}^{2}}{3} - \frac{c_4^{(1)}}{\bar{\Lambda}_3^3}\frac{   2\left(1+\bar{\omega }^2\right)\bar{\omega }^2\bar\sigma _0^4}{3 }-\frac{2c_4^{(1)}-d_4^{(-1)}}{\bar{\Lambda}_3^3}\frac{2\left(1-\bar{\omega}^2\right)^2\bar{\sigma}_0^2}{9}+\ldots\right]+\ldots,\label{eq.expansion.origin.3}
\end{align}
\end{subequations}
which is valid close to the origin. In order to get manageable expressions we have assumed that $\Lambda_3$ remains large in units of $M_{\textrm{Pl}}^{1/3}m^{2/3}$, scenario $i$. This condition does not necessarily hold in general, and actually the most interesting features associated to the higher derivative operators start at this scale, as we previously argued. The general expressions considering the higher operators are however cumbersome and not very illuminating, hence the reason why we have decided to write the truncated expressions of Eqs.~(\ref{eq.expansion.origin}). Intuitive as it may be from these equations the ellipsis make reference to factors of $\mathcal{O}(1/\bar{\Lambda}_3^6)$ and $\mathcal{O}(\bar{r}^4)$ that we have not included explicitly. In Figure~\ref{fig.Origen} we illustrate the validity of a series expansion near the origin for two representative examples. 

Even if limited in application, there is some useful information that we can infer from Eqs.~(\ref{eq.expansion.origin}). On the one hand, note that while the leading contribution of the operator $d_4^{(-1)}$ to the wave function is linear in $\bar{\sigma}_0$, Eq.~(\ref{eq.expansion.origin.1}), there are no factors of $c_4^{(1)}$ up to third order in the central amplitude. This is because to linear order in $\bar{\sigma}_0$ the spacetime line-element remains flat, $N^2(\bar{r})=g^2(\bar{r})=1$ in Eqs.~(\ref{eq.expansion.origin.2}) and~(\ref{eq.expansion.origin.3}), and in that limit the operator $c_4^{(1)}$ only contributes as a total derivative to the effective action, as we previously anticipated. We need to move away from Minkowski spacetime, i.e. from linear order in $\bar{\sigma}_0$, to see any effect of this term.

\begin{figure}[t]
\centering
	\scalebox{0.45}{
	\input{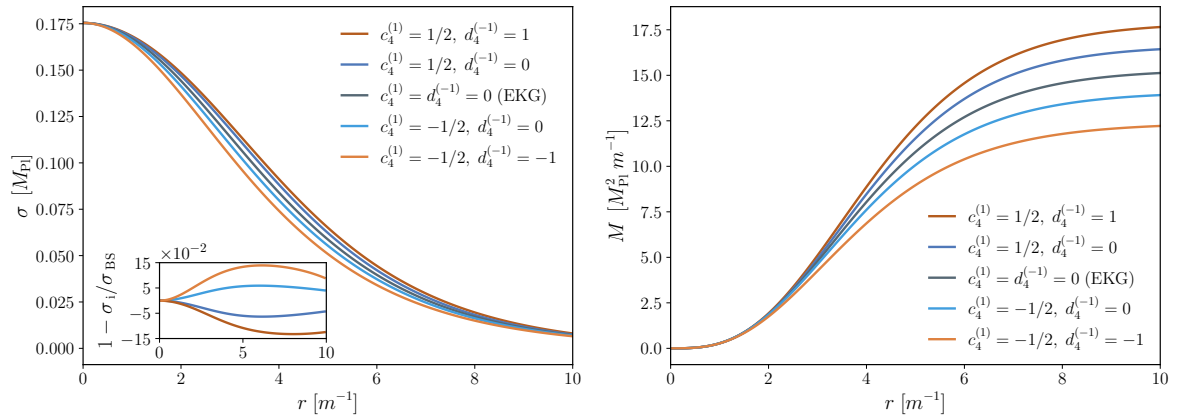}
	}  
	\caption{{\bf The profile of the star.} The radial component of the wave function $\sigma(r)$, Eq.~(\ref{ansat}) (left panel), and the mass profile $M(r)$, Eq.~(\ref{eq.mass.function}) (right panel), as functions of the radial coordinate in models with the low energy limit of Horndeski ($c_4^{(1)}=\pm 1/2$, $d_4^{(-1)}=0$), and ``viable'' beyond Horndeski ($c_4^{(1)}=\pm 1/2$, $d_4^{(-1)}=\pm 1$). For comparison we also show the case where $c_4^{(1)}=d_4^{(-1)}=0$, corresponding to a standard BS. The central amplitude $\sigma_0=0.175M_{\textrm{Pl}}$ was chosen to provide the maximum total mass in the $c_4^{(1)}=- 1/2$, $d_4^{(-1)}=- 1$ scenario (see Figure~\ref{MvswandR}), and we fixed $\Lambda_3=1.5M_{\textrm{Pl}}^{1/3}m^{2/3}$. The profile of the stars squeezes (expands) for negative (positive) coupling constants $c_4^{(1)}$ and $d_4^{(-1)}$, while their radii are not affected significantly. The inner figure shows the relative difference with respect to the BS limit, which is never larger than 15$\%$ for these configurations. The most significant properties of these objects are reported in Table~\ref{tabla1}.
	\label{Fig1}}
\end{figure}

\begin{figure}[t]
	\scalebox{0.45}{
	\input{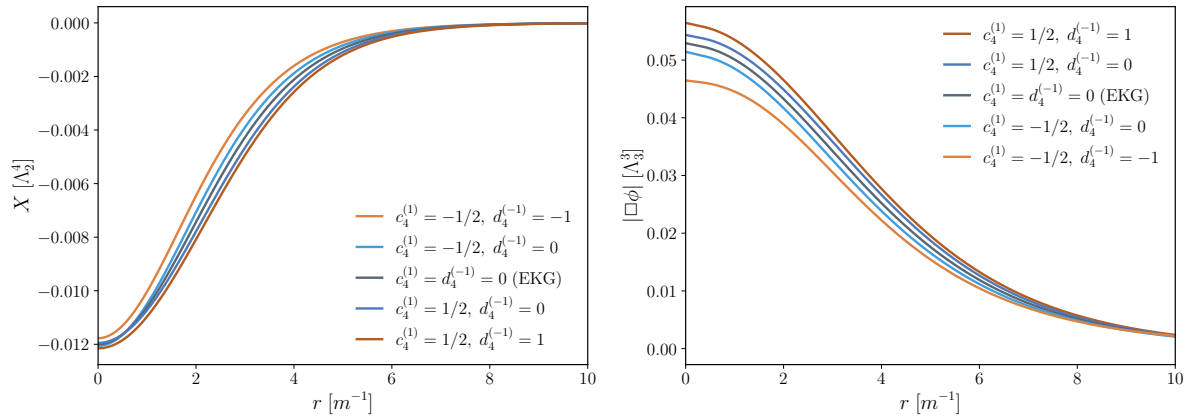}
	}
	\caption{{\bf The validity of the EFT.} The kinetic term, $X=\partial_{\mu}\phi\partial^{\mu}\bar{\phi}$ (left panel), and the modulus of the d'Alembert operator acting on the scalar field, $\Box\phi=\nabla_{\mu}\partial^{\mu}\phi$ (right panel), as functions of the radial coordinate for the same configurations as in Figure~\ref{Fig1}. All functions take a maximum absolute value at the origin, and then decay to zero at infinity. The derivative of $\Box\phi$ evaluated at the origin does not vanish, but this is consistent with the regularity conditions of Eqs.~(\ref{boundayC}). Note that these functions remain always lower than one in units of $\Lambda_2^4$ and $\Lambda_3^3$, respectively. This guarantees the insensibility of our configurations to high energy physics and quantum corrections. \label{fig.validity}}
\end{figure}

Note also that for ground state configurations (those that decay to zero without nodes) the second derivative of the wave function evaluated at the origin is negative (see the left panel of Figure~\ref{Fig1} as reference), and then $\bar{\omega}^2>1$, or equivalently $\omega>N_0m$ if we restore the physical units and the value of the lapse function at the origin. This implies that the coefficient in front of the coupling constant $c_4^{(1)}$ is positive definite. The same holds true for the coefficient in front of $d_4^{(-1)}$, once we assume that the central amplitude remains smaller than the Planck scale (there are no stable configurations for values of  $\bar{\sigma}_0\gtrsim 1$, as we discuss next). As a conclusion, positive values of the coupling constants $c_4^{(1)}$ and $d_4^{(-1)}$ will open the profile of the wave function to configurations that are broader in comparison with respect to a BS of same field amplitude, whereas negative values of these constants will close the profile to narrower configurations. In the next section we confirm this behaviour numerically for the general case of an arbitrary $\Lambda_3$ of order $M_{\textrm{Pl}}^{1/3}m^{2/3}$, see Figure~\ref{Fig1} as a preview. This suggests that positive (negative) couplings are associated to repulsive (attractive) self-interactions, although a more detailed analysis is necessary and beyond the scope of the present work.

To obtain localized configurations, the boundary condition at infinity must be the same as that for the vacuum state. That is why, at large distances, we force a flat metric by imposing: 
\begin{equation}\label{bounday2}
    \lim_{\bar{r}\to\infty}\bar{\sigma}(\bar{r})= 0,\quad
    \lim_{\bar{r}\to\infty}N(\bar{r})= N_{\infty},\quad 
    \lim_{\bar{r}\to\infty}g(\bar{r})= 1,
\end{equation}
where $N_{\infty}$ is an arbitrary and positive constant. From the equations of motion, and in the limit in which the scalar field remains small and the spacetime metric is flat, it is possible to identify the following asymptotic behaviour of the wave function:
\begin{equation}\label{eq.sigma.infinity}
    \bar{\sigma}(\bar r) \sim \frac{1}{\bar{r}}\exp \left(\sqrt{1-\frac{\bar{\omega}^2}{N_{\infty}^2}}\bar{r}\right),
\end{equation}
where the discrepancy of this formula with respect to the exact solution where the spacetime is curved is exponentially small. If one restores physical units it is not difficult to conclude that it is the mass term $m$ of the scalar field what makes possible the exponential decay of the wave function at spatial infinity, as long as $\omega<N_{\infty}m$, i.e. $\omega<m$ once we normalize the lapse function to one at infinity, as we do in Section~\ref{numerical}. This sets an upper bound on the range of the allowed frequencies by the mass $m$. Incidentally, Eq.~(\ref{eq.sigma.infinity}) is exactly the same expression that one obtains for the EKG model, where there are no higher derivative operators. This is by no means surprising since we already identified that our effective model is dominated by a canonical scalar field in the regime of weak coupling.

\subsection{Numerical profiles}\label{numerical}

In this section we proceed to solve  numerically the equations of motion~(\ref{eq.motion.spherical}). Starting from the ``initial'' conditions that we identified in~(\ref{boundayC}), and setting $N_0=1$ for convenience, we adjust the frequency eigenvalue $\bar{\omega}$ in such a way that the integrated solution from the origin matches the asymptotic behavior of Eq.~\eqref{bounday2} at infinity. We do so by a shooting method~\cite{Numerical, Dias:2015nua}. In principle, given a central amplitude $\bar{\sigma}_0$ there is an infinite number of frequencies that satisfy the conditions in~\eqref{bounday2}. In practice, however, we look for {\it ground state} configurations only, or equivalently, for scalar field profiles without nodes. For a given value of $\bar{\sigma}_0$ these solutions have associated the lowest possible magnitude of the frequency $\bar{\omega}$ that is compatible with the boundary conditions. Moreover, in the EKG limit the excited states are known to be unstable~\cite{Balakrishna:1997ej}, and this is the reason why we do not consider them here.

In Figure~\ref{Fig1} we show some representative examples of our numerical implementation. All the solutions that we present in this figure have the same values of $\bar{\Lambda}_3$ and $\bar{\sigma}_0$. For illustrative purposes we explore the following cases: $c_4^{(1)}=\pm1/2$, $d_4^{(-1)}=0$, corresponding to models with the low energy limit of Horndeski, and $c_4^{(1)}=1/2$, $d_4^{(-1)}= 1$ and $c_4^{(1)}=-1/2$, $d_4^{(-1)}=- 1$, corresponding to models that are beyond Horndeski. In the second case the parameter $d_4^{(-1)}$ has been chosen in such a way that the speed of propagation of gravitational waves coincide with that of light, $c_{\textrm{GW}}=1$. We use GR with a minimally coupled scalar field as a fiducial model to compare with, that in our parametrization corresponds to the case where  $c_4^{(1)}=d_4^{(-1)}=0$. The most relevant properties of these objects are summarised in Table~\ref{tabla1}. Although this preliminary analysis focuses on $\bar{\Lambda}_3=1.5$ and $\bar{\sigma}_{0}=0.175$, we later present results for a range of values of the EFT scale and the central amplitude. For completeness, in Figure~\ref{fig.validity} we show the two quantities that control the amplitude of loop corrections. As we can easily appreciate from the figure for the realizations that we have explored both functions remain much lower than one for all radii. This guarantees the validity of our solutions, which can be also trusted once quantum corrections are included. In the same way this is also consistent with the perturbative expansion of Eq.~(\ref{eq.interactions}), where higher order operators can be neglected in a first approximation (see the discussion of Appendix~\ref{app.higher.order} for a quantitative analysis).

By virtue of the boundary conditions, the scalar field vanishes asymptotically as the spacetime metric approaches the Schwarzschild solution. Therefore, if one chooses a sufficiently large radius $\bar{r}$, it is possible to use the Schwarzschild metric to asymptotically estimate the mass of one of these objects, $\bar{M}_T = \lim_{\bar{r}\to\infty} \bar{M}(\bar{r})$, where the function $\bar{M} (\bar{r})$ is given by\footnote{There are other possible mass definitions, see for example~\cite{Pugliese:2013gsa}.}
\begin{equation}\label{eq.mass.function}
    \bar{M} (\bar{r}) = 4\pi \bar{r}\left[ 1-\frac{1}{g^{2}(\bar{r})}\right],
\end{equation}
i.e. $M=\bar{M}M_{\textrm{Pl}}^2m^{-1}$ in physical units. For smaller radii, and as it is usually done, we refer to the equation above as the ``mass profile'' of the configuration, which is shown in Figure~\ref{Fig1}, right panel, for some of our solutions. We verify that the scalar field profile decreases monotonically as $\bar{r}$ increases, and $\bar{M}(\bar{r})$ tends to a constant value, as expected. However, the scalar field does not strictly vanish at finite $\bar{r}$, hence we define the effective radius\footnote{Note that this is not a physical border, in the sense of a surface that cannot be penetrated by particles or light rays. This will be relevant when discussing the phenomenology of these objects in Section~\ref{sec:ph}.} of the object, $\bar{R}_{95}$, as that where 95$\%$ of the total mass is contained, $\bar{M}_{95}=0.95\bar{M}_{T}$. We will discuss in detail the different configurations in the next section, but as an teaser notice that for $c_4^{(1)},\, d_4^{(-1)} > 0$, $\bar{M}(\bar{r})$ is bigger than in GR, and for $c_4^{(1)},\, d_4^{(-1)} < 0$, $\bar{M}(\bar{r})$ is smaller than in GR. The radius, however, remains almost insensitive to the new terms (see the numbers of Table~\ref{tabla1}), although they tend to increase (decrease) slightly for positive (negative) values of the coupling constants. Note that this is consistent with our previous discussion in Section~\ref{sec.boundary}.

\begin{figure}
    \scalebox{0.465}{
	\input{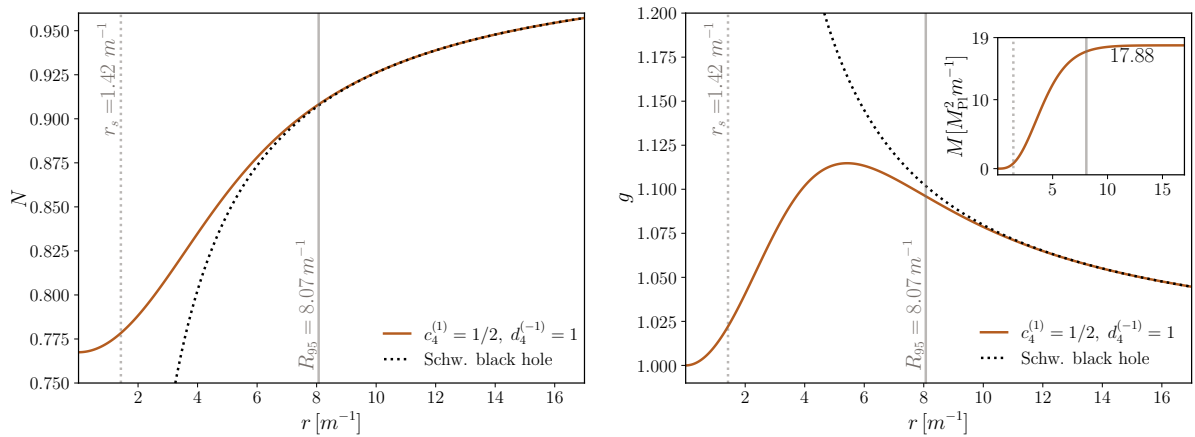}
	}
	\caption{{\bf The components of the metric tensor.} The lapse function $N(r)$ (normalized to one at spatial infinity, left panel), and the radial component of the metric tensor $g(r)$ (right panel), as functions of the radial coordinate for the configuration in the first line of Table~\ref{tabla1}. For reference we also show the Schwarzschild metric components $N_{\text{Schw.}}(r)=(1-r_{s}/r)^{1/2}$ and $g_{\text{Schw.}}(r)=N^{-1}_{\text{Schw.}}(r)$, where $r_{s}\equiv M/(4\pi M_{\textrm{Pl}}^2)$ is the Schwarzschild radius for an object of the same total mass $M=M_T=17.88 M_{\mathrm{Pl}}^{2} m^{-1}$ and negligible radius. Outside the star, $r\gtrsim R_{95}$, the scalar field decays exponentially and it is not possible to differentiate between the two objects. The mass profile associated to the HS is shown as a insert figure in the right plot.\label{figB}}
\end{figure}

At this point there is a remark that we need to make about the numerical solutions. One usually chooses the lapse function in such a way that $N_{\infty}=1$, but this is not respected by outwards integration starting from $N_0=1$, as we did. However, we can make use of the freedom to redefine the time coordinate, and the frequency accordingly, $(N,\bar{\omega})\mapsto x (N,\bar{\omega})$, in such a way that this condition is satisfied at infinity, where $x$ is an arbitrary constant factor greater than zero. To this effect, we first obtain the corresponding solution associated to a central amplitude $\bar{\sigma}_{0}$, which results in $\bar{\omega}$ as the frequency eigenvalue and $N(\bar{r})$ as the profile of the lapse function, with the property that it asymptotes a value that is different from one at infinity. To meet the condition $N_\infty=1$, we now redefine the time coordinate as $N^{\textrm{new}}(\bar{r})=xN(\bar{r})$, hence a new frequency $\bar{\omega}^{\textrm{new}}=x\bar{\omega}$, in such a way that
\begin{equation}
   x N(\bar{r}_{\textrm{max}})=\frac{1}{g(\bar{r}_{\textrm{max}})},\label{redef}
\end{equation}
with $\bar{r}_{\textrm{max}}$ the maximum radius of integration in the numerical code. From now on and to simplify the notation we will not write the super-index ``new'' explicitly, and it is understood that only rescaled values are reported. The corresponding profiles of the metric functions associated to one of these configurations once the rescaling has been carried out are shown in Figure~\ref{figB}.

\renewcommand{\tabcolsep}{11pt}
\renewcommand {\arraystretch}{1.5}
\begin{table}[t]
	\centering
	
\scalebox{0.94}{
	\begin{tabular}{l c c c c c c}
		\hline\hline
		&$\Lambda_{3}$ & $\sigma_{0}$ & $\omega$&$M_{T}$&$R_{95}$& $C$\\[-0.1cm]
		&$[M_{\mathrm{Pl}}^{1/3} m^{2/3}]$ & $[M_{\mathrm{Pl}}]$ & $[m]$&$[M_{\mathrm{Pl}}^{2} m^{-1}]$&$[m^{-1}]$& \\[0.1cm]
		\hline
		$c_{4}^{(1)}=1/2,\;d_{4}^{(-1)}=1$ &$1.5$& $0.175$&$0.880272$&$17.878$ &$8.068$&$0.084 $\\[0.1cm]
		$c_{4}^{(1)}=1/2,\, d_{4}^{(-1)}=0$ &$1.5$& $0.175$&$0.889535$&$16.633$ &$7.977$&$ 0.079$\\[0.1cm]
		 $c_{4}^{(1)}=d_{4}^{(-1)}=0$ \;(EKG) &---&$0.175$&$0.896136$&$15.314$&$7.959$ &$0.073 $\\[0.1cm]
		 $c_{4}^{(1)}=-1/2,\, d_{4}^{(-1)}=0$ &$1.5$& $0.175$&$0.902360$&$14.090$&$7.945$&$0.067 $\\[0.1cm]
		 $c_{4}^{(1)}=-1/2,\;d_{4}^{(-1)}=-1$ &$1.5$& $0.175$&$0.916858$&$12.413$&$8.033$&$0.058 $\\[0.1cm]
		\hline
	\end{tabular}
	}
	\caption{{\bf The parameters of the star.} The central amplitude $\sigma_0$, frequency $\omega$ (with the lapse normalized to one at spatial infinity, $N_{\infty}=1$), total mass $M_{T}$, effective radius $R_{95}$, and compactness $C$ [as defined in Eq.~(\ref{eq.compacness})] for the same configurations as in Figures~\ref{Fig1} and~\ref{fig.validity}. Positive (negative) values of the coupling constants $c_4^{(1)}$ and $d_4^{(-1)}$ increase (decrease) the mass $M_{T}$ in comparison with respect to a BS of same field amplitude $\sigma_0$. The radius $R_{95}$, however, is much less sensitive to the higher derivative operators. This is also evident from the profile of the stars that are modified only at intermediate radii, see left panel in Figure~\ref{Fig1}.}\label{tabla1}
\end{table}

\section{\label{sec:ph} Phenomenology}

We intend to take the first steps towards the study of phenomenological aspects of presented solutions that could be relevant for testing more realistic models. Due to the recent observational input from near-horizon astronomy, we concentrate on the possible role of HSs as exotic compact objects with potential astrophysical signals. For concreteness we consider the deflection of light rays passing near the surface of the stars, the emitted spectrum of accretion disks surrounding the objects, and the GW signal produced in binary collisions. Before doing that, however, we provide a general description of HSs using BSs and Schwarzschild black holes as the fiduacial models to compare with.

\subsection{Horndeski stars at the scale $\Lambda_3\sim M_{\textrm{Pl}}^{1/3}m^{2/3}$: general properties}\label{sec.HSproperties}

To complement the presentation of Section~\ref{sec3} we provide a more exhaustive analysis of the solutions to the Horndeski Eqs.~(\ref{eq.motion.spherical}) that are consistent with the boundary conditions~(\ref{boundayC}) and~(\ref{bounday2}). For different choices of the parameters $c_4^{(1)}$ and $d_4^{(-1)}$, and letting the scale of the EFT fixed to $\bar{\Lambda}_3=1.5$, in Section~\ref{numerical} we constructed the solutions that result for a specific value of the central amplitude $\bar{\sigma}_0$, that at that time we set for convenience to $\bar{\sigma}_0=0.175$. In practice we restricted to solutions without nodes. These objects present some differences with respect to their EKG counterpart, see e.g. Table~\ref{tabla1}, which we also reviewed for comparison. However, the differences are never larger than 20$\%$ in the case of the total mass, and hardly visible for the effective radius, which seems to be less sensitive to the higher derivative operators. The small deviations are by no means surprising and were expected from the scale of the EFT at the border of $\Lambda_3\sim M_{\textrm{Pl}}^ {1/3}m^{2/3}$.

\begin{figure}
\centering	
	\scalebox{0.455}{
	\input{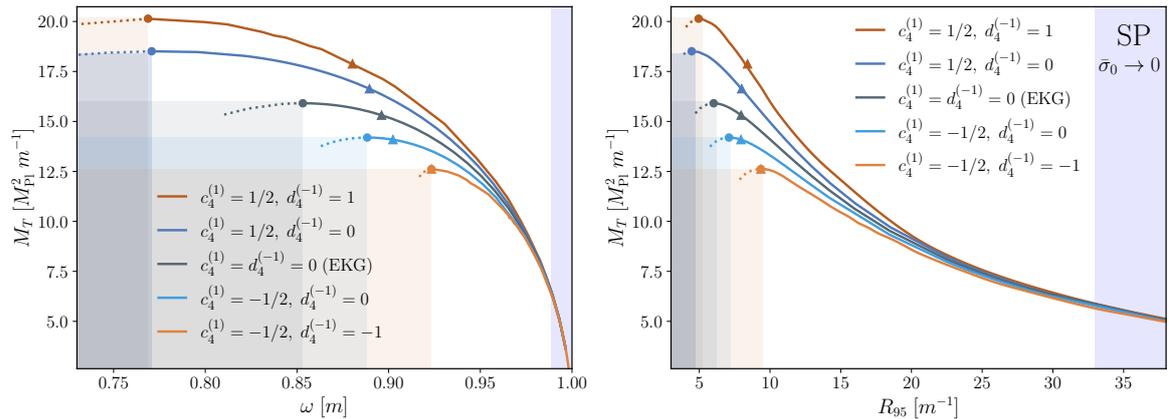}
	}
\caption{{\bf The $M_{T}$ vs $\omega$ and $M_{T}$ vs $R_{95}$ profiles.} The total mass $M_T$ as a function of the frequency, $\omega$ (left panel), and the effective radius, $R_{95}$ (right panel), for the same models as discussed in Figures~\ref{Fig1} and~\ref{fig.validity}, and Table~\ref{tabla1}. Each curve corresponds to a smooth line connecting a set of configurations computed varying the central amplitude $\sigma_0$, with the EFT scale fixed to $\Lambda_3=1.5M_{\textrm{Pl}}^{1/3}m^{2/3}$ at all times. Notice that configurations in the positive (negative) coupling branches always have higher (lower) masses than the equivalent BS. The points at the maximum of the curves make reference to the last stable configuration of the theory, such that the solutions to the left (dotted lines) in both panels are unstable under small radial perturbations. The triangles denote the configurations of Figures~\ref{Fig1} and~\ref{fig.validity}, and Table~\ref{tabla1}, obtained for $\sigma_0=0.175 M_{\textrm{Pl}}$. The area in purple to the right of these figures contains configurations that can be described in terms of the non-relativistic weak field approximation, where the relative difference with respect to the SP configurations is less than 1\% (see Appendix~\ref{app.SP} for details).\label{MvswandR}}
\end{figure}

To understand in greater detail what the higher derivatives lead to, we explore a larger range of the parameters $\bar{\sigma}_0$ and $\bar{\Lambda}_3$. In Figure~\ref{MvswandR}, and with the scale of the EFT still fixed to $\bar{\Lambda}_3=1.5$, we extend the solutions of Section~\ref{numerical} to include different values of the central amplitude $\bar{\sigma}_0$. In this instance we do not show explicitly the profiles of the wave function, which are not very different from those of Figure~\ref{Fig1}, and rather we plot the total mass $\bar{M}_{T}$ as a function of the frequency $\bar{\omega}$ (left panel), and the effective radius $\bar{R}_{95}$ (right panel). Each point in these curves corresponds to a different configuration, all of them characterized by a given value of $\bar{\sigma}_0$ which for the purpose of illustration we vary in the range $0.01<\bar{\sigma}_0<0.2$. 

As we decrease the value of the central amplitude, $\bar{\sigma}_0\to 0$, the normalized frequency approaches unity (in units of $m$), whereas the size of the corresponding object grows indefinitely. In that limit the effect of the higher derivative operators becomes negligible, up to the point that it is not necessary to solve the Horndeski Eqs.~(\ref{eq.motion.spherical}) and the non-relativistic weak field approximation that we describe in Appendix~\ref{app.SP} is sufficient for practical purposes. Under those circumstances the profile of the wave function can be expressed as a rescaling of a numerical function, Eq.~(\ref{eq.profile.sigma}), with other quantities of interest related in a simple way to this expression. The success of this approximation can be easily appreciated in Figure~\ref{fig.SP}, and also from the well known SP limit $\bar{M}_{T}\approx 213/\bar{R}_{95}$, where all curves converge at large radii/small masses in the $\bar{M}_{T}$ vs $\bar{R}_{95}$ plot. For reference, in both panels of Figure~\ref{MvswandR} we have shaded in purple the degenerated region where the relative differences in the total mass are always smaller than 1\%, no matter the choice of the coupling constants. 

Only as we increase the central amplitude and approach the regime where the scalar field couples strongly, $\sigma_0\sim \sigma_0^{\textrm{s.f.}}$, Eq.~(\ref{scale.self}), the deviations with respect to the SP system become evident. For a choice of the scales in which $\Lambda_3\sim M_{\textrm{Pl}}^ {1/3}m^{2/3}$ the onset of the strong field and the strong gravity regimes coincide, $\sigma_0\sim \sigma_0^{\textrm{s.g.}}\sim \sigma_0^{\textrm{s.f.}}~\sim M_{\textrm{Pl}}$, and this is why the different curves of Figure~\ref{MvswandR} start to deviate one from each other only as $\bar{M}_{T}$ approaches its maximum value, where GR effects are already important. We anticipated this behaviour in Section~\ref{subsec.eq.motion}, scenario $ii$ (see also Figure~\ref{fig.SP} in the Appendix for additional details). 

\subsubsection{Stability}
The state of maximum mass of a BS is known to represent the last stable configuration of the star, the solutions to the left of this point in the $\bar{M}_T$ vs $\bar{\omega}$ and the $\bar{M}_T$ vs $\bar{R}_{\textrm{95}}$ curves being unstable under small radial perturbations. The existence of a family of BSs which are linearly stable under small radial perturbations was proven in Refs.~\cite{Gleiser:1988rq, Gleiser:1988ih}, and the argument is as follows. Consider an equilibrium solution to the EKG equations. The evolution of the small radial deviations about the equilibrium configuration is described in terms of a linear system with coefficients that depend on the background fields, and in particular on the central amplitude $\bar{\sigma}_0$. At this point it is not important to specify the details of this system, but the interested reader may go over Eqs.~(34) and~(35) of Ref.~\cite{Gleiser:1988ih}. Moving to Fourier space we can express these equations as an eigenvalue problem for the square of the oscillation frequencies $\bar{\chi}_i$, where the bar indicates that the frequencies have been normalized with the mass $m$. If the eigenvalues are all positive, $\bar{\chi}_i^2>0$, the configuration is stable under small radial perturbations. However, if an eigenvalue turns negative, even if it is one only, then a generic perturbation will grow exponentially in time, destabilizing the equilibrium configuration.

The eigenvalues are functions of the central amplitude, $\bar{\chi}_i^2(\bar{\sigma}_0)$, and their signs may change as we move along the $\bar{M}_T(\bar{\omega})$ and $\bar{M}_T(\bar{R}_{\textrm{95}})$ curves. The existence and identification of zero-frequency modes $\bar{\chi}_i^2(\bar{\sigma}_{0c})=0$ play a central role in establishing the boundaries between the stable and unstable regimes. Zero-frequency modes connect two equilibrium configurations of nearby amplitudes $\bar{\sigma}_0$ and $\bar{\sigma}_0+\delta\bar{\sigma}_0$ through a (time-independent) perturbation. If we restrict to perturbations that preserve the total charge $Q_T$, zero-frequency modes can exist only if there exist two neighboring equilibrium configurations $\bar{\sigma}_{0c}$ and $\bar{\sigma}_{0c}+\delta\bar{\sigma}_0$ of the same charge, that is, if $d\bar{Q}_T/d\bar{\sigma}_0|_{\bar{\sigma}_{0}=\bar{\sigma}_{0c}}=0$; see the discussion of Section~4, Ref.~\cite{Gleiser:1988ih}, for details. For a BS the states of (local) maximum/minimum charge $\bar{Q}_T$ are known to coincide with those of (local) maximum/minimum mass $\bar{M}_T$, see e.g. Figure 1 of~\cite{Gleiser:1988ih}, hence the eigenvalues $\bar{\chi}_i^2(\bar{\sigma}_0)$ can change sign only at the extrema $\bar{\sigma}_0=\bar{\sigma}_{0c}$ of the curve $\bar{M}_T(\bar{\sigma}_0)$.

As the eigenvalues form an infinite discrete ordered sequence, $\bar{\chi}_{i}^2(\bar{\sigma}_0)<\bar{\chi}_{i+1}^2(\bar{\sigma}_0)$, we can concentrate on the fundamental mode, $\bar{\chi}_0^2(\bar{\sigma}_0)$, i.e. the one with the lowest frequency. Solving numerically the eigenvalue problem the authors of~\cite{Gleiser:1988ih} proved that the fundamental mode is stable for configurations of small field amplitude, i.e. $\bar{\chi}_0^2(\bar{\sigma}_0)>0$ in the limit where $\bar{\sigma}_0\to 0$, and that $\bar{\chi}_0^2(\bar{\sigma}_0)$ changes sign after crossing the first maximum of the curve $\bar{M}_{T}(\bar{\sigma}_{0})$, located at $\bar{\sigma}_0=\bar{\sigma}_0^{(1)}$, i.e. $\bar{\chi}_0^2(\bar{\sigma}_0)<0$ for $\bar{\sigma}_0>\bar{\sigma}_0^{(1)}$, signaling the onset of an instability. This state is represented by a black point in the two panels of Figure~\ref{MvswandR} (remember that the amplitude $\bar{\sigma}_0$ grows from the right to the left in these plots). The appearance of an unstable mode at the maximum of the curves $\bar{M}_{T}(\bar{\omega})$ and $\bar{M}_{T}(\bar{R}_{95})$ was confirmed subsequently using numerical evolutions in Ref.~\cite{Hawley2000}.

\begin{figure}
\centering	
	\scalebox{0.455}{
	\input{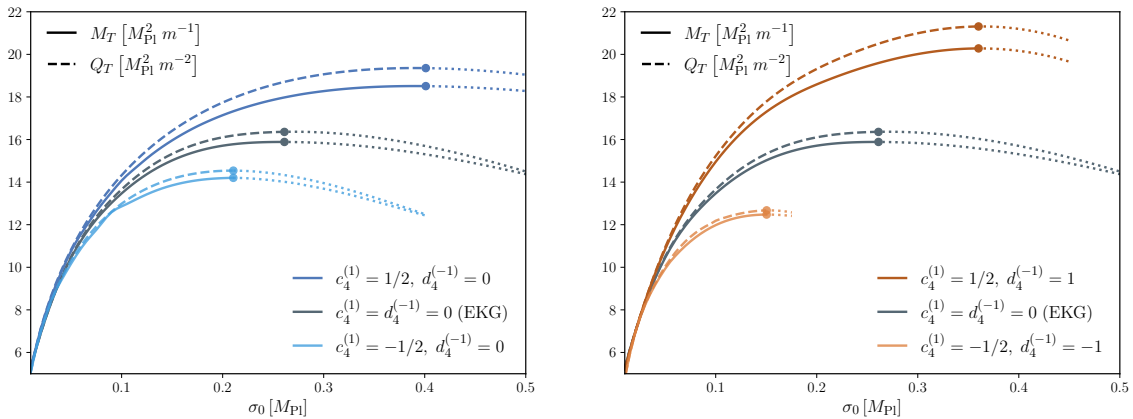}
	}
\caption{{\bf The $M_{T}$ and $Q_{T}$ vs $\sigma_0$ profiles.} The total mass, $M_{T}$, and charge, $Q_T$, as functions of the central amplitude, $\sigma_0$, for the same theories as discussed in Figures~\ref{Fig1} and~\ref{fig.validity}, and Table~\ref{tabla1} (compare also with Figure~\ref{MvswandR}). Note that the state of maximum mass coincides with that of maximum charge, as it also happens for BSs, see e.g. Figure~1 in Ref.~\cite{Gleiser:1988ih} for comparison. The solutions to the right of the first maximum of the $\bar{M}_T(\bar{\sigma}_0)$ and $\bar{Q}_T(\bar{\sigma}_0)$ curves (dotted lines) correspond to unstable configurations.
\label{fig.M_TandQ_T}}
\end{figure}

This conclusion can be extended to the case of HSs, the reason why is that our effective model is connected continuously with the EKG system in the regime in which the scalar field is coupled weakly, then the eigenvalues associated to the fundamental mode are necessarily positive definite  to the left of the first maximum of the curve $\bar{Q}_{T}(\bar{\sigma}_{0})$, where the configurations are stable. As for the BS, the local extrema of the total charge $\bar{Q}_T$ coincide with those of the total mass $\bar{M}_T$, see Figure~\ref{fig.M_TandQ_T}, and only after crossing these points unstable modes can appear (note that if we continue these curves to the right there is an infinite number of maxima and minima).\footnote{For theories with second order derivatives in the action the conserved current is given by
\begin{equation}
 j^{\mu}= -i\left[\frac{\partial\mathcal{L}}{\partial(\nabla_{\mu}\nabla_{\nu}\phi)}\nabla_{\nu}\phi-\nabla_{\nu}\left(\frac{\partial\mathcal{L}}{\partial(\nabla_{\mu}\nabla_{\nu}\phi)}\right)\phi  
 +\frac{\partial\mathcal{L}}{\partial(\nabla_{\mu}\phi)}\phi-c.c.\right],
\end{equation}
where $c.c.$ makes reference to complex conjugation. Associated to this current there is a conserved charge
\begin{equation}
    Q_T=\int_{\Sigma_t}j^{\mu}n_\mu d\gamma = -4\pi\int_0^\infty N g r^2j^0dr,
\end{equation}
where $n^{\mu}$ is the future-directed time-like unit normal vector to the Cauchy hypersurface $\Sigma_t$, $d\gamma=\sqrt{\textrm{det}(\gamma_{ij})} d^3x$ denotes the volume element on the hypersurface, and the second integral assumes a static spherically symmetric spacetime line-element of the form~(\ref{metric}).
}
Away from the BS limit, when $\Lambda_3\ll M_{\textrm{Pl}}^{1/3}m^{2/3}$, it could be possible, in principle, that no mode changes sign after the first maximum of the curve 
$\bar{M}_{T}(\bar{\sigma}_0)$,
but we do not expect this to happen for the region of the parameter space that we are exploring in this paper, where $\Lambda_3\gtrsim M_{\textrm{Pl}}^{1/3}m^{2/3}$.
A more detailed analysis is beyond the scope of the present work.

The configurations of Figure~\ref{fig.M_TandQ_T} have negative binding energy, $E_B = M_T-mQ_T$, and if unstable (i.e. to the right of the first maximum) they will migrate to the stable branch, or collapse to a black hole. If we extend these curves to the right in the figure, however, there appear unstable states of positive binding energy that when perturbed will explode and disperse to infinity (with is some cases a black hole remnant). In the BS limit this behaviour has been confirmed using non-linear numerical evolutions in e.g.~\cite{Seidel:1990jh,Balakrishna:1997ej,Guzman:2009xre}.

\subsubsection{Compactness}

Back to Figure~\ref{MvswandR}, positive (negative) values of the coupling constants $c_4^{(1)}$ and $d_4^{(-1)}$ move the $\bar{M}_{T}(\bar{\omega})$ and $\bar{M}_{T}(\bar{R}_{95})$ curves upwards (downwards), using the plot of the BS as reference. This is consistent with our discussion at the end of Section~\ref{numerical}, according to which the mass function $\bar{M}(\bar{r})$, and thus also the total mass $\bar{M}_T$, is bigger (smaller) for positive (negative) values of the coupling constants, although it is important to stress that the comparison is not exactly the same:  we previously compared configurations of the same field amplitude, $\bar{\sigma}_0=0.175$, whereas now configurations of the same effective radius $\bar{R}_{95}$, while the function $\bar{R}_{95}(\bar{\sigma}_0)$ is different from theory to theory. Notice that in both cases, i.e. the positive and the negative branches, the difference in the masses is larger once we include the beyond Horndeski operator $d_4^{(-1)}$.

The mass and radius of a HS can be read from the $\bar{M}_{T}(\bar{R}_{95})$ curves of Figure~\ref{MvswandR}, right panel, using the expressions $M_{T}=\bar{M}_{T}M_{\textrm{Pl}}^2/m$ and $R_{95}=\bar{R}_{95}/m$ for their normalizations. To recover the right units for the physical quantities we simply use the conversion factors
\begin{equation}\label{eq.mass.radius.HS}
    M_{T} =  \frac{5.31\bar{M}_{T}}{m[\textrm{eV}]}\times 10^{-12}\,M_{\odot} ,\quad
    R_{95} =  \frac{1.97\bar{R}_{95}}{m[\textrm{eV}]}\times 10^{-10}\,\textrm{km} .
\end{equation}
Bear in mind that these curves are only valid when $\Lambda_3 = 1.5 M_{\textrm{Pl}}^ {1/3}m^{2/3}$, and for some particular choices of the coupling constants $c_4^{(1)}$ and $d_4^{(-1)}$. Similar plots can be obtained for other combinations of the parameters, as long as we do not move away from $\Lambda_3\gtrsim M_{\textrm{Pl}}^ {1/3}m^{2/3}$. This is not only a technical issue but a physical condition since our effective theory losses predictability as the scalar field approaches the strong coupling scale $\sigma_{0}^{\textrm{s.f.}}$, Eq~(\ref{scale.self}). Actually, for values of $\Lambda_3$ lower than $M_{\textrm{Pl}}^ {1/3}m^{2/3}$ the effective action~(\ref{eq.interactions}) breaks down before we can reach the state of maximum mass that divides the stable and unstable branches and a more careful model-dependent examination of the EFT in the strong coupling regime is necessary.

\begin{figure}
 \centering	
 	\scalebox{0.45}{
	\input{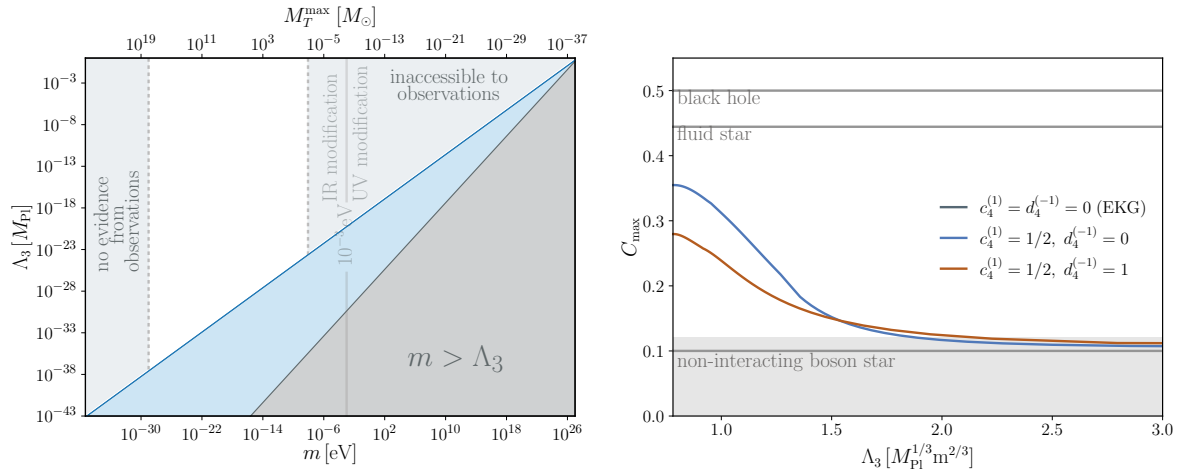}
	}
 	\caption{{\bf Horndeski stars as exotic compact objects.} {\it Left panel:} Same as Figure~\ref{fig.parameter.space}, left panel, with the upper horizontal axis indicating the mass of the most compact stable object for a given combination of the parameters of the model. This upper bound was obtained using Eq.(\ref{eq.mass.radius.HS}), where we fixed $\bar{M}_T$ to the maximum value of the BS, $\bar{M}_T^{\textrm{max}}=15.91$, and this is the reason why $M_T^{\textrm{max}}$ appears as independent of the EFT scale. This conversion is appropriate for values of $\Lambda_3\gg M_{\textrm{Pl}}^{1/3}m^{2/3}$, although it gradually begins to fall as we approach the transverse line that divides the white and the blue regions at $\Lambda_3~\sim M_{\textrm{Pl}}^{1/3}m^{2/3}$. Large deviations are expected in the blue region, where the theory is strongly coupled and our effective theory is not predictable. The white window shows the region of the parameter space that is well described by the EFT and is accessible to current astrophysical observations. {\it Right panel:} The maximum compactness $C_{\textrm{max}}$ allowed in stable HSs as a function of the EFT scale $\Lambda_3$ for some choices of the parameters of the model. For comparison we also show the following maximum compactness limits: Schwarzschild black hole ($C=1/2$), fluid star  ($C_{\textrm{max}}=4/9$)~\cite{Buchdahl:1959zz}, and BS without self-interactions ($C_{\textrm{max}}=0.1$). Above the shaded area we obtain super-emitters in the terminology of~\cite{Palenzuela:2017kcg} (see the discussion of Section~\ref{sec.GW} for details).\label{fig5}}
 \end{figure}

In practice, only stable configurations are of interest for phenomenological purposes. For theories with $\Lambda_3\gtrsim M_{\textrm{Pl}}^{1/3}m^{2/3}$, scenarios $i$ and $ii$ of Section~\ref{subsec.eq.motion}, the stable states are limited to the region of the $\bar{M}_T(\bar{R}_{95})$ plots where $\bar{M}_{T}\lesssim 10-20$ and $\bar{R}_{95} \gtrsim 5-10$. More specifically, in the limit where $\Lambda_3\gg M_{\textrm{Pl}}^{1/3}m^{2/3}$, we recover the same values as for the BS, which are given by $\bar{M}_T<\bar{M}_T^{\textrm{max}}=15.91$ and $\bar{R}_{95}>\bar{R}_T^{\textrm{min}}=6.01$. If gravity is modified in the UV, $m\gtrsim 10^{-3}\,$eV, the resulting objects are always lighter than about $10^{-7}\,M_{\odot}$, inaccessible to current astrophysical observations (although they could contribute to the matter content leaving signatures on cosmological observables). If on the contrary gravity is modified in the infrared, $m\lesssim 10^{-3}\,$eV, the objects can be of astrophysical interest, although in this case we need to implement a successful mechanism to screen fifth forces at local scales, which is not a problem in principle for our effective theory. In particular, for masses of the scalar particle in the range $10^{-13}\,\textrm{eV}\lesssim m \lesssim 10^{-10}\,\textrm{eV}$ we obtain objects of a maximal mass between $1000\,M_\odot \gtrsim M_T^{\textrm{max}} \gtrsim 1\,M_\odot$, see the left panel of Figure~\ref{fig5}. It is also interesting to stress that for masses lower than $m\lesssim 10^{-29}\,$eV the resulting configurations are always larger than $R_{95}\gtrsim 10\,$Mpc, the typical size of a galaxy cluster. We do not have evidence of objects of this size in the Universe. This is of particular interest for modified gravity models which are motivated by the cosmological constant problem, where typically $m \sim 10^{-33}\,$eV. This does not mean, of course, that these models are excluded by observations, but only that scalar particles of these masses cannot be clustered in compact configurations. The situation may change for theories where $\Lambda_3\ll M_{\textrm{Pl}}^{1/3}m^{2/3}$, blue region in Figure~\ref{fig5}, but a proper study of this scenario is beyond the present paper.

In analogy to stellar objects, we define the compactness of a HS as the ratio between the $95\%$ of the total mass of the configuration, and the radius where that mass is contained, namely,\footnote{For objects with a sharp border (e.g. a fluid star) it is usual to define the compactness in terms of the total mass $M_T$ and the unambiguous, well-defined radius $R$, as we do next when reporting the Buchdahl limit. For a black hole it is common to consider the event horizon as the border of the object.}$^{,}$\footnote{In Ref.~\cite{Cardoso:2019rvt} the authors introduce the {\it closeness} parameter $\epsilon\equiv 1-M_{95}/(4\pi M_{\textrm{Pl}}^2 R_{95})$, such that we recover a black hole in the limit where $\epsilon$ goes to zero. Both the compactness and the closeness parameters are easily related through the expression $\epsilon=1-2C$.}
\begin{equation}\label{eq.compacness}
 C\equiv \frac{M_{95}}{8\pi M_{\textrm{Pl}}^2 R_{95}},
\end{equation}
where $\bar{M}_{95}=0.95\bar{M}_T$ and the extra constant factor of $8\pi M_{\textrm{Pl}}^2$ is included in such a way that the resulting number is dimensionless. With this definition the compactness of a Schwarzschild black hole is $C=1/2$, whereas that of a fluid star in the Buchdahl limit $C=4/9$~\cite{Buchdahl:1959zz}. In GR neutron stars may reach compactness in the range of $C \approx 0.1 - 0.2$~\cite{Xtreme}, and the precise value depends on the equation of state of the matter component, which under such extreme conditions is an active research area~\cite{Haensel:2007yy}. For a BS with no self-interactions the compactness can be as large as $C=0.1$~\cite{Liebling:2012fv}. However, this number grows up to $C=0.158$ if we include an attractive $\lambda\phi^4$ self-interaction term, and higher compactness, around 
$C \approx 0.33$, are possible considering different potentials for the scalar field~\cite{PhysRevD.35.3658,Cardoso:2016oxy,Palenzuela:2017kcg}.

We now compare these numbers with those of HSs. In particular we focus on the most compact stable object that exists for a given choice of the parameters of the theory, which lies at the maximum of the $\bar{M}_T(\bar{R}_{95})$ curve (see the colored dots in the right panel of Figure~\ref{MvswandR}). Moreover, given that we are interested in obtaining objects of large compactness we concentrate on positive values of the coupling constants. With this purpose we vary the EFT scale $\Lambda_3$ and tune the central density $\bar{\sigma}_0$ to find the maximum of the $\bar{M}_T(\bar{R}_{95})$ curve in each corresponding case. Figure~\ref{fig5}, right panel, shows the most stable compact objects for a range of values of $\Lambda_3$ at the scale of $M_{\textrm{Pl}}^{1/3}m^{2/3}$, where for comparison we have also included values of $C_{\textrm{max}}$ for other compact object solutions (making no difference if it is the total or 95\% of the mass in the definition of the compactness). To the right of that figure, $\Lambda_3>M_{\textrm{Pl}}^{1/3}m^{2/3}$, the curves approach the BS limit where $C_{\textrm{max}}=0.1$. However, they start to deviate from the BS value as we approach $\Lambda_3\sim M_{\textrm{Pl}}^{1/3}m^{2/3}$. As we can appreciate from the figure a similar enhancement to that of the BSs with self-interactions, $C_{\textrm{max}}\approx 0.33$, is possible if we remove the beyond Horndeski term. This is because the inclusion of the operator $d_{4}^{(-1)}$ leads to more massive objects than their counterparts in Horndeski, but they are also more extended. Notice that even if the resulting numbers are not very different from those obtained in previous works~\cite{PhysRevD.35.3658,Cardoso:2016oxy,Palenzuela:2017kcg}, an important difference is that our model appears as the low energy limit of a well defined EFT, and as a consequence our results are stable under quantum corrections, i.e. the inclusion of loop contributions to the effective action do not affect these numbers. The question as to whether lower values of the scale $\Lambda_3$ could reach compactness beyond the of a fluid star, or even a black hole, remains open (notice that the upper bound at $C=1/2$ can in principle be violated in theories beyond GR).

\subsubsection{Anatomy of Horndeski stars}\label{sec.anatomy}

It would be interesting to explore how the properties of a black hole are recovered as we increase the compactness of an object up to the limit in which $C \to 1/2$. Given the large amount of mass (relative to its radius) that a HS may accommodate, it is natural to inquire whether these configurations might develop {\it innermost stable circular orbits} (ISCOs), the smallest stable circular trajectories in which a massive test particle can stably orbit a star,\footnote{The regularity conditions that we imposed in Eqs.~(\ref{boundayC}) guarantee that there exist stable circular orbits of arbitrarily small radius inside the star. However, if for certain interval of the radial coordinate the stable circular orbits are not allowed we still refer to the smallest possible stable circular trajectory outside the star as the ISCO. See Figure~\ref{Fig_ISCO} for details.} or even {\it photon spheres} (PSs), a region of the space where gravity is so strong that light rays are forced to follow circular orbits. They both exist in black hole systems, where apart form the horizon at $r=r_s=M/(4\pi M_\textrm{Pl}^2)$ we also have  $r_{\textrm{isco}}=3r_s$ and $r_{\textrm{ps}}=\frac{3}{2}r_s$ in the case of Schwarzschild. As reference, notice that the smallest compactness expected to develop an ISCO is of the order of $C\approx M/(8\pi r_{\textrm{isco}})=1/6$, and for a PS $C\approx M/(8\pi r_{\textrm{ps}})=1/3$, which are higher than the $C\approx 0.1$ obtained in standard neutron and BS scenarios but possible for HSs. The existence of an ISCO may have a relevance in determining the inner part of an accretion disk, see the discussion of Section~\ref{sec.accretion}, whereas PSs may leave their imprint in the ringdown of the GW signals excited during mergers, see Section~\ref{sec.GW}. Such is the phenomenological impact of a PS that the authors of Ref.~\cite{Cardoso:2019rvt} use the appearance of  this region of space as a criterion to discriminate between compact and ultra-compact stars.

Let us analyze our configurations to understand if these closed curves form. The weak equivalence principle, encoded in the minimal coupling of matter of Eq.~(\ref{eq:lag}), states that test particles move in geodesic motion. To proceed, we generalize the analysis of e.g. Chapter~25, Ref.~\cite{Misner:1974qy} (although with a slightly different notation) to the case of an arbitrary static and spherically symmetric spacetime line-element~(\ref{metric}). The motion of a test particle is described by the equation
\begin{equation}
   \left(\frac{dr}{d\lambda}\right)^2 + V^{\varepsilon}_{\textrm{geo}}(r) = E^2,\label{eq.orbit1}
\end{equation}
where 
\begin{equation}
  V^{\varepsilon}_{\textrm{geo}}(L^2,E^2;r) \equiv \frac{1}{g^2}\left[\frac{L^2}{r^2}+E^2\left(g^2-\frac{1}{N^2}\right)+\varepsilon\right]  
\end{equation}
is an effective potential that determines the change of the particle's $r$-coordinate,
and
\begin{equation}\label{eq.orbit2}
    E \equiv -p_0 = N^2\frac{dt}{d\lambda}, \quad
    L \equiv \pm p_\varphi = \pm r^2\frac{d\varphi}{d\lambda},
\end{equation}
are two quantities that remain conserved in the evolution; see Eq.~(5) in Ref.~\cite{Cardoso:2019rvt}. Here $p^{\mu}$ is the 4-momentum of the particle, $p_\mu= g_{\mu\nu}p^{\nu}$, and given the spherical symmetry we have restricted our attention to the equatorial plane, where $\theta=\pi/2$. For practical purposes it is necessary to distinguish between massive and massless test particles, which is encoded in the parameter $\varepsilon$. For massive particles ($\varepsilon=1$) the parameter $\lambda$ represents the proper time measured by a comoving observer, while for massless particles ($\varepsilon=0$) it is an affine parameter along the geodesic. Moreover, $E$ and $L$ are the total energy and angular momentum of the masless particles, respectively, whereas in the massive case these quantities must be understood per unit rest mass. Note that the second term in the square-bracket of $V^{\varepsilon}_{\textrm{geo}}$ disappears in the case of the Schwarzschild metric (also outside a star), where $N_{\textrm{Schw.}}^2g_{\textrm{Schw.}}^2=1$ and $V^{\varepsilon}_{\textrm{geo}}=V^{\varepsilon}_{\textrm{geo}}(L^2;r)$. 

The existence of a stable circular orbit at a radius $r=r_0$ demands:
\begin{equation}\label{eqs.circular}
  V_{\textrm{geo}}^{\varepsilon}(r_0)=E, \quad   
  V_{\textrm{geo}}^{\varepsilon\prime}(r_0)=0, \quad
  V_{\textrm{geo}}^{\varepsilon\prime\prime}(r_0)>0, \quad
\end{equation}
where the inequality on the second derivative of $V^{\varepsilon}_{\textrm{geo}}(r)$ is a condition for the stability of the orbit: if this condition is not satisfied we can conclude that the circular orbit is unstable (the case with $V^{\varepsilon\prime\prime}_{\textrm{geo}}(r_0)=0$ requires a more careful analysis but this is not going to be necessary here).

\begin{figure}
\centering	
	\scalebox{0.455}{
	\input{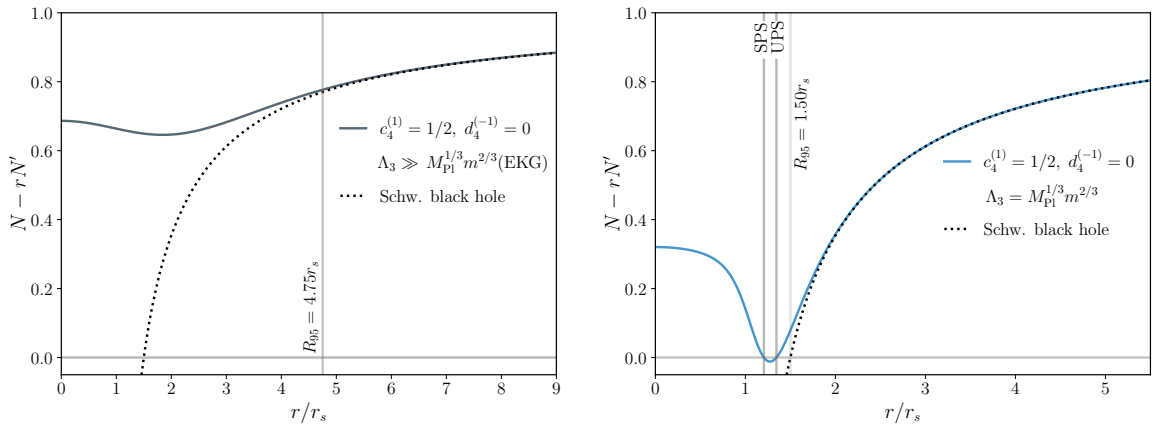}
	}
	\caption{{\bf Massless test particles in circular orbits.}
The left hand side $N-rN'$ of Eq.~(\ref{eq.condition.massless}) as function of the radial coordinate. {\it Left panel: }The prediction for the most compact stable HS in a theory with $c_4^{(1)}=1/2$, $d_4^{(-1)}=0$ and $\Lambda_3\gg M_{\textrm{Pl}}^{1/3}m^{2/3}$ (which is almost indistinguishable from a BS of same field amplitude). {\it Right panel:} Similar as in the left panel but for a lower value of the EFT scale, $\Lambda_3=M_{\textrm{Pl}}^{1/3}m^{2/3}$. In both cases we have included for reference the prediction for a Schwarzschild black hole of same total mass as the star, $M_T=15.9M_{\textrm{Pl}}^2/m$ (left panel), $M_T=27.7 M_{\textrm{Pl}}^2/m$ (right panel). BSs are so diluted that the function $N-rN'$ never vanishes and they do not develop PSs. As we decrease the EFT scale, however, the star becomes more compact and the minimum of the function $N-rN'$ moves down. The compactness $C=0.32$ of a HS with $\Lambda_3= M_{\textrm{Pl}}^{1/3}m^{2/3}$ is high enough to develop two PSs: one of them is almost coincident with that of Schwarzschild, $r_{\textrm{ups}}=1.34r_s$, and is also unstable, whereas the other one is smaller, $r_{\textrm{sps}}=1.21r_s$, and stable. \label{Fig_PS}}
\end{figure}

\begin{figure}
\centering	
	\scalebox{0.455}{
	\input{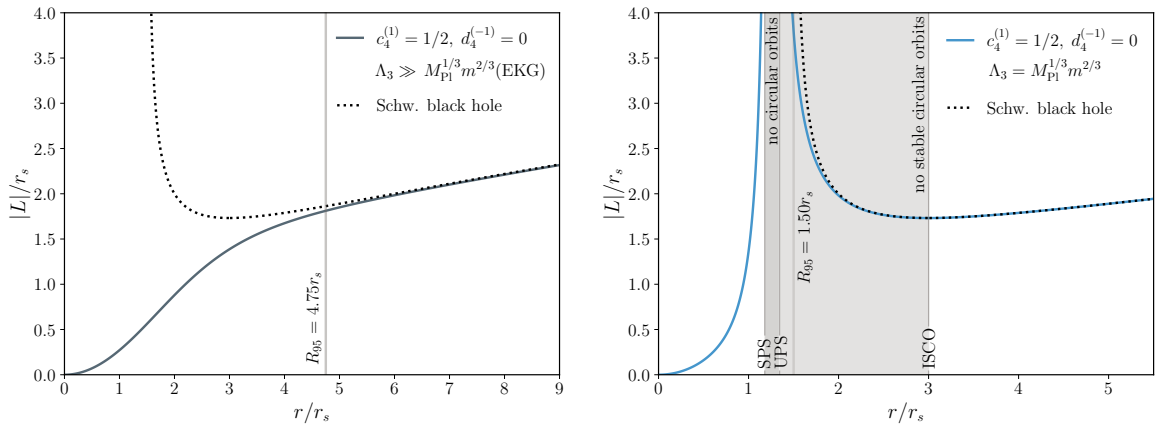}
	}
\caption{{\bf Massive test particles in circular orbits.}
The magnitude of the angular momentum per unit rest mass $|L|$ as function of the radial coordinate, Eq.~(\ref{eqs.EandL}), for the same objects as in Figure~\ref{Fig_PS}. {\it Left panel:} Configurations with $\Lambda_3\gg M_{\textrm{Pl}}^{1/3}m^{2/3}$ are so diluted that $|L|$ grows monotonically with $r$ and stable circular orbits are permitted everywhere. {\it Right panel:} The classification of the circular orbits for the case where $\Lambda_3=M_{\textrm{Pl}}^{1/3}m^{2/3}$ is more subtle, though. As in the Schwarzschild black hole case there are stable circular orbits at radii $r>r_{\textrm{isco}}=3r_s$. Circular orbits with $r_{\textrm{ups}}=1.34r_s<r<r_{\textrm{isco}}$ are possible, but unstable, whereas there are no circular orbits between the two PSs, $r_{\textrm{sps}}=1.21r_s<r<r_{\textrm{ups}}$. Finally, as a consequence of the regularity at the origin there is an internal region inside the stable PS where stable circular orbits are again possible. Compare the right panel with Figure~1 of Ref.~\cite{Cardoso:2019rvt}.\label{Fig_ISCO}}
\end{figure}

If the particles are massless, $\varepsilon =0$, the two equalities of~(\ref{eqs.circular}) can be combined into 
\begin{equation}\label{eq.condition.massless}
    N(r_0)-r_0N'(r_0)=0.
\end{equation}
Exploring the asymptotic behaviour of the function $N-rN'$ close to the origin, one concludes that $N-rN'\approx N_0 - (N_2/N_0) r^2+\ldots$, where $N_0,N_2>0$. Meanwhile, at spatial infinity $N-rN'\approx 1-2M/r+\ldots$. Moreover, we notice that $N''$ (i.e. the first derivative of $N-rN'$) vanishes at a single point of the interval $r\in (0,\infty)$ close to the border of the star, $r\sim R_{95}$.\footnote{The skeptical reader may look at the left panel of Figure~\ref{figB} as an illustration.} Under these conditions it is possible to convince oneself that there are two possible scenarios. One possibility is that the star is so diluted that its radius is much larger than the PS of a Schwarzschild black hole of same total mass, $R_{95}\gg r_{\textrm{ps}}=\frac{3}{2}r_s$, where $r_{\textrm{ps}}$ satisfies the condition $N_{\textrm{Schw.}}(\frac{3}{2}r_s)-\frac{3}{2}r_sN_{\textrm{Schw.}}'(\frac{3}{2}r_s)=0$, and there is not any solution to the Eq,~(\ref{eq.condition.massless}). The other alternative is that the star is sufficiently compact so that $R_{95}\lesssim \frac{3}{2}r_s$ and there are two roots to the Eq.~(\ref{eq.condition.massless}), $r_0=r_{\textrm{ps}}^{\textrm{in}}$ and $r_{\textrm{ps}}^{\textrm{out}}$, $r_{\textrm{ps}}^{\textrm{in}}<r_{\textrm{ps}}^{\textrm{out}}$, corresponding to the two PSs of an ultra-compact {\it regular} object (a Schwarzschild black hole has only one but it is singular at the origin). Furthermore, the stability condition of the orbit $r_{\textrm{ps}}^i$, $i=\textrm{in, out}$, is determined by the sign of the first derivative of the function $N-rN'$ evaluated at $r_{\textrm{ps}}^i$, in such a way that if $N-rN'$ grows at $r=r_{\textrm{ps}}^i$ then the orbit is unstable, whereas if it decreases it is stable. It is then not difficult to prove that the inner PS is always stable, $r_{\textrm{ps}}^{\textrm{in}}\equiv r_{\textrm{sps}}$, whereas the outer one unstable, $r_{\textrm{ps}}^{\textrm{out}}\equiv r_{\textrm{ups}}$. This behaviour is illustrated in Figure~\ref{Fig_PS}.

If on the contrary the particles are massive, $\varepsilon =1$, the two equalities of~(\ref{eqs.circular}) lead to the condition
\begin{equation}\label{eq.condition.massive}
    \frac{L^2}{r^2_0}N(r_0)-\left[\frac{L^2}{r^2_0}+1\right]r_0N'(r_0)=0.
\end{equation}
Notice that in the limit in which $L\gg r_0$ we recover the same expression as in the null case, cf.~Eq.~(\ref{eq.condition.massless}). The factor of 1 in the square-bracket of Eq.~(\ref{eq.condition.massive}), however, even if negligible in the limit $L\gg r_0$, makes a big difference, because the angular momentum does not cancel out from this equation. Actually, contrary to what happens in the massless case, it is possible, in principle, to obtain a continuous family of circular orbits at different radii $r_0$, each labeled by a different value of $L$. The energy, $E$, and angular momentum, $L$, per unit rest mass of a particle in one of these orbits are given by the expressions
\begin{equation}\label{eqs.EandL}
    E=  \sqrt{\frac{N^{3}}{N-r N'}},\quad
    |L|= \sqrt{\frac{r^3N'}{N-r N'}},
\end{equation}
where from now on we omit the sub-index $0$ on the radial coordinate when it does not lead to confusion. The equation for $|L|(r)$ was obtained by inverting the condition~(\ref{eq.condition.massive}), whereas that for $E(r)$ is a consequence of the previous one and the fact that $V^{\varepsilon}_\textrm{geo}(r)=E$ for a circular orbit. These expressions simplify the study, since we can now guarantee that there exists a solution to the Eq.~(\ref{eq.condition.massive}) for every radius in the domain of definition of the functions $E(r)$ and $|L|(r)$. Given that the functions $N^3$ and $r^3N'$ are positive definite, the domain of $E(r)$ and $|L|(r)$ is determined by the inequality $N-rN'>0$; see Figure~\ref{Fig_PS} and the discussion on the massless particles. As a consequence, there are no circular trajectories within the interval between two PSs, whereas they are permitted at all radii in configurations with no circular photon trajectories. The question of the stability can be traced back to the properties of $|L|(r)$: the circular orbits are stable wherever this function grows with $r$, whereas they are unstable where it decreases. This behaviour is nicely illustrated in Figure~1 of Ref.~\cite{Cardoso:2019rvt}.

Our results are shown in Figures~\ref{Fig_PS} and~\ref{Fig_ISCO}, where we plot the left hand side $N-rN'$ of Eq.~(\ref{eq.condition.massless}), and the magnitude of the angular momentum per unit rest mass $|L|$, Eq.~(\ref{eqs.EandL}), as functions of the radial coordinate $r$ for the most compact stable HSs resulting in different realizations of an EFT with $c_4^{(1)}=1/2$, $d_4^{(-1)}=0$.  On the one hand, we have considered the limit in which the EFT scale is large  when compared to the other characteristic scales of the model, namely $\Lambda_3\gg  M_{\textrm{Pl}}^{1/3}m^{2/3}$, leading to a HS that is indistinguishable from the most compact possible stable object resulting in the EKG model, left panels in both figures. On the other hand we have lowered this scale down to $\Lambda_3= M_{\textrm{Pl}}^{1/3}m^{2/3}$, leading to a HS that is more compact than any BS, right panels. In both cases the results are compared to the analytic expressions obtained for a Schwarzschild black hole of the same total mass. For the BS, which is equivalent to the HS with $\Lambda_3\gg  M_{\textrm{Pl}}^{1/3}m^{2/3}$, the compactness is $C= 0.1$, so it is not possible to develop a PS nor an ISCO. In contrast, for the HS with $\Lambda_3=  M_{\textrm{Pl}}^{1/3}m^{2/3}$ the compactness grows up to $C=0.32$, implying a richer phenomenology. The border of the star approaches the black hole's PS, $R_{95}=1.50r_s$, which remains (almost) a valid prediction also in our own case, $r_{\textrm{ups}}=1.34r_s$. To the right of this point the HS is virtually indistinguishable from a Schwarzschild black hole, with $r_{\textrm{isco}}=3r_s$ signaling the smallest circular orbit that remains stable outside the star. Note, however, that since the configuration is regular there appears a second PS, which in this case is stable, at $r_{\textrm{sps}}=1.21r_s$. As for the Schwarzschild black hole there is a region between $r_{\textrm{ups}}<r<r_{\textrm{isco}}$ where circular orbits are unstable, whereas no circular orbits are permitted between the PSs. However, in this case, and again due to the regularity of the object, there appears an internal region $r<r_{\textrm{sps}}$ where stable circular orbits are permitted (note that in the case of Schwarzschild there are no circular orbits, neither stable nor unstable, inside the PS).

\subsection{Gravitational lensing}\label{sec.lensing}
A potential observational signature of the existence of HSs comes from the deflection of light rays passing near its surface. For this purpose we leave for a moment the circular orbits aside and turn our attention to the case of a light ray that approaches the star from a very large distance. Combining Eqs.~(\ref{eq.orbit1}) and~(\ref{eq.orbit2}) so that we can get rid of the affine parameter $\lambda$, and following the same steps as in Section~5 of Chapter~8, Ref.~\cite{Weinberg:1972kfs}, we obtain that the deflection angle of the orbit from the straight line is given by
\begin{equation}\label{eq.deflection.weinberg}
    \Delta\varphi (r_0)= 2\int_{r_0}^{\infty}g\left[\left(\frac{r}{r_0}\right)^2\frac{N_0^2}{N^2}-1\right]^{-1/2}\frac{dr}{r} -\pi ,
\end{equation}
where $r_0$ is the closest distance of the photon's trajectory to the center of the star and $N_0\equiv N(r_0)$. In flat spacetime $N^2(r)=g^2(r)=1$ it is easy to convince oneself that $\Delta\varphi (r_0)=0$, independently of the value of $r_0$. In the more interesting case of the Scharzschild metric a well known analytic result states that, to fist order in a series expansion in powers of $r_s/r_0$, the deflection angle is given by $\Delta\varphi (r_0)=2r_s/r_0$~\cite{Weinberg:1972kfs}. As we approach the horizon, however, this expression breaks down, but there still exists an analytic solution in terms of elliptic integrals which was obtained by Darwin in Refs.~\cite{doi:10.1098/rspa.1959.0015, doi:10.1098/rspa.1961.0142}. In more general situations the integral of Eq.~(\ref{eq.deflection.weinberg}) cannot be performed analytically and one must resort to a numerical implementation, as we have done in our case.
 
\begin{figure}
    \centering
    \scalebox{0.46}{
	\input{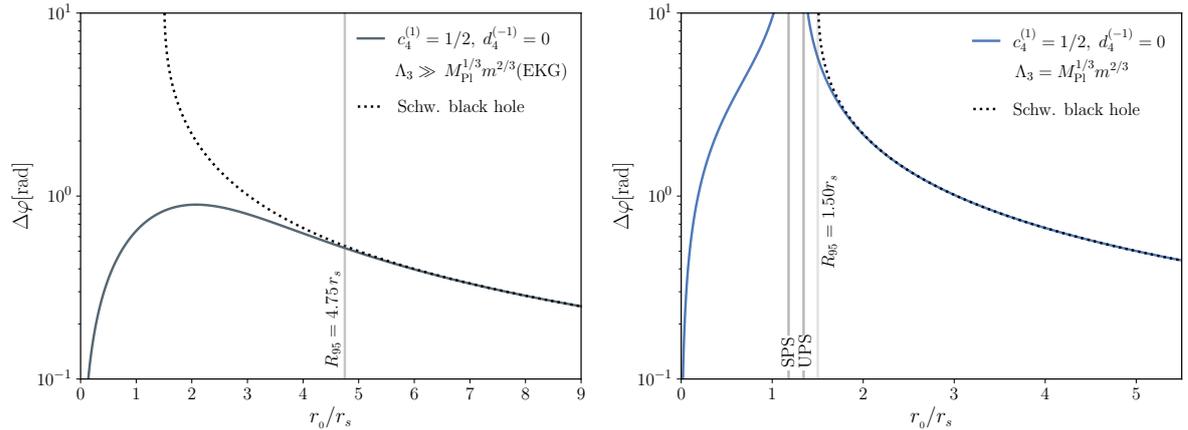}
	}
    \caption{{\bf Gravitational lensing.} The deflection angle $\Delta\varphi$ of light rays as function of the distance $r_0$ of closest approximation for the same configurations as in Figures~\ref{Fig_PS} and~\ref{Fig_ISCO}.
    Vertical lines are for indicative purposes only and correspond to the effective radius of the HS, and the two PSs that appear in the case with $\Lambda_3=M_{\textrm{Pl}}^{1/3}m^{2/3}$.
    \label{fig:def}}
\end{figure}

Our results are shown in Figure~\ref{fig:def}, where we plot the deflection angle $\Delta\varphi$ as a function of the distance of closest approximation $r_0$ for the same configurations as in Figures~\ref{Fig_PS} and~\ref{Fig_ISCO}. For large values of $r_0$ (when compared to the radius of the star, $R_{95}$) the deflection angle is indistinguishable from the prediction of a Schwarzschild black hole of same total mass. This is not surprising in any way given the exponential decay of the scalar field outside the star, where GR is recovered, see Figure~\ref{figB} for details. The discrepancies  with respect to Schwarzschild appear as we approach the border of the star. There are two possible scenarios that we must consider, depending on the compactness of the resulting object, and in particular on the appearance, or not, of PSs. In the limit where $\Lambda_3\gg  M_{\textrm{Pl}}^{1/3}m^{2/3}$, left panel, the deflection angle grows up to a maximum value inside the star, and then decreases to zero as $r_0$ vanishes (given that there is no a physical surface we can extend this curve up to the origin). Compare this with the analytic Schwarzschild black hole prediction, that diverges on the PS, $r=\frac{3}{2}r_s$, and is ill-defined for shorter radii (trajectories that approach the black hole closer than $\frac{3}{2}r_s$ will unavoidably fall into the horizon), or with the neutron star case, where the deflection angle coincides with that of a black hole although it is only defined outside the star. The lower (in units of $M_{\textrm{Pl}}^{1/3}m^{2/3}$) the EFT scale $\Lambda_3$ is, the larger is the maximum of the deflection angle, which of course remains always below the Schwarzschild prediction. If we decrease sufficiently the scale $\Lambda_3$, to the point where the configuration develops a couple of PSs, the $\Delta\varphi(r_0)$ plot develops two asymptotes. This can be appreciated on the right panel of Figure~\ref{fig:def} for the particular case in which $\Lambda_3= M_{\textrm{Pl}}^{1/3}m^{2/3}$. To the right of the unstable PS, $r_{\textrm{ups}}=1.34$, the HS is almost indistinguishable from a black hole. However, given the regularity conditions, the distance of closest approximation can be extended to the left of $r_{\textrm{sps}}=1.21r_s$. Contrary to what happens for black hole and neutron stars, the deflection angle of a HS does not decrease monotonically with $r_0$, and this may facilitate the appearance of multiple images.

The strongest observational constraints on the gravitational deflection angle come from analysing radio waves passing near the Sun with the Very Long Baseline Interferometry experiment, and have an uncertainty of about $0.04\%$~\cite{Shapiro:2004zz}. However, for more distant objects measurements of the deflection angle become less precise and will not reach the precision needed for a direct observation of the deviation from a black hole or a neutron star, unless we have access to sources with small impact parameters, where the deviations are large. Our expectation is that further studies of light deflection in a large object sample, in the spirit of weak lensing, will statistically enhance the signal to noise sufficiently to asses this HS prediction.

\subsection{Accretion disks}\label{sec.accretion}
Another possible observational test to discriminate between HSs and other astrophysical objects, such as black holes and neutron stars, is the study of the emitted power spectrum of an accretion disk surrounding the vicinity of the configuration. For simplicity we assume the geometrically thin, optically thick, disk approximation~\cite{Novikov:1973kta,Page:1974he,Thorne:1974ve}, that is, the radius of the disk is considered to be much bigger than its maximal thickness, and the disk is assumed to radiate efficiently, keeping its temperature constant in time. More sophisticated models of the accretion disk are currently available~\cite{2002apa..book.....F}, but for our purposes this description is sufficient. For concreteness, we use the same methodology as in Refs.~\cite{Torres:2002td, Harko:2009gc,Bambi:2011jq, Guzman:2009zz,Staykov:2016dzn}, that we describe briefly for completeness.

Because the disk is assumed thin, it is possible to consider that all the particles move in the equatorial plane, $\theta=\pi/2$. Further, we also assume that the accretion is a slow process, with particles following a series of stable circular orbits as they lose energy and angular momentum in the radiation process. These intermediate orbits have been described in Section~\ref{sec.anatomy}, and consist on time-like geodesics that are labeled in terms of the specific energy, $E$, and angular momentum, $L$, of the particles per unit rest mass, see Eq.~(\ref{eqs.EandL}).

Following the same arguments as in~\cite{Page:1974he,Harko:2009gc,Bambi:2011jq, Staykov:2016dzn}, the electromagnetic flux per unit area generated by a thin accretion disk that rotates around a central object is given by
\begin{equation}
    F(r)=-\frac{\dot{M}_{0}}{4\pi \sqrt{-g^{(3)}}}\frac{\partial_{r}\Omega}{(E-\Omega L)^2}\int_{r_{i}}^{r}(E-\Omega L)\partial_{r} L dr,\label{ecpot}
\end{equation}
where $g^{(3)}=-N^2 g^2 r^2$ is the determinant of the induced metric on the equatorial plane, $\theta=\pi/2$, $\Omega$ is the angular velocity of the particles, defined in the standard way as
\begin{equation}
   \Omega\equiv \dfrac{d\varphi}{dt}=\pm \frac{\sqrt{r N N'}}{r},
\end{equation}
and $\dot{M}_0$ and $r_i$ are the accretion mass rate of the disk and its inner radius, respectively, which are external parameters that should be determined observationally. The disk is assumed to be in a steady state, that is, $\dot{M}_0$ is constant in time and independent of the radial coordinate. Regarding the internal radius $r_i$, in some cases it is estimated using theoretical arguments. In particular, for black holes the value of $r_i$ is usually taken as the radius of the last stable circular orbit, $r_i=r_{\textrm{isco}}= 3r_s$, whereas for non-baryonic stars with no physical surfaces the choice is usually $r_i=0$. This assumption is well motivated in the case of e.g. non-interacting BSs, however, if the object is sufficiently compact to develop an ISCO, as it may be the case of a HS in a model with a low EFT scale, a more careful choice is necessary, as we do next.

Assuming thermodynamical equilibrium, one can make use of Eq.~\eqref{ecpot} to infer the radial temperature distribution $T(r)$ of the disk via the Stefan-Boltzmann law $F(r)=\sigma T^{4}(r)$, where $\sigma=5.670\times10^{-5}\,\text{erg}\cdot\text{s}^{-1}\cdot\text{cm}^{-2}\cdot\text{K}^{-4}$ is the Stefan-Boltzmann constant. This result, together with the assumption that each disk's element radiates as a black body, can be used to compute the luminosity per unit frequency in the range between $\nu$ and $\nu+d\nu$~\cite{Bambi:2011jq}, obtaining
\begin{eqnarray}
	L(\nu)&=&\frac{16\pi^{2}h (\cos i)\nu^{3}}{c^{2}}\int_{r_{i}}^{r_{\textrm{out}}}\frac{Ngr}{e^{h\nu/[k_BT(r)]}-1}dr,
\end{eqnarray}
where $i$ is the disk's inclination angle with respect to the plane of the celestial sphere and $r_{\textrm{out}}$ is the radial coordinate at the exterior border. Note that in this last expression we have restored the physical units, where $h=6.626\times10^{-27}\, \textrm{erg}\cdot\textrm{s}$ and $k_B=1.381\times10^{-16}\,\textrm{erg}\cdot\textrm{K}^{-1}$ are the Planck and Boltzmann constants, respectively.

\begin{figure}
    \centering
    \scalebox{0.45}{
	\input{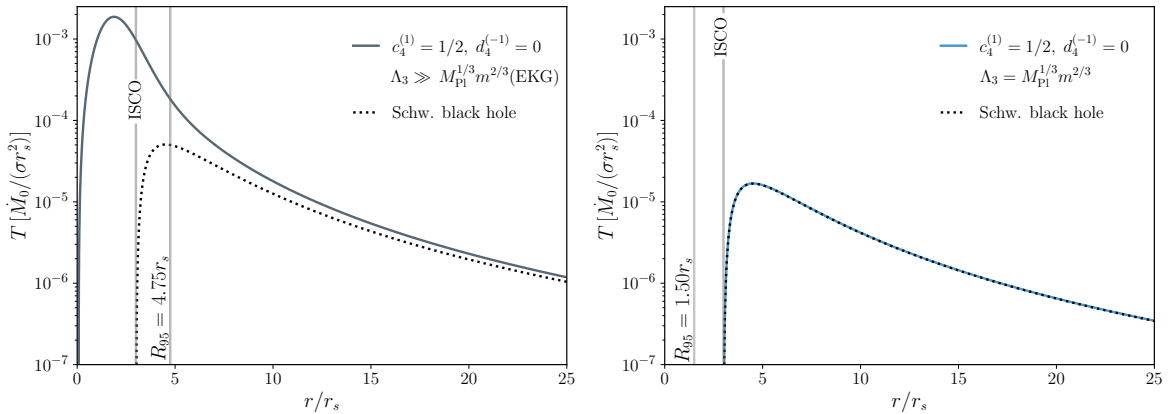}
	}
    \caption{{\bf Thin and thick disk approximation.} The radial temperature distribution for the same configurations as in Figures~\ref{Fig_PS},~\ref{Fig_ISCO} and~\ref{fig:def}. Large values of $\Lambda_3$ allow disk configurations that can extend to the centre of the object, leading to an inner region that is very different form the Schwarzschild case with an internal border at $r_{\textrm{isco}}=3 r_s$, left panel. This difference disappears as we decrease the EFT scale, right panel. Vertical lines are for reference and correspond to the effective radius of the HSs, and the location of the ISCO that appears in the case of $\Lambda_3=M_{\textrm{Pl}}^{1/3}m^{2/3}$, which for this particular choice of the parameters is superimposed on that of the corresponding black hole.
    }
    \label{fig:temperature}
\end{figure}

In Figure~\ref{fig:temperature} we show, for the same configurations as in Figures~\ref{Fig_PS},~\ref{Fig_ISCO} and~\ref{fig:def}, the disk's temperature as a function of the radial coordinate. At large distances (when compared to the radius of the star, $R_{95}$) the temperature is insensitive to the nature of the central object, which can be approximated in terms of the Scwhazrschild metric, and the temperature decays as $1/r^{3/4}$ with increasing $r$. The interior side of the disk, by contrast, depends strongly on the properties of the star. It is not hard to imagine why this happens. As we penetrate the configuration, the gravitational field of a HS and a Schwarzschild black hole differ one from each other, and test particles orbit the central region in a different way, see e.g. Figure~\ref{Fig_ISCO}. If the scale of the EFT is large in units of $M_{\textrm{Pl}}^{1/3}m^{2/3}$, this causes a large difference in the trajectories of the particles with respect to the Schwarzschild case, see the left panel of Figure~\ref{Fig_ISCO}, which translates in the large difference in the temperatures that we appreciate in the right panel of Figure~\ref{fig:temperature}. This is motivated mainly by the fact that test particles cannot follow stable orbits around a Schwarzschild black hole at radii smaller than $r_{\textrm{isco}}=3r_s$, leading to an empty space for $r<3r_s$ which does not occur for the HS configuration. As we decrease the scale of the EFT, the HS becomes more and more compact and this difference starts to disappear. In the case of $\Lambda_3=  M_{\textrm{Pl}}^{1/3}m^{2/3}$, for instance, the disks around a HS or a Schwarzschild black hole are almost indistinguishable, as we can appreciate from the right panel of Figure~\ref{fig:temperature}.

\begin{figure}
    \centering
    \scalebox{0.45}{
	\input{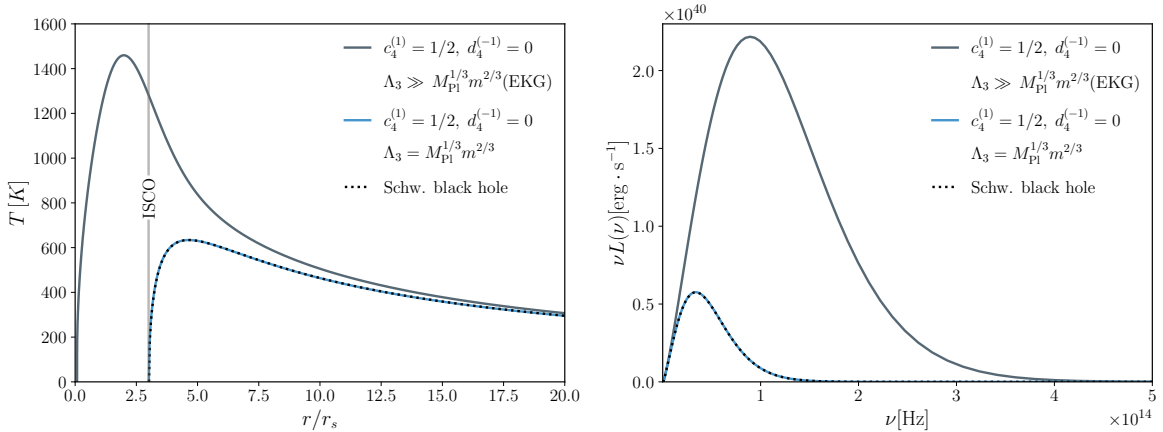}
	}
    \caption{{\bf Super-massive black hole mimickers.} The radial temperature distribution of the disk, $T$ (left panel), and its emission spectrum, $\nu L(\nu)$ (right panel), for three  configurations of same total mass $M_T = 3\times 10^9\;M_{\odot}$ appearing in models with $\Lambda_3\gg M_{\textrm{Pl}}^{1/3}m^{2/3}$ and $m=2.82\times 10^{-20}\,$eV, $\Lambda_3= M_{\textrm{Pl}}^{1/3}m^{2/3}$ and $m=4.89\times 10^{-20}\,$eV, and the Schwarzschild black hole case. As we decrease the scale of the EFT the emission spectrum of a HS is indistinguishable from that of a black hole.
    }
    \label{fig:acrec}
\end{figure}

To make a more quantitative comparison between the models, we compute the emission spectra of an object of total mass $M_T=3\times 10^{9}\,M_\odot$ that mimics the order of magnitude of the super-massive black hole candidates~\cite{2019Natur.570...83M, 2020AA...644A.105H} in  Sagittarius A*~\cite{Eckart:1997em} and M87~\cite{Gebhardt_2011}.\footnote{It is important to stress that in this part of the analysis the mass of the scalar field changes from configuration to configuration: $m=2.82\times 10^{-20}\,$eV ($m=4.89\times 10^{-20}\,$eV) if $\Lambda_3\gg  M_{\textrm{Pl}}^{1/3}m^{2/3}$ ($\Lambda_3=  M_{\textrm{Pl}}^{1/3}m^{2/3}$). This is not a problem, however, because in practice we do not have access to the value of this number, and our purpose is to show how close the prediction of a HS can be to that of a Schwarzschild black hole as we decrease the scale of the EFT.} For this purpose, we keep fixed the properties of the disk apart from the inner radius, which is determined using theoretical considerations. In particular, we fix the accretion rate $\dot{M}_0=2\times 10^{-6}\,M_\odot/\textrm{yr}$, the outer radius $r_{\textrm{out}}=100r_s$ (although the result is not sensitive to this choice once the radius of the disk is much bigger than that of the object, $r_{\textrm{out}}\gg R_{95}$; this is because the luminosity is dominated by the inner region), and the inclination angle $i=0$ (note that this angle contributes as a global factor of $\cos i$ that can be always re-scaled at the end of the calculation). Our results are shown in Figure~\ref{fig:acrec}, where for comparison we have also included the radial temperature of the disk. As can be appreciated from the figure, right panel, the emission increases with the scale $\Lambda_3$, and this is due to the large temperatures that are reached in the inner regions of the disk that are absent when an ISCO is present, left panel. If we decrease the EFT scale, on the contrary, the spectrum is indistinguishable from that of a black hole.

\subsection{Gravitational radiation}\label{sec.GW}
The detection of GWs from the collision of compact objects have revolutionized gravitational physics, opening the possibility to test fundamental theories that are otherwise inaccessible to current experiments at the Earth. Over the past years collisions of highly compact BSs have been shown to produce signals that make them clearly distinguishable from those of other astrophysical objects, such as black holes and neutron stars, see e.g.~\cite{Palenzuela:2017kcg, Bezares:2018qwa, Pacilio:2020jza}, especially due to its tidal deformations in spiriling binary systems~\cite{Pacilio:2020jza, Cardoso:2017cfl, Sennett:2017etc}. Since many properties of the gravitational signal depend on how compact are the objects involved in the collision, it is interesting to remark that for certain choices of the parameters of the effective model the compactness of a HS can be increased substantially with respect to that of neutron and BSs. With this idea in mind we proceed to study the gravitational emission from the collision of two HSs.

Although we do not present a full dynamical analysis, we can learn something purely from the compactness of the configurations. Following a simple model for estimating the gravitational radiation from compact binary systems~\cite{Hanna:2016uhs}, it is possible to obtain a rough upper bound for the total energy radiated by the collision of two equal mass HSs~\cite{Palenzuela:2017kcg}, given by
\begin{equation}
    \xi_\textrm{{rad}} \approx 0.48 C M_S , \label{eq:xirad}
\end{equation}
where $M_S=M_{T1}+M_{T2}=2M_T$ is the initial total mass of the system and $C$ is the compactness of the objects. This result was derived originally for a minimally coupled scalar field with a solitonic potential~\cite{PhysRevD.35.3658,Cardoso:2016oxy,Palenzuela:2017kcg}, but the considerations in~\cite{Hanna:2016uhs} are generic enough that the details of the model do not affect the result. Actually Eq.~\eqref{eq:xirad} arises from energy conservation arguments of an irrotational system, where two constant density objects with equal mass and radius merge into a single configuration of mass approximately equal to $2M_T$ and vanishing angular momentum, assuming that all other forms of radiation apart from gravitational are negligible (e.g. electromagnetic or scalar radiation). 

Using~\eqref{eq:xirad} as an upper bound, and since black hole collisions radiate approximately $5\%$ of their total mass in GWs,\footnote{Numerical simulations of the collision of two non-rotating black holes with equal mass predict that the gravitational radiation emitted during the merging process is less than $4\%$ of the mass of the binary system~\cite{Pollney:2009yz}. These results are in concordance with the observation of GW150914, which corresponds to a binary system of black holes that emitted approximately $5\%$ of its mass in the form of GWs~\cite{Abbott:2016blz}.} it is possible that HSs can emit more than a black hole system of similar mass; for this to happen, it is only necessary that their compactness is greater than $C\gtrsim 0.12$. As shown in Figure~\ref{fig5}, right panel, many of the configurations reach a compactness higher than $C=0.12$, even for moderate values of $\Lambda_3$, leading to super-emitters in the terminology of~\cite{Palenzuela:2017kcg}. For instance, if $\Lambda_{3} = 1.26 M_{\textrm{Pl}}^{1/3}m^{2/3}$, we have $C_\textrm{max}\approx 0.20$, and the total energy radiated in the form of GWs could be as large as $10\%$ of the total mass of the system. This is only one example, larger values are even possible if we decrease further the scale  $\Lambda_3$ of the EFT, but it is a clear indicative that it would be interesting to extend the study of the GW radiation to a dynamical perspective, which is beyond the present analysis.

\section{\label{sec5} Discussion}

Compact objects are an excellent laboratory to test fundamental physics. Besides the well known neutron star and black hole systems, astrophysical observations can also be used to search and constrain more exotic objects, such as the vacuum self-gravitating configurations that are inherent to certain modified gravity models. 

In this work we focus on scalar-tensor theories that present a weak breakdown of the galileon symmetry. The low energy effective model that describes these theories is characterized by three energy scales: the Planck mass $M_\textrm{Pl}$, the characteristic scale of the EFT $\Lambda_3$ (that also measure the strength of the three galileon interactions), and the mass of the scalar degree of freedom $m$. These scales control the symmetry breaking operators, which are suppressed by powers of $\Lambda_2=(M_{\textrm{Pl}}\Lambda_3^3)^{1/4}$, apart from the mass term, which is in principle independent. This model possesses the peculiarity that includes higher derivative operators that are not present in simpler realizations, such as the Brans-Dicke theory, and are part of the (beyond) Horndeski action. It also contains operators that are not included in the covariant galileons, and make possible to evade a generalization of Derrick's theorem that forbids the existence of soliton solutions in this scenario. We refer to these objects as Horndeski stars.

To proceed, we restrict ourselves to the static and spherically symmetric regime of the EFT, although a more general analysis is possible. For theories where $\Lambda_3\gg M_{\textrm{Pl}}^{1/3}m^{2/3}$, the resulting configurations are indistinguishable from BSs. This is a consequence of the fact that our effective model is connected continuously to the EKG theory in the regime in which the scalar field is coupled weakly. If the scale of the EFT is lower than $\Lambda_3\lesssim M_{\textrm{Pl}}^{1/3}m^{2/3}$, however, there is a threshold value of the central field that triggers the higher derivative operators of the effective action and couples the scalar strongly. We identified this scale in Eq.~(\ref{scale.self}). In practice we concentrate on the subspace of the parameters in which $\Lambda_3\gtrsim M_{\textrm{Pl}}^{1/3}m^{2/3}$, where the first deviations with respect to BSs manifest themselves. These deviations can be parametrized in a model-independent way in terms of two operators of mass dimension six that lie on the quartic sectors of the Horndeski, $c_4^{(1)}$, and the beyond Hornseki, $d_4^{(-1)}$, theories, respectively. It is interesting to note that the two operators can be chosen in the appropriate way such that the speed of propagation of GWs coincides with that of light, $d_4^{(-1)}=2c_4^{(1)}$, although this is not necessarily true from first principles.

Once we have assumed that $\Lambda_3\gtrsim M_{\textrm{Pl}}^{1/3}m^{2/3}$, and depending on the value of the mass of the scalar field, we distinguish between two different phenomenological realizations of the model. Firstly, and if gravity is modified in the infrared, $m\lesssim 10^{-3}\,$eV, HSs may present the typical mass and size of an astrophysical object. Moreover, for certain choices of the parameters the resulting objects may be as compact as $C\approx 0.35$ (compare this with the usual $C\approx 0.1$ of a neutron star~\cite{Xtreme}). In presence of matter the higher derivative operators can result in a Vainshtein screening, which is indeed necessary to account for Earth based experiments and Solar System observations, although a more exhaustive study requires a model-dependent analysis from what we have presented here. Towards the end of the work, we focus on possible observational signatures that may help to identify HSs. We discuss their GWs signal (leading to super-emitters in the terminology of~\cite{Palenzuela:2017kcg}), the deflection of light rays passing near their surface (producing large deviations in the deflection angle with respect to the Schwarzschild black hole case for small impact parameters), and the emission spectra that are expected from accretion disks surrounding this type of objects.

A different behaviour is found for masses of the scalar field that are higher than $m\gtrsim 10^{-3}\,$eV. In this case the mass term screens the fifth forces at accessible scales, and we do not need to rely on the higher derivative operators to this purpose. However, the resulting individual objects, which have masses that are always smaller than about $M_T\lesssim 10^{-5}\,M_\odot$, are inaccessible to current astrophysical observations, although it is still possible that they could leave their impact at cosmological scales in the form of a dark matter component. We leave this analysis for a future paper.

\begin{acknowledgments}
We thank Max Dohse and Eliana Chaverra for early discussions that lead to some parts of this paper, and Shuang-Yong Zhou for pointing out the positivity bounds of~\cite{Tolley:2020gtv}. This work was partially supported by CONACyT grants CB-259228, CB-286897 and Ciencia de Frontera No. 102958. We thank the Instituto Avanzado de Cosmolog\'{\i}a A. C. and the DCI-UG DataLab for academic and computational resources. A.A.R. was partially supported by CONACyT graduate scholarship No. 57032, and Ayudante de Investigador SNI, EXP. AYTE. 19650. G.N. also appreciates the grants of UG-DAIP.
\end{acknowledgments}

\appendix

\section{Field Equations}\label{apendix1}

In this Appendix we report the functions that are needed to fully define the equations of motion that we introduced in Section~\ref{subsec.eq.motion}. For the case of Eq.~(\ref{fieleq}), we have the following definitions:
\begin{allowdisplaybreaks}
\begin{subequations}\label{eqs.app1}
\begin{eqnarray}
t_{\mu\nu} &=& \bar{\phi}_{\mu}\phi_{\nu}+\phi_{\mu}\bar{\phi}_{\nu}-g_{\mu\nu}(X+m^2\bar{\phi}\phi)\,, \label{eq.Cov} \\
a_{\mu\nu}&=& 2\left[ \bar{\phi}_{\mu\nu}\Phi+\phi_{\mu\nu}\bar{\Phi}-G_{\mu\alpha}\left(\bar{\phi}_{\nu}\phi^{\alpha}+\phi_{\nu}\bar{\phi}^{\alpha}\right) \right]-R\phi_{\mu}\bar{\phi}_{\nu}-\bar{\phi}_{\mu}\left( R\phi_{\nu}+2G_{\nu\alpha}\phi^{\alpha}\right)\nonumber\\
&& -\,2\Big[ G_{\nu\alpha}\phi_{\mu}\bar{\phi}^{\alpha}+\bar{\phi}_{\nu\alpha}{\phi_{\mu}}^{\alpha}+\bar{\phi}_{\mu\alpha}{\phi_{\nu}}^{\alpha}+\left( R_{\mu\alpha\nu\eta}+R_{\mu\eta\nu\alpha}\right) \phi^{\alpha}\bar{\phi}^{\eta}-g_{\mu\nu}\Big(X R-\Phi\bar{\Phi}\nonumber\\
&&+\,2G_{\alpha\eta}\phi^{\alpha}\bar{\phi}^{\eta}+\bar{\phi}_{\alpha\eta}\phi^{\alpha\eta}\Big) \Big],\\
b_{\mu\nu}&=&2\bar{\Phi}\big(X \phi_{\mu\nu}-\phi_{\mu}\phi_{\nu}\bar{\Phi}\big)+2\Phi\left[X\bar{\phi}_{\mu\nu}-\bar{\phi}_{\mu}\bar{\phi}_{\nu}\Phi-X g_{\mu\nu}\bar{\Phi}-\big( \bar{\phi}_{\mu}\phi_{\nu}+\phi_{\mu}\bar{\phi}_{\nu}\big)\bar{\Phi}\right]-2 X g_{\mu\nu}\bar{\phi}_{\alpha\eta}\phi^{\alpha\eta}\nonumber\\
&&+\,\bar{\phi}_{\mu}\big[ 2\bar{\phi}_{\nu}\phi_{\alpha\eta}\phi^{\alpha\eta}+\phi_{\nu}\big(4\bar{\phi}_{\alpha\eta}\phi^{\alpha\eta}-2 X R\big)\big] +2\phi_{\mu}\big[\bar{\phi}_{\nu}\big(2\bar{\phi}_{\alpha\eta}\phi^{\alpha\eta}-X R\big)+\phi_{\nu}\bar{\phi}_{\alpha\eta}\bar{\phi}^{\alpha\eta}\big]\nonumber\\
&&+\,\bar{\phi}^{\alpha}\Big\lbrace 2X\phi_{\mu\nu\alpha}+\bar{\phi}_{\nu}\big( 3\phi_{\mu\alpha}\Phi+\phi_{\mu}{{\phi_{\alpha}}^{\eta}}_{\eta}-3\phi_{\alpha\eta}{\phi_{\mu}}^{\eta}\big)+3\bar{\phi}_{\mu}\big(\phi_{\nu\alpha}\Phi-\phi_{\alpha\eta}{\phi_{\nu}}^{\eta}\big)+\phi_{\nu}\Big[2\bar{\phi}_{\mu\alpha}\Phi\nonumber\\
&&-\,2X G_{\mu\alpha}+\phi_{\mu\alpha}\bar{\Phi}
+\bar{\phi}_{\mu}{{\phi_{\alpha}}^{\eta}}_{\eta}-\phi_{\mu\alpha\eta}\bar{\phi}^{\eta}+\phi_{\mu}\big( R\bar{\phi}_{\alpha}+2G_{\alpha\eta}\bar{\phi}^{\eta}\big)-3\bar{\phi}_{\alpha\eta}{\phi_{\mu}}^{\eta}-2\bar{\phi}_{\mu\eta}{\phi_{\alpha}}^{\eta}\Big]\nonumber\\
&&+\,\phi_{\mu}\big(2\bar{\phi}_{\nu\alpha}\Phi-2X G_{\nu\alpha}+\phi_{\nu\alpha}\bar{\Phi}-\phi_{\nu\alpha\eta}\bar{\phi}^{\eta}-3\bar{\phi}_{\alpha\eta}{\phi_{\nu}}^{\eta}-2\bar{\phi}_{\nu\eta}{\phi_{\alpha}}^{\eta}\big)-2\Big[\big( \phi_{\mu\alpha}\phi_{\nu\eta}-\phi_{\mu\nu}\phi_{\alpha\eta}\big)\bar{\phi}^{\eta}\nonumber\\
&&+\,g_{\mu\nu}\Big( \bar{\phi}^{\eta}\big( \phi_{\alpha\eta}\Phi-\phi_{\eta\gamma}{\phi_{\alpha}}^{\gamma}\big)+X{{\phi_{\alpha}}^{\eta}}_{\eta}\Big)\Big]\Big\rbrace+\phi^{\alpha}\Big\lbrace\bar{\phi}_{\nu}\Big[\bar{\phi}_{\mu\alpha}\Phi+2\phi_{\mu\alpha}\bar{\Phi}-2X G_{\mu\alpha}+\phi_{\mu}\bar{\phi}_{\alpha}{}^{\eta}{}_{\eta}\nonumber\\
&&-\,\bar{\phi}_{\mu\alpha\eta}\phi^{\eta}+\bar{\phi}_{\mu}\big(R\phi_{\alpha}+2 G_{\alpha\eta}\phi^{\eta}\big)-\phi_{\mu\alpha\eta}\bar{\phi}^{\eta}-2\bar{\phi}_{\alpha\eta}{\phi_{\mu}}^{\eta}-3\bar{\phi}_{\mu\eta}{\phi_{\alpha}}^{\eta}+R_{\mu\alpha\eta\gamma}\phi^{\eta}\bar{\phi}^{\gamma}\Big]+\bar{\phi}_{\mu}\Big(\bar{\phi}_{\nu\alpha}\Phi\nonumber\\
&&-\,2X G_{\nu\alpha}+2\phi_{\nu\alpha}\bar{\Phi}-\bar{\phi}_{\nu\alpha\eta}\phi^{\eta}-\phi_{\nu\alpha\eta}\bar{\phi}^{\eta}-2\bar{\phi}_{\alpha\eta}{\phi_{\nu}}^{\eta}-3\bar{\phi}_{\nu\eta}{\phi_{\alpha}}^{\eta}+R_{\nu\alpha\eta\gamma}\phi^{\eta}\bar{\phi}^{\gamma}\Big)+\phi_{\nu}\Big(3\bar{\phi}_{\mu\alpha}\bar{\Phi}\nonumber\\
&&+\,\bar{\phi}_{\mu}\bar{\phi}_{\alpha}{}^{\eta}{}_{\eta}-\bar{\phi}_{\mu\alpha\eta}\bar{\phi}^{\eta}-3\bar{\phi}_{\alpha\eta}\bar{\phi}_{\mu}{}^{\eta}-2R_{\mu\eta\alpha\gamma}\bar{\phi}^{\eta}\bar{\phi}^{\gamma}\Big)+\phi_{\mu}\Big(3\bar{\phi}_{\nu\alpha}\bar{\Phi}-\bar{\phi}_{\nu\alpha\eta}\bar{\phi}^{\eta}-3\bar{\phi}_{\alpha\eta}\bar{\phi}_{\nu}{}^{\eta}\nonumber\\
&&-2R_{\nu\eta\alpha\gamma}\bar{\phi}^{\eta}\bar{\phi}^{\gamma}\Big)+2\Big[X\bar{\phi}_{\mu\nu\alpha}-\big( \bar{\phi}_{\mu\alpha}\bar{\phi}_{\nu\eta}-\bar{\phi}_{\mu\nu}\bar{\phi}_{\alpha\eta}\big)\phi^{\eta}-\Big(\phi_{\nu\alpha}\bar{\phi}_{\mu\eta}+\phi_{\mu\alpha}\bar{\phi}_{\nu\eta}-\bar{\phi}_{\mu\nu}\phi_{\alpha\eta}\nonumber \\
&&-\,\phi_{\mu\nu}\bar{\phi}_{\alpha\eta}\Big)\bar{\phi}^{\eta}+g_{\mu\nu}\Big\lbrace X( R\bar{\phi}_{\alpha}-\bar{\phi}_{\alpha}{}^{\eta}{}_{\eta})+\bar{\phi}^{\eta}\Big(2X G_{\alpha\eta}-\bar{\phi}_{\alpha\eta}\Phi-\phi_{\alpha\eta}\bar{\Phi}+\phi_{\alpha\eta\gamma}\bar{\phi}^{\gamma}+3\bar{\phi}_{\eta\gamma}{\phi_{\alpha}}^{\gamma}\nonumber\\
&&+\,\bar{\phi}_{\alpha\gamma}{\phi_{\eta}}^{\gamma}\Big)+\phi^{\eta}\big(\bar{\phi}_{\alpha\eta\gamma}\bar{\phi}^{\gamma}-\bar{\phi}_{\alpha\eta}\Phi+\bar{\phi}_{\eta\gamma}{\phi_{\alpha}}^{\gamma}+R_{\alpha\gamma\eta\psi}\bar{\phi}^{\gamma}\bar{\phi}^{\psi}\big) \Big\rbrace\Big]\Big\rbrace,\\
c_{\mu\nu}&=&2X\left[\bar{\phi}_{\nu}\left( \bar{\phi}_{\mu\alpha}\phi^{\alpha}+\phi_{\mu\alpha}\bar{\phi}^{\alpha}\right)\Phi+\phi_{\mu\nu}\left( \bar{\phi}^{\alpha}\phi_{\alpha\eta}+\phi^{\alpha}\bar{\phi}_{\alpha\eta}\right)\bar{\phi}^{\eta}+\bar{\phi}_{\mu\nu}\phi^{\alpha}\left( \bar{\phi}_{\alpha\eta}\phi^{\eta}+\phi_{\alpha\eta}\bar{\phi}^{\eta}\right) \right]\nonumber\\
&&-\,\bar{\phi}_{\nu}\Big\lbrace 2X\bar{\phi}^{\alpha}\phi_{\alpha\eta}{\phi_{\mu}}^{\eta} +\phi^{\alpha}\Big[\bar{\phi}_{\mu\alpha}\phi^{\eta}\bar{\phi}_{\eta\gamma}\phi^{\gamma}+\left(3\bar{\phi}_{\mu\eta}\phi^{\eta}+\phi_{\mu\eta}\bar{\phi}^{\eta}\right) \phi_{\alpha\gamma}\bar{\phi}^{\gamma}+\bar{\phi}_{\alpha\eta}\Big( 2X{\phi_{\mu}}^{\eta}\nonumber\\
&&-\,\phi^{\eta}\phi_{\mu\gamma}\bar{\phi}^{\gamma}\Big) \Big] \Big\rbrace+\bar{\phi}_{\mu}\Big\lbrace 2X\left( \bar{\phi}_{\nu\alpha}\phi^{\alpha}+\phi_{\nu\alpha}\bar{\phi}^{\alpha}\right) \Phi-2X\bar{\phi}^{\alpha}\phi_{\alpha\eta}{\phi_{\nu}}^{\eta}-\phi^{\alpha}\Big[\bar{\phi}_{\nu\alpha}\phi^{\eta}\bar{\phi}_{\eta\gamma}\phi^{\gamma}\nonumber\\
&&+\,\left(3\bar{\phi}_{\nu\eta}\phi^{\eta}+\phi_{\nu\eta}\bar{\phi}^{\eta}\right) \phi_{\alpha\gamma}\bar{\phi}^{\gamma}+\bar{\phi}_{\alpha\eta}\left[2X{\phi_{\mu}}^{\eta}+\phi^{\eta}\left( 2\bar{\phi}_{\nu}\Phi-\phi_{\nu\gamma}\bar{\phi}^{\gamma}\right) \right]+2\bar{\phi}_{\nu}\Big( \phi_{\alpha\eta}\bar{\phi}^{\eta}\Phi\nonumber\\
&&-\,\left(\bar{\phi}^{\eta}\phi_{\eta\gamma}+\phi^{\eta}\bar{\phi}_{\eta\gamma}\right){\phi_{\alpha}}^{\gamma}\Big)\Big] \Big\rbrace+\phi_{\mu}\Big(2X \left[\left( \bar{\phi}_{\nu\alpha}\phi^{\alpha}+\phi_{\nu\alpha}\bar{\phi}^{\alpha}\right) \bar{\Phi}+\bar{\phi}_{\nu}\bar{\phi}_{\alpha\eta}\phi^{\alpha\eta}\right]\nonumber\\
&&-\,\bar{\phi}^{\alpha}\left( 2X\bar{\phi}_{\nu\eta}{\phi_{\alpha}}^{\eta}+\phi_{\nu\alpha}\bar{\phi}^{\eta}\phi_{\eta\gamma}\bar{\phi}^{\gamma}\right)-\phi^{\alpha}\Big\lbrace 2X\bar{\phi}_{\alpha\eta}\bar{\phi}_{\nu}{}^{\eta}+\Big[3\phi_{\nu\eta}\bar{\phi}^{\eta}\bar{\phi}_{\alpha\gamma}+\bar{\phi}_{\nu\alpha}\Big(\phi^{\eta}\bar{\phi}_{\eta\gamma}\nonumber\\
&&-\,\bar{\phi}^{\eta}\phi_{\eta\gamma}\Big) \Big]\bar{\phi}^{\gamma}\Big\rbrace+\bar{\phi}_{\nu}\Big[\bar{\phi}^{\alpha}\bar{\phi}^{\eta}\phi_{\eta\gamma}{\phi_{\alpha}}^{\gamma}-2X\Phi\bar{\Phi}+\phi^{\alpha}\Big\lbrace 2\bar{\phi}_{\alpha\eta}\bar{\phi}^{\eta}\Phi+2\phi_{\alpha\eta}\bar{\phi}^{\eta}\bar{\Phi}\nonumber\\
&&+\,\bar{\phi}_{\eta\gamma}\left(\phi^{\eta}\bar{\phi}_{\alpha}^{\;\;\gamma}-2\bar{\phi}^{\eta}{\phi_{\alpha}}^{\gamma}\right)\Big\rbrace \Big]\Big) +\phi_{\nu}\Big\lbrace 2X\left( \bar{\phi}_{\mu\alpha}\phi^{\alpha}+\phi_{\mu\alpha}\bar{\phi}^{\alpha}\right) \bar{\Phi}-\bar{\phi}^{\alpha}\Big( 2X\bar{\phi}_{\mu\eta}{\phi_{\alpha}}^{\eta}\nonumber\\
&&+\,\phi_{\mu\alpha}\bar{\phi}^{\eta}\phi_{\eta\gamma}\bar{\phi}^{\gamma}\Big)-\phi^{\alpha}\left( 2X\bar{\phi}_{\alpha\eta}\bar{\phi}_{\mu}{}^{\eta}+\left[ 3\phi_{\mu\eta}\bar{\phi}^{\eta}\bar{\phi}_{\alpha\gamma}+\bar{\phi}_{\mu\alpha}\left( -\bar{\phi}^{\eta}\phi_{\eta\gamma}+\phi^{\eta}\bar{\phi}_{\eta\gamma}\right) \right]\bar{\phi}^{\gamma}\right)\nonumber\\
&&+\,2\phi_{\mu}\bar{\phi}^{\eta}\left[ \bar{\phi}^{\alpha}\left(-\phi_{\alpha\eta}\bar{\Phi}+\bar{\phi}_{\eta\gamma}{\bar{\phi}_{\alpha}}^{\gamma}\right)+\phi^{\alpha}\left( -\bar{\phi}_{\alpha\eta}\bar{\Phi}+\bar{\phi}_{\eta\gamma}\bar{\phi}_{\alpha}{}^{\gamma}\right) \right]+\bar{\phi}_{\mu}\Big[-2X \Phi\bar{\Phi}+2X\bar{\phi}_{\alpha\eta}\phi^{\alpha\eta}\nonumber\\
&&+\,\bar{\phi}^{\alpha}\bar{\phi}^{\eta}\phi_{\eta\gamma}{\phi_{\alpha}}^{\gamma}+\phi^{\alpha}\big\lbrace 2\bar{\phi}_{\alpha\eta}\bar{\phi}^{\eta}\Phi+2\phi_{\alpha\eta}\bar{\phi}^{\eta} \bar{\Phi}+\bar{\phi}_{\eta\gamma}\big(-2\bar{\phi}^{\eta}{\phi_{\alpha}}^{\gamma}+\phi^{\eta}\bar{\phi}_{\alpha}{}^{\gamma}\big) \big\rbrace \Big]\Big\rbrace\nonumber\\
&&-\,g_{\mu\nu}\Big\lbrace 2X\bar{\phi}^{\alpha}\phi_{\alpha\eta}\bar{\phi}^{\eta}\Phi+2\phi^{\alpha}\Big[+\phi_{\alpha\eta}\bar{\phi}^{\eta}\big(X\bar{\Phi}-\bar{\phi}^{\gamma}\phi_{\gamma\psi}\bar{\phi}^{\psi}\big)-2\phi^{\eta}\phi_{\alpha\gamma}\bar{\phi}^{\gamma}\bar{\phi}_{\eta\psi}\bar{\phi}^{\psi}\nonumber\\
&&+\,\bar{\phi}_{\alpha\eta}\big\lbrace X\bar{\phi}^{\eta}\Phi+\phi^{\eta}\big(X \Phi-\phi^{\gamma}\bar{\phi}_{\gamma\psi}\bar{\phi}^{\psi}\big) \big\rbrace \Big] \Big\rbrace , \label{eq.Cov2}
\end{eqnarray}
\end{subequations}
\end{allowdisplaybreaks}
whereas for Eq.~(\ref{scalareq}) we need to specify the following terms:
\begin{allowdisplaybreaks}
\begin{subequations}\label{eqs.app2}
\begin{eqnarray}
a&=& 2\big(R^{\mu\nu}\bar{\phi}_{\mu\nu}+\bar{\phi}^{\mu}\nabla_{\nu}{{R_{\mu}}^{\nu}}\big)-\nabla_{\mu}\big(R\bar{\phi}^{\mu}\big),\\
b&=&X R^{\mu\nu} \bar{\phi}_{\nu\mu}+2\Phi \bar{\Phi}^{2}+4 \phi^{\mu\nu}\left(\bar{\phi}_{\nu\gamma} \bar{\phi}_{\mu}{}^{\gamma}-\bar{\phi}_{\nu\mu} \bar{\Phi}\right)-2\Phi \bar{\phi}_{\nu\gamma} \bar{\phi}^{\nu\gamma}+\bar{\phi}^{\mu}\Big[X {R^{\nu}}_{\mu\; ; \nu}+2\bar{\phi}^{\nu}\Big({R_{\mu}}^{\gamma} \phi_{\nu\gamma}\nonumber\\
&&-\,R_{\mu\nu} \Phi+R_{\mu\gamma\nu\sigma} \phi^{\gamma\sigma}\Big)\Big]+\phi^{\mu} \bar{\phi}^{\nu} \big({R_{\nu}}^{\gamma} \bar{\phi}_{\mu\gamma}+{R_{\mu}}^{\gamma} \bar{\phi}_{\nu\gamma}-3 R_{\mu\nu} \bar{\Phi} -R_{\mu\nu\; ; \gamma} \bar{\phi}^{\gamma}+2 R_{\mu\gamma\nu\sigma} \bar{\phi}^{\gamma\sigma}\big),\\
c&=&3\bar{\phi}^{\mu}\bar{\phi}^{\nu}\big[\phi_{\mu\nu}\big(\Phi\bar{\Phi}-\bar{\phi}_{\psi\rho}\phi^{\rho\psi}\big) -\bar{\phi}_{\nu\psi}{\phi_{\mu}}^{\psi}\Phi+{\phi_{\mu}}^{\psi}\big(\bar{\phi}{}_{\psi\rho}{\phi{}_{\nu}}^{\rho}+\bar{\phi}{}_{\nu\rho}{\phi_{\psi}}^{\rho}-\phi_{\nu\psi}\bar{\Phi}\big)\big]+X\Big[2 {R_{\nu}}^{\psi}\phi^{\mu}\bar{\phi}^{\nu}\bar{\phi}_{\mu\psi}\nonumber\\
&&+\,2{R_{\mu}}^{\psi}\bar{\phi}^{\mu}\bar{\phi}^{\nu}\phi_{\mu\psi}+\Phi\bar{\Phi}^{2}-R_{\mu\nu}\big(\bar{\phi}^{\mu}\Phi+\phi^{\mu}\bar{\Phi}\big)\bar{\phi}^{\nu}-3\bar{\phi}_{\mu\nu}\phi^{\mu\nu}\bar{\Phi}+{{\bar{\phi}_{\mu}}^{\;\psi}}{}_{\psi}\big(\Phi\bar{\phi}^{\mu}-\bar{\Phi}\phi^{\mu}\big)\nonumber \\ 
&&+\,2\phi^{\mu\nu}\bar{\phi}_{\nu\psi}\bar{\phi}_{\mu}{}^{\psi}-\bar{\phi}^{\mu}\bar{\phi}_{\mu\nu\psi}\phi^{\nu\psi}+\phi^{\mu}\bar{\phi}_{\mu\nu\psi}\bar{\phi}^{\nu\psi}+R_{\mu\psi\nu\rho}\big(\bar{\phi}^{\mu}\phi^{\psi\rho}+\phi^{\mu}\bar{\phi}^{\psi\rho}\big)\bar{\phi}^{\nu}\Big]\nonumber\\
&&-\,\phi^{\mu}\Big\lbrace \bar{\phi}^{\nu}\Big[\bar{\phi}_{\mu\nu\psi}\bar{\phi}^{\psi}\Phi-\bar{\phi}_{\nu\psi}{\phi_{\mu}}^{\psi}\bar{\Phi}+2\bar{\phi}_{\mu\psi}{\phi_{\mu}}^{\psi}\bar{\Phi}+\phi_{\mu\nu}\bar{\phi}^{\psi}\bar{\phi}_{\psi}{}^{\rho}{}_{\rho}+R_{\mu\nu}\bar{\phi}^{\psi}\phi_{\psi\rho}\bar{\phi}^{\rho}-\bar{\phi}^{\psi}\bar{\phi}_{\nu\psi\rho}{\phi_{\mu}}^{\rho}\nonumber \\
&&+\,3\Phi\bar{\phi}_{\nu\rho}{\bar{\phi}_{\mu}}^{\rho}-3{\phi_{\nu}}^{\psi}\bar{\phi}_{\psi\rho}\bar{\phi}_{\mu}{}^{\rho}-\bar{\phi}^{\psi}\bar{\phi}_{\mu\psi\rho}{\phi_{\nu}}^{\rho}-\bar{\phi}_{\mu\psi}\bar{\phi}_{\nu\rho}\phi^{\psi\rho}+\bar{\phi}_{\mu\nu}\big(\bar{\phi}_{\psi\rho}\phi^{\psi\rho}-2\Phi\bar{\Phi}\big)+\phi_{\mu\nu}\bar{\phi}_{\psi\rho}\bar{\phi}^{\psi\rho}\Big]\nonumber\\
&&+\,\phi^{\nu}\Big[2\bar{\phi}_{\nu\psi}\bar{\phi}_{\mu}{}^{\psi}\bar{\Phi}+\bar{\phi}^{\psi}\bar{\phi}_{\nu\psi\rho}\bar{\phi}_{\mu}{}^{\rho}-\bar{\phi}_{\mu\nu\psi}\bar{\phi}^{\psi}\bar{\Phi}-\bar{\phi}_{\mu\psi}\bar{\phi}^{\psi}\bar{\phi}_{\nu}{}^{\rho}{}_{\rho}-2{\bar{\phi}_{\mu}}^{\,\psi}\bar{\phi}_{\psi\rho}\bar{\phi}_{\nu}{}^{\rho}+\bar{\phi}^{\psi}\bar{\phi}_{\mu\nu\rho}\bar{\phi}_{\psi}{}^{\rho}\nonumber \\
&&\,+\bar{\phi}_{\mu\nu}\big(R_{\psi\rho}\bar{\phi}^{\psi}\bar{\phi}^{\rho}+\bar{\phi}_{\psi\rho}\bar{\phi}^{\psi\rho}-\bar{\Phi}^{2}\big)+2R_{\nu\psi\rho\epsilon}\bar{\phi}^{\psi}\bar{\phi}^{\rho}\bar{\phi}_{\mu}{}^{\epsilon}\Big]\Big\rbrace ,\\
d&=&X\Big\lbrace \bar{\phi}^{\mu}\bar{\phi}^{\nu}\big[{\phi_{\mu}}^{\psi}\big(\bar{\phi}_{\psi\rho}{\phi_{\nu}}^{\rho}-\phi_{\nu\psi}\bar{\Phi}\big)+\phi_{\mu\nu}\big(\Phi\bar{\Phi}-\bar{\phi}_{\psi\rho}\phi^{\psi\rho}\big)\big]+\phi^{\mu}\Big[2\bar{\phi}^{\nu}{\phi_{\nu}}^{\psi}\big(\bar{\phi}_{\psi\rho}{\bar{\phi}_{\mu}}^{\;\rho}-\bar{\phi}_{\mu\psi}\bar{\Phi}\big)\nonumber\\
&&+\,\phi^{\nu}\bar{\phi}_{\mu}{}^{\psi}\left(\bar{\phi}_{\psi\rho}{\bar{\phi}_{\nu}}^{\;\rho}-\bar{\phi}_{\nu\psi}\bar{\Phi}\right)+\bar{\phi}_{\mu\nu}\bar{\phi}^{\nu}\big(\Phi\bar{\Phi}-\bar{\phi}_{\psi\rho}\phi^{\psi\rho}\big)\Big] \Big\rbrace-\phi^{\mu}\bar{\phi}^{\psi}\bar{\phi}^{\rho}\Big\lbrace \phi^{\nu}\big[\bar{\phi}_{\mu\psi}\bar{\phi}_{\nu\rho}\Phi\nonumber\\
&&-\,\bar{\phi}_{\nu\psi}\bar{\phi}_{\rho\epsilon}{\phi_{\mu}}^{\epsilon}+\phi_{\psi\rho}\bar{\phi}_{\nu\epsilon}\bar{\phi}_{\mu}{}^{\epsilon}+\phi_{\mu\psi}\bar{\phi}_{\rho\epsilon}\bar{\phi}_{\nu}{}^{\epsilon}-2\bar{\phi}_{\mu\rho}\bar{\phi}_{\nu\epsilon}{\phi_{\psi}}^{\epsilon}-\bar{\phi}_{\mu\nu}\big(\phi_{\psi\rho}\bar{\Phi}-\bar{\phi}_{\rho\epsilon}{\phi_{\psi}}^{\epsilon}\big)\big]\nonumber\\
&&-\,\bar{\phi}^{\nu}\big[\phi_{\nu\psi}\big(\bar{\phi}_{\rho\epsilon}{\phi_{\mu}}^{\epsilon}+\bar{\phi}_{\mu\rho}\Phi\big)+\big(\bar{\phi}_{\mu\rho}{\phi_{\nu\epsilon}}-\phi_{\mu\nu}\bar{\phi}_{\rho\epsilon}\big){\phi_{\psi}}^{\epsilon}\big]\Big\rbrace+\phi^{\mu}\bar{\phi}_{\nu\mu}\phi^{\nu}\phi^{\psi}\bar{\phi}^{\rho}\big(\bar{\phi}_{\psi\rho}\bar{\Phi}-\bar{\phi}_{\rho\epsilon}\bar{\phi}_{\psi}{}^{\epsilon}\big).\label{eq.app2.Cov2}
\end{eqnarray}
\end{subequations}
\end{allowdisplaybreaks}

As usual $g_{\mu\nu}$ is the metric tensor, $\bar{\phi}$ denotes the complex conjugate of the scalar field, $X\equiv\phi_{\mu}\bar{\phi}^{\mu}$ is the kinetic term, and we have introduced: $\phi_{\mu}\equiv\nabla_{\mu}\phi$, $\phi_{\mu\nu}\equiv\nabla_{\mu}\nabla_{\nu}\phi$, $\Phi\equiv{\phi^{\alpha}}_{\alpha}$, $\bar{\Phi}\equiv{\bar{\phi}^{\alpha}}_{\;\;\alpha}$, $\phi^{2}\equiv\phi_{\mu} \phi^{\mu}$, and $\bar{\phi}^{2}\equiv\bar{\phi}_{\mu} \bar{\phi}^{\mu}$. We obtained these equations using~\emph{xAct}, an open-source Mathematica package for abstract tensor calculus developed by J.~M.~Martin-Garcia. It is available under a GNU Public Licence from~\cite{xact}.

Finally, the deviations with respect the static and spherically symmetric EKG system of Eqs.~(\ref{eq.motion.spherical}) are parametrized in terms of the dimensionless functions
\begin{allowdisplaybreaks}
\begin{subequations}\label{eq.parametersEKG}
\begin{eqnarray}
M_{\textrm{Pl}}\Lambda_3^3\alpha &=&  2c_4^{(1)}\left(3\frac{\sigma'^2}{g^2}-\frac{\omega^2\sigma^2}{N^2}\right)+2d_4^{(-1)}\frac{1}{X^2}\left(3\frac{\sigma'^4}{g^4}-6\frac{\sigma'^2}{g^2}\frac{\omega^2\sigma^2}{N^2}-\frac{\omega^4\sigma^4}{N^4}\right)\frac{\sigma'^2}{g^2},\\
M_{\textrm{Pl}}\Lambda_3^3\beta &=& 2c_4^{(1)}\left(\frac{\sigma'^2}{g^2}+\frac{\omega^2\sigma^2}{N^2}\right),\\
\frac{\Lambda_3^3}{M_{\textrm{Pl}}} r^2g^2 \gamma_1 &=& 4c_4^{(1)}+6d_4^{(-1)}\frac{1}{X^2}\frac{\sigma'^4}{g^4},\\
\frac{\Lambda_3^3}{M_{\textrm{Pl}}} r^2g^2\gamma_2 &=& 4c_4^{(1)}+2d_4^{(-1)}\frac{1}{X^2}\frac{\sigma'^4}{g^4}+ 8\left(c_4^{(1)}+d_4^{(-1)}\frac{1}{X^2}\frac{\sigma'^4}{g^4}\right)\frac{r\sigma''}{\sigma'},\\
\frac{\Lambda_3^3}{M_{\textrm{Pl}}} r^2g^2\delta_1 &=& \left(8c_4^{(1)}+12d_4^{(-1)}\frac{1}{X}\frac{\sigma'^2}{g^2}\right)\frac{r\sigma'}{\sigma}-4d_4^{(-1)}\frac{1}{X}\frac{\sigma'^2}{g^2}\frac{r\sigma''}{\sigma'}\nonumber\\
&&- 2d_4^{(-1)}\frac{1}{X^2}\left(4\frac{\sigma'^2}{g^2}-\frac{\omega^2\sigma^2}{N^2}\right)\frac{\sigma'^2}{g
^2},\\
\frac{\Lambda_3^3}{M_{\textrm{Pl}}} r^2g^2\delta_2 &=&
-4d_4^{(-1)}\frac{1}{X^2}\left(5\frac{\sigma'^2}{g^2}+\frac{\omega^2\sigma^2}{N^2}\right)\frac{\sigma'^2}{g^2}\frac{r\sigma''}{\sigma'}\nonumber\\
&&+4d_4^{(-1)}\frac{1}{X^2}\left(3\frac{\sigma'^2}{g^2}-\frac{\omega^2\sigma^2}{N^2}\right)\frac{\sigma'^2}{g^2}\frac{r\sigma'}{\sigma}-2d_4^{(-1)}\frac{1}{X^2}\frac{\omega^2\sigma^2}{N^2}\frac{\sigma'^2}{g
^2},\\
\frac{\Lambda_3^3}{M_{\textrm{Pl}}} r^2g^2\varepsilon &=& 
2c_4^{(1)}\left(2\frac{rN'}{N}+(1-g^2)\right)+2d_4^{(-1)}\frac{1}{X^3}\left[\left(1+2\frac{rN'}{N}\right)\frac{\sigma'^6}{g^6}\right.\nonumber\\
&&\left.-\left(4+5\frac{rN'}{N}-\frac{rg'}{g}-2\frac{r\sigma'}{\sigma}\right)\frac{\sigma'^4}{g^4}\frac{\omega^2\sigma^2}{N^2}+\left(3+18\frac{rN'}{N}-8\frac{r\sigma'}{\sigma}\right)\frac{\sigma'^2}{g^2}\frac{\omega^4\sigma^4}{N^4}\right.\nonumber\\
&&\left.+\left(\frac{rN'}{N}-\frac{rg'}{g}+6\frac{r\sigma'}{\sigma}\right)\frac{\omega^6\sigma^6}{N^6}\right],\\
\frac{\Lambda_3^3}{M_{\textrm{Pl}}} r^2g^2\zeta &=& 
2c_4^{(1)}\left(\frac{r^2N''}{N}-3\frac{rN'}{N}\frac{rg'}{g} \right)-2d_4^{(-1)}\frac{1}{X^3}\left(\frac{\sigma'^4}{g^4}-\frac{\omega^4\sigma^4}{N^4}\right)\frac{\omega^2\sigma^2}{N^2}\frac{r^2\sigma'^2}{\sigma^2}, \nonumber\\
&&-d_4^{(-1)}\frac{1}{X^3}\left[\left(5-14\frac{rg'}{g}\right)\frac{\sigma'^4}{g^4}-4\left(2-\frac{rg'}{g}\right)\frac{\sigma'^2}{g^2}\frac{\omega^2\sigma^2}{N^2}+3\left(1-2\frac{rg'}{g}\right)\frac{\omega^4\sigma^4}{N^4}\right]\frac{\omega^2\sigma^2}{N^2}\frac{r\sigma'}{\sigma} \nonumber\\
&& +d_4^{(-1)}\frac{1}{X^3}\left\{2\left(3\frac{rN'}{N}\frac{rg'}{g}-\frac{r^2N''}{N}\right)\frac{\sigma'^ 6}{g^6} \right.\nonumber\\
&&\left.+\left[4\frac{rg'}{g}-2\left(\frac{r^2N'^2}{N^2}+\frac{r^2g'^2}{g^2}\right)-20\frac{rN'}{N}\frac{rg'}{g}+\left(7\frac{r^2N''}{N}+\frac{r^2g''}{g}\right)\right]\frac{\sigma'^4}{g^4}\frac{\omega^2\sigma^2}{N^2}\right.\nonumber\\
&&\left.-\left[4\frac{rg'}{g}-2\left(7\frac{r^2N'^2}{N^2}-3\frac{r^2g'^2}{g^2}\right)+26\frac{rN'}{N}\frac{rg'}{g}-2\left(2\frac{r^2N''}{N}+\frac{r^2g''}{g}\right)\right]\frac{\sigma'^2}{g^2}\frac{\omega^4\sigma^4}{N^4}\right.\nonumber\\
&&\left.+\left[4\frac{rg'}{g}\left(\frac{rN'}{N}-\frac{rg'}{g}\right)+\left(\frac{r^2N''}{N}-\frac{r^2g''}{g}\right)\right]\frac{\omega^6\sigma^6}{N^6}\right\}, \label{eq.zeta} \\
\frac{\Lambda_3^3}{M_{\textrm{Pl}}} r^2g^2\eta &=& 
2c_4^{(1)}(3-g^2)+2d_4^{(-1)}\frac{1}{X^3}\left(7\frac{\sigma'^4}{g^4}-12\frac{\sigma'^2}{g^2}\frac{\omega^2\sigma^2}{N^2}-3\frac{\omega^4\sigma^4}{N^4}\right)\frac{\omega^2\sigma^2}{N^2}\frac{r\sigma'}{\sigma}\nonumber\\
&&-2d_4^{(-1)}\frac{1}{X^3}\left(3\frac{\sigma'^6}{g^6}-\frac{\sigma'^4}{g^4}\frac{\omega^2\sigma^2}{N^2}-7\frac{\sigma'^2}{g^2}\frac{\omega^4\sigma^4}{N^4}+\frac{\omega^6\sigma^6}{N^6}\right), \\
\frac{\Lambda_3^3}{M_{\textrm{Pl}}} r^2g^2\theta &=& 
2c_4^{(1)}\left(2\frac{rg'}{g}-(1-g^2)\right).
\end{eqnarray}
\end{subequations}
\end{allowdisplaybreaks}

\section{Higher order operators}\label{app.higher.order}

\begin{figure}[t]
	\scalebox{0.445}{
	\input{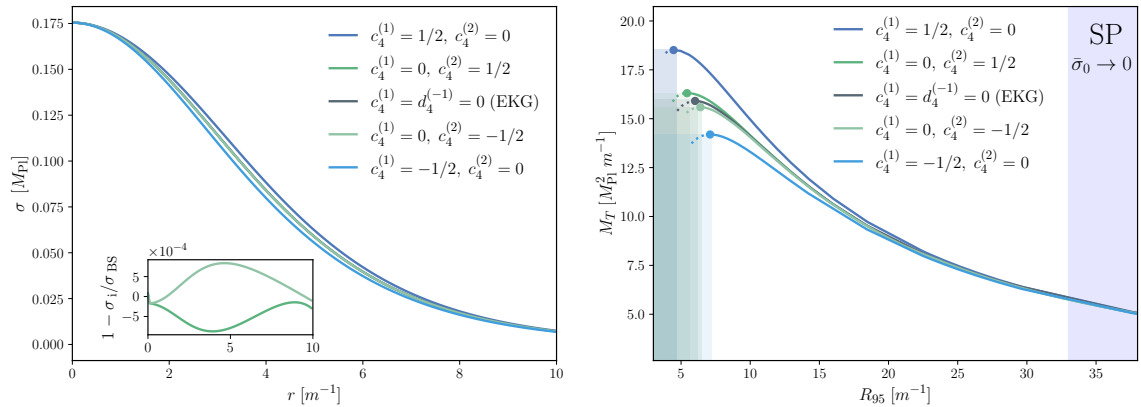}
	}
	\caption{{\bf The validity of the EFT.} Similar as in Figures~\ref{Fig1}, left panel, and~\ref{MvswandR}, right panel, but for a model where the first relevant operator is the quadratic covariant galileon $c_4^{(2)}$. As in the previous figures we have fixed the amplitude and the EFT scale to $\sigma_0=0.175M_{\textrm{Pl}}$ and $\Lambda_3=1.5M_{\textrm{Pl}}^{1/3}m^{2/3}$, respectively. For reference we have also included the $c_4^{(1)}=1/2$, $c_4^{(2)}=0$, the $c_4^{(1)}=c_4^{(2)}=0$ (EKG), and the $c_4^{(1)}=-1/2$, $c_4^{(2)}=0$ cases. Notice that the differences with respect to the BS are less evident when $c_4^{(1)}=0$. For simplicity we have ignored the operators that are beyond Horndeski, $d_4^{(i)}=0$.
	\label{fig.higher_order}}
\end{figure} 

If $c_4^{(1)}=d_4^{(-1)}=0$, the second line of Eq.~(\ref{eq.interactions}) vanish identically and it is necessary to write this expression to the next order in the expansion series. This is particularly important as we approach the strong field scale, Eq.~(\ref{scale.self}). If for simplicity we assume a discrete $\mathbb{Z}_2$ mirror symmetry, we obtain the following expression that includes the next-to-leading order terms:\footnote{Previous to the two operators in the second line there is a term in the quadratic sector of Horndeski that is of mass dimension eight and dominates at low energies. However, it does not modify gravity and for that reason we have not included it here by setting $c_2^{(2)}=0$.}
\begin{eqnarray}
 \mathcal{L}_{\textrm{grav}} &=& \frac{1}{2}M_{\textrm{Pl}}^2 R - X -m^2\phi\bar{\phi} \\
 && + \frac{1}{\Lambda_3^6}\left[c_4^{(2)}X^2R - 4c_4^{(2)}X[\Box\phi\Box\bar{\phi}-\phi^{\mu\nu}\bar{\phi}_{\mu\nu}]
 +d_4^{(0)}{\epsilon^{\mu\nu\rho}}_{\sigma}\epsilon^{\mu'\nu' \rho'\sigma}\phi_{\mu}\bar{\phi}_{\mu'}\phi_{\nu\nu'}\bar{\phi}_{\rho\rho'}
 \right]+ \ldots. \nonumber
\end{eqnarray}
As in~(\ref{eq.interactions}), the low energy regime is parametrized in terms of two coupling constants, $c_4^{(2)}$ and $d_4^{(0)}$. The first operator (together with the Einstein-Hilbert term) is the quartic covariant galileon~\cite{Deffayet:horndeski} [c.f. Eq.~(\ref{eq.covariant.galileons}) with $c_4=8c_4^{(2)}$], whereas the second one is in the quartic sector of the beyond Horndeski theory. With no loss of generality we can absorb the coupling constant $c_4^{(2)}$ into the scale $\Lambda_3$ and fix it to $c_4^{(2)}=\pm 1/2$, whereas the value of $d_4^{(0)}$ remains arbitrary. Moreover, the choices $c_4^{(2)}= 1/2$, $d_4^{(0)}=2$ and $c_4^{(2)}= -1/2$, $d_4^{(0)}= -2$ fix the speed of propagation of GWs equal to that of light, although we do not explore this scenario here where for simplicity we set all the parameters beyond Horndeski equal to zero, $d_4^{(i)}=0$.

Figure~\ref{fig.higher_order} is analogous to Figures~\ref{Fig1}, left panel, and~\ref{MvswandR}, right panel, where for the same amplitude $\sigma_0=0.175M_{\textrm{Pl}}$ and EFT scale $\Lambda_3=1.5M_{\textrm{Pl}}^{1/3}m^{2/3}$ as in those figures, we show the profile of the wave function, $\sigma(r)$, left panel, and the $\bar{M}_T$ vs $\bar{R}_{95}$ plot, right panel, for different realizations of the effective model. On the one hand, we have considered the cases $c_4^{(1)}=1/2$, $c_4^{(2)}=0$ and $c_4^{(1)}=-1/2$, $c_4^{(2)}=0$, which actually were also reported in Figures~\ref{Fig1} and~\ref{MvswandR}. Nonetheless, we have extended the analysis by including the models with $c_4^{(1)}=0$, $c_4^{(2)}=1/2$ and $c_4^{(1)}=0$, $c_4^{(2)}=-1/2$, in such a way that the first non-trivial contribution to the action comes from the operator $c_4^{(2)}$. For completeness, we have also included the EKG model as reference. As we can appreciate from the figures when $c_4^{(1)}=0$ the deviations with respect to the BS are smaller. This is in no sense surprising given that we are dealing with operators of mass dimension ten. This also guarantees that to lowest order we can work with the operators $c_4^{(1)}$ and $d_4^{(-1)}$ alone, as long as neither is different from zero.

\section{The Schr\"odinger-Poisson limit}\label{app.SP}

\begin{figure}
\centering	
	\scalebox{0.45}{
	\input{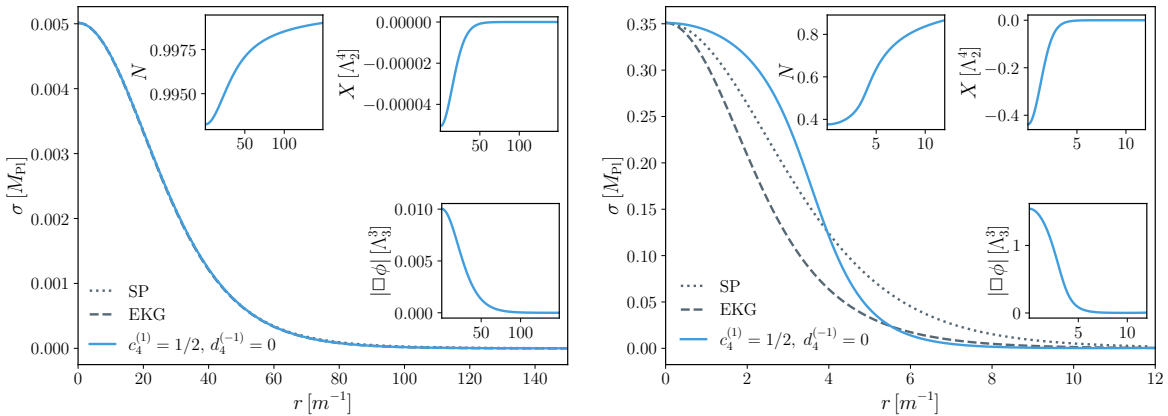}
	}
\caption{{\bf The Schr\"odinger-Poisson limit.} 
The profile of the wave function $\sigma(r)$, Eq.~(\ref{ansat}), for two different amplitudes in a model with the low energy limit of Horndeski, $c_4^{(1)}=1/2$, $d_4^{(-1)}=0$. The central field was chosen to $\sigma_0=5\times 10^{-3}M_{\textrm{Pl}}$ (left panel), and $\sigma_0=0.35M_{\textrm{Pl}}$ (right panel), and we have fixed the scale of the EFT to $\Lambda_3=0.8M_{\textrm{Pl}}^{1/3}m^{2/3}$. When the amplitude is low the higher derivative operators are negligible and we obtain the same result as with the EKG model, which actually can be captured in terms of the SP approximation, left panel. As we increase the central amplitude, however, and since we are dealing with a model in scenario $ii$ of Section~\ref{subsec.eq.motion}, both the scalar and the tensor modes approach the strongly coupled regime as the central amplitude approaches the Planck mass, and not only the SP system, but also the EKG one, stop working, right panel. The inner panels show the behaviour of the lapse function $N$, the kinetic term $X$, and the ``boxed field'' $\Box\phi$. Since $\Lambda_3\sim M_{\textrm{Pl}}^{1/3}m^{2/3}$ both $X/\Lambda_2^4$ and $\Box\phi/\Lambda_3^3$ become of order one simultaneously and a more careful study requires a model-dependent analysis.}
\label{fig.SP}
\end{figure}	

In the infrared Eqs.~(\ref{eq.motion.spherical}) reduce to the more familiar SP system~\cite{Guzman:2003kt}:
\begin{subequations}\label{eq.SP}
\begin{eqnarray}
   & \Delta \mathcal{U} = 4\pi G \rho,\\
   & E\sigma = -\frac{1}{2m}\Delta\sigma + m\mathcal{U}\sigma,
\end{eqnarray}
\end{subequations}
where $\Delta f \equiv f''+\frac{2}{r}f'$ is the Laplace operator in spherical symmetry and $\rho \equiv 2m^2\sigma^2$ denotes the mass density of scalar particles. These expressions appear in the non-relativistic, $\omega=m+E$, weak field $N^2(r) = 1+2\mathcal{U}(r)$, $g^2(r) = 1+2r\mathcal{U}'(r)$ limit of Eqs.~(\ref{eq.motion.spherical}), where $E\ll m$ represents the energy of the individual particles (once we have removed the contribution from the rest mass) and $\mathcal{U}(r)\ll 1$ is the Newtonian gravitational potential. We have also assumed that $\sigma_0\ll \sigma_0^{\textrm{s.f.}}$, Eq.~(\ref{scale.self}), so that we can neglect higher derivative operators.

Regularity conditions at the origin demand [c.f. Eqs.~(\ref{boundayC})]:
\begin{subequations}\label{}
\begin{align}
    & \sigma(r=0)=\sigma_0,\hspace{.53cm} \sigma'(r=0)=0,\\
    &\mathcal{U}(r=0)=\mathcal{U}_0,\hspace{0.5cm} \mathcal{U}'(r=0)=0.
\end{align}
\end{subequations}
If in addition we want to obtain localized configurations that remain bounded in space we need to impose further that [c.f. Eq.~(\ref{bounday2})] $\sigma(r\to \infty)=0$ and $\mathcal{U}(r\to \infty)=\mathcal{U}_{\infty}$. In analogy to the lapse function (see the discussion of the last paragraph of Section~\ref{numerical}), the gravitational potential can be fixed arbitrarily to one at the origin, $\mathcal{U}_0=1$, and then shifted to $\mathcal{U}(r)\mapsto\mathcal{U}(r)-\mathcal{U}_\infty$, $E\mapsto E-\mathcal{U}_\infty$ at the end of the calculation, in such a way that it goes to zero at spatial infinity, where $\mathcal{U}_\infty$ is the value of the potential at infinity before performing the transformation.

These equations are invariant under the scaling transformation (see e.g.~\cite{Guzman:2003kt}): 
\begin{equation}\label{eq.scaling}
    r\to\lambda^{-1}r,\quad E\to \lambda^2E,\quad \mathcal{U}\to \lambda^2\mathcal{U},\quad \sigma\to\lambda^2\sigma,
\end{equation}
where $\lambda$ is an arbitrary dimensionless parameter greater than zero. This allows us to solve Eqs.~(\ref{eq.SP}) for a given central amplitude, $\sigma_0^{(1)}$, and then rescale the solution to an arbitrary value of the field at the origin, $\sigma_0^{(1)}\mapsto\sigma_0^{(2)}$. In this way the profile of the wave function can be expressed in the form
\begin{equation}\label{eq.profile.sigma}
    \sigma(r) = \sigma_0 F\left[\frac{\sigma_0^{1/2}m}{M_{\textrm{Pl}}^{1/2}}r\right],
\end{equation}
where $F(r)$ is the solution to Eqs.~(\ref{eq.SP}) for $M_{\textrm{Pl}}=m=1$ and unit central field, $\sigma_0=1$. This function can be computed numerically using the shooting method~\cite{Numerical, Dias:2015nua} (see e.g.~\cite{Schive:2014dra} for a polynomial fitting). Using the solution in Eq.~(\ref{eq.profile.sigma}), one obtains the following analytical expressions for the total mass, $M_T$, and the effective radius, $R_{95}$, as a function of the central amplitude, $\sigma_0$:
\begin{equation}
    M_{T} = 54.6 \frac{M_{\textrm{Pl}}^{3/2}\sigma_0^{1/2}}{m},\quad R_{95} = 3.9\frac{M_{\textrm{Pl}}^{1/2}}{m\sigma_0^{1/2}},
\end{equation}
where we have used the same notation as in the main text. From these expressions one can identify the relation $\bar{M}_T\approx 213/\bar{R}_{95}$ that describes the asymptotic behaviour of the $\bar{M}_T$ vs $\bar{R}_{95}$ plots in the limit where $\bar{\sigma}_0\to0$, toward which all the curves converge in the right panel of Figure~\ref{MvswandR}. Furthermore, the compactness, as defined in Eq.~(\ref{eq.compacness}), grows up to order unity as we approach $\sigma_0\sim M_{\textrm{Pl}}$ (where we have omitted some numerical factors of order one), when the tensor modes couple strongly and the SP approximation lose predictability. 

This behaviour is illustrated in Figure~\ref{fig.SP}, where for a model in which the EFT scale is chosen of order $M_{\textrm{Pl}}^{1/3}m^{2/3}$, scenario $ii$ in Section~\ref{subsec.eq.motion}, we plot the profile of the wave function for two different values of the central field: one which is much lower than the Planck mass, and the other that starts to approximate this mass scale. When the amplitude is low, left panel, the higher derivative operators are negligible, and solving numerically the full system of Eqs.~(\ref{eq.motion.spherical}) results in essentially the same profile as the one that we obtain by solving the much simpler EKG system, which actually can be captured in terms of the SP approximation, Eqs.~(~\ref{eq.SP}). This is because in this limit not only the higher derivative operators, but also the higher order terms e.g. $M_{\textrm{Pl}}^{-n}h^2(\partial h)^2$ that are codified in the Ricci scalar are also small. As we increase the central amplitude, however, the higher order operators turn on. Nonetheless, and since we are dealing with a theory in scenario~$ii$, both the scalar and the tensor modes couple strongly as we approach the central amplitude to the Planck scale, $\sigma_0^{\textrm{\textrm{s.f.}}}\sim\sigma_0^{\textrm{\textrm{s.g.}}}\sim M_{\textrm{Pl}}$, and not only the SP system, but also the EKG one, stop reproducing the right physics, right panel.

If $\Lambda_3\gg M_{\textrm{Pl}}^{1/3}m^{2/3}$, scenario $i$, the state of maximum mass $\sigma_0\sim \sigma_0^{\textrm{s.g.}}\sim M_{\textrm{Pl}}$ that divides the solution curve into the stable and unstable branches appears before we can appreciate any effect of the higher derivative operators, expected at $\sigma_0\sim \sigma_0^{\textrm{s.f.}}\sim M_{\textrm{Pl}}^{1/2}\Lambda_3^{3/2}/m$, and for all practical purposes we can always use the EKG equations. If $\Lambda_3\ll M_{\textrm{Pl}}^{1/3}m^{2/3}$, scenario $iii$, the higher derivative operators turn on earlier, when $\sigma_0\sim \sigma_0^{\textrm{s.f.}}\sim \Lambda_3^3/m^2$, a scale that is much lower than that of strong gravity, $\sigma_0^{\textrm{s.g}}\sim M_{\textrm{Pl}}$, and then the SP approximation breaks down before we can reach the relativistic EKG regime.

\bibliographystyle{jhepmod}

\bibliography{Bibliografia.bib} 


\end{document}